\documentclass[12pt]{article}
\usepackage{graphicx}
\usepackage{hyperref}

\usepackage{latexsym,amssymb,amsmath,epsfig}
\usepackage[usenames,dvipsnames]{pstricks}
\usepackage{epsfig}
\usepackage{epstopdf} 
\usepackage{epstopdf}
\usepackage{graphicx}
\usepackage{pstricks}
\usepackage{epsfig}
\usepackage{float}
\usepackage{epstopdf}
\usepackage{amsmath,amssymb}
\usepackage{epsfig}
\usepackage{amsthm}
\usepackage{cite}
\usepackage{caption}
\usepackage{authblk}
\usepackage{mathtools}
\usepackage{amsmath}

\usepackage{amsthm} 
\newtheorem{theorem}{Theorem}

\usepackage[a4paper, margin=1in]{geometry}
\usepackage{changepage}

\usepackage{tabularx,ragged2e,booktabs,caption}
\newcolumntype{C}[1]{>{\Centering}m{#1}}

\usepackage{lscape}
\usepackage{pdflscape}
\usepackage{float,lscape}
\usepackage{qtree}

\author[1,2]{Ali Raza}
\author[3]{F.M. Mahomed}
\author[4]{F.D. Zaman}
\author[4]{A.H. Kara}
\affil[1]{Centre for Mathematics and Statistical Sciences, Lahore School of Economics, Lahore, Pakistan}
\affil[2]{Department of Mathematical Sciences, Stellenbosch University, Stellenbosch, South Africa}
\affil[3]{School of Computer Science and Applied Mathematics, University of the Witwatersrand, Johannesburg, Wits 2050, South Africa}
\affil[4]{School of Mathematics, University of the Witwatersrand, Wits 2050, South Africa}

\date{}

\begin{document}

\title{General Classification, Invariance and Conservation Laws Analyses of  Nonlinear Fourth Order Wave and Nerve Membrane Equations with Dissipation}
\maketitle


\begin{abstract} 
We study the nonlinear wave equation for arbitrary function with fourth order dissipation. A special case that is analysed exclusively is the model of nerve membranes; we consider this model, both, in the presence and absence of the fourth order dissipation. The equivalence transformations, Lie symmetries and a complete classification is presented. We also discuss the one dimensional optimal system in each case obtained via classification. The reduction of the partial differential equations (PDEs) is carried out and the forms of invariant solutions are presented. The study also include the construction of conservation laws using the direct method. The invariant solutions and some special type of solutions including solitions are presented with their graphical illustrations.   
\end{abstract}


\section{Introduction} \label{sec_a0}
A wave phenomenon is modelled and studied with a range of variations taking into account the nature of the physical models. These variations can introduce nonlinearities or supplementary extremal terms such as force or damping, which may exhibit nonlinear or high-order properties. Some early studies using the Lie symmetry approach was undertaken by Ovsiannikov \cite{st1}, N.H. Ibragimov \cite{st2}, Ames et. al. \cite{s1}, G.W. Bluman and S. Kumei \cite{st3} among others. In a recent work, the study of one dimensional wave equation was extended by A. Raza et. al. \cite{s2} in which presented a class of wave equations and discussed the optimal system and higher order conservation laws.

In this study, we consider the wave equation in which the dissipation term is proportional to the fourth order derivative of $u$ with respect to $x$. This term would represent the dissipation of energy due to factors such as friction or viscosity. The resulting equation is given by 
\begin{eqnarray} \label{Eq_2}
u_{tt} = ( \Phi(u)  u_x )_x + h_1 u_{xxxx}, 
\end{eqnarray}
where $h_1$ is a constant that represents the strength of the dissipation and $\Phi(u)$ is an arbitrary function that arises in gas dynamics and $u$ represents displacement of the wave in spaces $x$ and time $t$. Dissipation terms in differential equations describe the loss of energy or dissipative processes in a system. These terms can arise in a variety of physical systems, for example mechanical systems and electrical circuits. Dissipation terms are typically modeled as a function of the system state or the rate of change of the state.

The form of the dissipation term can vary depending on the specific system being modeled and the physical processes at play. In a differential equation, a dissipative term can involve mixed derivatives, meaning that it involves both time derivatives and spatial derivatives. For example, in a partial differential equation that describes the evolution of a physical quantity over both space and time, the dissipative term may involve both spatial derivatives such as the gradient of the quantity and time derivatives such as the rate of change of the quantity over time. In order to guarantee finite phase velocity for higher frequencies by adding higher terms, equation (\ref{Eq_2}) should be modified to
\begin{eqnarray} \label{Eq_3}
u_{tt} = ( \Phi(u)  u_x )_x  + h_1 u_{xxxx} + h_2 u_{xxtt}.
\end{eqnarray}
The wave equation and its modifications are studied frequently \cite{s2,s3,s4}  and used widely to describe various physical phenomena. To study nonlinear DEs, many methods failed to provide the exact solution as finding an exact solution of a nonlinear equation is a challenging problem in nonlinear science. However, the Lie symmetry method is generally effective in finding exact solutions to nonlinear models.

A special case of the above is the Hodgkin–Huxley model  \cite{b1} which is used in the study of nerve pulse propagation particularly for studying the characteristics of biomembranes. The mechanical wave equation in terms of density change $\Delta \rho = u$ is given by 
\begin{eqnarray} \label{bio1}
u_{tt} = ( (p u^2 + qu + r) u_x)_x,
\end{eqnarray}
where velocity $\Phi(u)=c^2(u)=p u^2 + qu + r$ builds the nonlinearity because of its dependence on density $\Delta \rho = u$ and $p,q,r$ are coefficients. Heimburg and Jackson \cite{b2} proposed the model to explain longitudinal waves in biomembranes that possesses nonlinearity and higher order dispersion term given by
\begin{eqnarray} \label{bio2}
u_{tt} = ( (p u^2 + qu + r) u_x)_x - h_1 u_{xxxx}, 
\end{eqnarray}
where $r$ is the sound velocity of a membrane. The term $h_1 u_{xxxx}$ is introduced to represent the dispersion and $h_1$ is the dispersion coefficient. Engelbrecht et al. \cite{b3} further modified the model in order to guarantee finite phase velocity for higher frequencies. The modified equation was proposed to be
\begin{eqnarray} \label{bio3}
u_{tt} = ( (p u^2 + qu + r) u_x)_x - h_1 u_{xxxx} + h_2 u_{xxtt}.
\end{eqnarray}
It should be noted that the term with coefficient $h_1$ represents elastic effects, whereas the one with coefficient $h_2$ explains the inertial effects of the lipids that make up the biomembranes. Further modifications, physical properties of biomembranes, soliton solutions and other characteristic of equations (\ref{bio1}), (\ref{bio2}) and (\ref{bio3}) are studied in references \cite{b4,b5,b6,b7,b8}. 

The main focus of this study is on classifying Lie symmetries, determining one-dimensional optimal systems, invariant solutions admitted by PDEs (\ref{Eq_2}) and (\ref{Eq_3}) and constructing their local conservation laws. Lie symmetry analysis is an algebraic approach used to find symmetries of differential equations that can be used to construct optimal systems that further lead us to invariant solutions. The technique involves finding a group of transformations that leave the equation invariant. Under one parameter Lie group of transformations, the optimal system is a set of transformations that preserve the structure of the Lie algebra and forms its subalgebras. The method to construct the optimal system is discussed in detail in \cite{s5,s6,s7,s8,s9}. 

In the case of PDEs, the use of conservation laws play a role in understanding the behaviour of the system and identify any underlying physical principles at play. Conservation laws are quantities that are conserved over space and time in a physical system described by a differential equation. Many methods has been developed \cite{s10,s11,s12,s13,s14} for the construction of conservation laws, such as Noether’s theorem \cite{s13} for variational problems, Lie-Bäcklund method \cite{s13,s14} and multiplier approach  \cite{s11,s12,s2}. We adopt the multiplier approach to find conservation laws for the equation (\ref{Eq_3}).

Our main focus is the study of equation (\ref{Eq_3}) and (\ref{bio3}), rest are the particular cases arises in this study. The above equation is pertinent in the study of nerve pulse propagation particularly for studying the characteristic of bio membranes. The methodology employed is a novel one dealing with the Lie symmetry classification of the classes that arise. The work is new, innovative and has real world applications particularly nerve pulse propagation. 

In this study we present the complete Lie symmetry classification depending on smooth function $\Phi(u)$. We construct the equivalence transformations that transform the equation (\ref{Eq_3}) into simpler forms.  For constructing the family of non-equivalent symmetry generators, we deduce all the one-dimensional optimal systems admitted by equation (\ref{Eq_3}). Reduction and forms of invariant solutions are presented. Furthermore, we examine the conservation laws.

\section{ General Classification and Optimal System}  \label{sec_a1}
In this section, we study the symmetries of the class of equations (\ref{Eq_3}) and outline some cases for the arbitrary function $\Phi(u)$. Lie theory, its background, procedure and manifold applications are well described in the
literature, see e.g. \cite{s8,s9,s17}. To apply Lie symmetry analysis to the fourth order wave equation (\ref{Eq_3}), we first need to write the given PDE in the form of a system of determining equations by using invariance criterion. This can usually be done by introducing new dependent variables for the higher order derivatives of the original dependent variable. To this end, a Lie symmetry generator $X$ given by 
\begin{eqnarray}
X= \xi^1(x,t,u) \frac{\partial}{\partial x} + \xi^2(x,t,u) \frac{\partial}{\partial x} + \eta(x,t,u) \frac{\partial}{\partial u},
\end{eqnarray}
is prolonged to $X^{[4]}$ and when applied to (\ref{Eq_3}), we obtain 
\begin{eqnarray} \label{IC4th}
X^{[2]} [u_{tt} - \Phi^{'} u^2_x - \Phi u_{xx}] + X^{[4]}[-h_1 u_{xxxx} - h_2 u_{xxtt}]  
{\LARGE|_{(\ref{Eq_3})}} = 0 .
\end{eqnarray}
If the above criterion is satisfied, then $X$ is said to be a Lie symmetry of the given PDE (\ref{Eq_3}). The detailed explanations for Lie symmetries classification based on the value of arbitrary function using the standard approach is presented in the Appendix. The Lie symmetry classification based on the values of function $\Phi(u)$ provide us the following cases.      

\subsection{ $\boldsymbol{\Phi = \epsilon (u+k_1)^{k_2} }$, where $\boldsymbol{\epsilon = \pm 1}$}
The infinitesimals for this case are given by
\begin{eqnarray}
\xi^1 = c_1 + \frac{k_2}{2} c_3 t , \quad
\xi^2 = c_5, \quad
\eta = c_3 u + k_1 c_3.
\end{eqnarray}
From the infinitesimals, we obtain three symmetries which form a three dimensional algebra spanned by 
\begin{eqnarray*}
X_1 = \frac{\partial}{\partial t}, \quad
X_2 = \frac{\partial}{\partial x} , \quad
X_3 = \frac{k_2}{2} t \frac{\partial}{\partial t} + (k_1 + u )  \frac{\partial}{\partial u}.
\end{eqnarray*}
Under the equivalence transformation, function $\Phi = \epsilon (u+k_1)^{k_2}$ is simplified to ${ \Phi = \epsilon u^{k_2} }$, where $\boldsymbol{\epsilon = \pm 1}$. Equivalence transformations are given by 
\begin{eqnarray}
\bar{t}=a_{1} t+a_{2}, \ \  \ \bar{x}=b_{1} x+b_{2}, \ \  \ \bar{u}=c_{1} u +c_{2}, \\
\bar{h}_{1}=\frac{b_{1}^{2}}{a_{1}^{2}} h_{1}, \ \  \  \bar{h}_{2} =b_{1}^{2} h_{2}, \ \  \ \bar{\Phi}=\frac{b_{1}^{2}}{a^{2}} \Phi.
\end{eqnarray}
Using these transformations, we obtain the equivalent set of symmetries given by 
\begin{eqnarray*}
X_1 = \frac{\partial}{\partial t},  \quad
X_2 = \frac{\partial}{\partial x} ,  \quad
X_3 = \frac{k_2}{2} t \frac{\partial}{\partial t} + u \frac{\partial}{\partial u}.
\end{eqnarray*}

\subsection{ $\boldsymbol{\Phi = \epsilon \exp (ku) }$, where $\boldsymbol{\epsilon = \pm 1}$}
The set of infinitesimals obtained in this case takes the form given by
\begin{eqnarray}
\xi^{1} = c_{1}+c_{2} t,  \quad
\xi^{2} = c_{5} ,  \quad
\eta = c_4. 
\end{eqnarray}
And the symmetries we obtain in this case are 
\begin{eqnarray*}
X_1 = \frac{\partial}{\partial t},  \quad
X_2 = \frac{\partial}{\partial x},  \quad
X_3 = \frac{k}{2} t \frac{\partial}{\partial t} +  \frac{\partial}{\partial u}.
\end{eqnarray*}

\subsection{ $\boldsymbol{\Phi = \epsilon (u)^{k_1} }$, where $\boldsymbol{\epsilon = \pm 1}$}
In this case we receive the set of infinitesimals given by 
\begin{eqnarray}
\xi^1 = -k_{1} c_{4} t+c_{2} ,  \quad
\xi^{2} = -\frac{1}{2} k_{1} c_{4} x + c_{3},  \quad
\eta = c_{4} u + k_{2}.
\end{eqnarray}
Further, we obtain the set of symmetries given by 
\begin{eqnarray*}
X_1 = \frac{\partial}{\partial t},  \quad
X_2 = \frac{\partial}{\partial x},  \quad
X_3 = -k_1 t \frac{\partial}{\partial t}  - \frac{1}{2} k_1 x \frac{\partial}{\partial x} + (u+k_2) \frac{\partial}{\partial u}.
\end{eqnarray*}
By equivalence transformation, symmetries simplify to  
\begin{eqnarray*}
X_1 =\frac{\partial}{\partial t},  \quad
X_2 = \frac{\partial}{\partial x} ,  \quad
X_3 = -k_1 t \frac{\partial}{\partial t}  - \frac{1}{2} k_1 x \frac{\partial}{\partial x} + u \frac{\partial}{\partial u}.
\end{eqnarray*}
\subsection{ $\boldsymbol{\Phi = \epsilon \exp (u) }$, where $\boldsymbol{\epsilon = \pm 1}$}
For this case, the set of infinitesimals is given by 
\begin{eqnarray}
\xi^1 = -k c_{5} t + c_{2},  \quad
\xi^{2} = -\frac{1}{2} k c_5 x + c_{3},  \quad
\eta = c_5 .
\end{eqnarray}
This provides us the set of symmetries 
\begin{eqnarray*}
X_1 = \frac{\partial}{\partial t},  \quad
X_2 = \frac{\partial}{\partial x},  \quad
X_3 =-k t \frac{\partial}{\partial t}  - \frac{1}{2}  x \frac{\partial}{\partial x} +  \frac{\partial}{\partial u}.
\end{eqnarray*}

\subsection{ $\boldsymbol{\Phi^{\prime} = 0}$}
The infinitesimals in this case are 
\begin{eqnarray}
\xi^1 = c_2,  \quad
\xi^2 = c_3,  \quad
\eta = c_1 u + \beta(x,t).
\end{eqnarray}
Thus, the symmetries we obtain are
\begin{eqnarray*}
X_1 = \frac{\partial}{\partial t},  \quad
X_2 = \frac{\partial}{\partial x},  \quad
X_3 =  u \frac{\partial}{\partial u},  \quad
X_{\beta} = \beta  \frac{\partial}{\partial u}.
\end{eqnarray*}
For this case, we have infinite symmetries, where $\beta(x,t)$ satisfies the wave equation.

\subsection{Principal Case}
For arbitrary $\Phi$, we obtain the infinitesimals which are given by 
\begin{eqnarray}
\xi^1 = c_2,  \quad
\xi^2 = c_3,  \quad
\eta = 0.
\end{eqnarray}
and the corresponding symmetries are 
\begin{eqnarray*}
X_1 = \frac{\partial}{\partial t},  \quad
X_2 = \frac{\partial}{\partial x} .
\end{eqnarray*}

\subsection{Discussion on Optimal System} \label{sec_a2}
In 1960, a Russian mathematician L. V. Ovsyannikov \cite{st1} developed the methods of symmetry analysis for differential equations in a very systematic way and initially presented the idea of optimal system of subalgebra. In this section, we present the one-dimensional optimal systems; we follow the approach outlined by Olver \cite{s17} and Ibragimov \cite{s13}. For a partial differential equations $E$ and the one parameter Lie group of transformations $\bar{x}$ admitted by $E$,
\begin{eqnarray}
\bar{x}=T_i(x, \epsilon), i=1,2, \ldots, \mathrm{n},
\end{eqnarray}
solutions of the $E$ has the property that any solution of the $E$ can be transformed to a certain solution of the same differential equation $E$ by any transformation $T$ of the group $G$. Consider, $\Phi$ is a solution of $E$ and $T \in G$, then $\Phi^{\prime}$ is another solution of $E$ given by,
\begin{eqnarray}
\Phi^{\prime}=  \boldsymbol{T} \Phi.
\end{eqnarray}
Therefore, two solutions of $E$ are said to be different with respect to $G$ if they are not transformed to each other by any transformation of the group $G$. If $H$ is a subgroup of the group $G$ then it is obvious that solutions different with respect to $G$, are also different with respect to $H$. However, solutions different with respect to $H$ do not necessarily have the same property with respect to $G$. Let us assume that $\Phi_1$ is a solution invariant with respect to $H_1$ and $\Phi_2$ is a solution invariant with respect to $H_2$, where $H_1, H_2$ are two subgroups of $G$. If there exists a transformation $T \in G$ such that
\begin{eqnarray}
H_2=T H_1 T^{-1},
\end{eqnarray}
then there is a one-to-one correspondence between $\boldsymbol{H}_1$-solutions and $\boldsymbol{H}_2$-solutions given by the formula
\begin{eqnarray}
\Phi_2=T \Phi_1.
\end{eqnarray}
Due to one-to-one correspondence between subgroups and subalgebras, we will identify optimal systems of subgroups with optimal systems of sub-algebras. Since a Lie group (or Lie algebra) usually contains infinitely many subgroups (or subalgebras) of the same dimension. That is why, a classification up to some equivalence relation is necessary. 

\subsection{Optimal System of Fourth Order Wave}

The summary of the Lie symmetry algebra is presented in this section for all the cases arising in this study depending on function $\Phi$.  \\

\noindent\textbf{1. $\boldsymbol{\Phi^{\prime} = 0}$}\\
Symmetry algebra is infinite dimensional and spanned by  
\begin{eqnarray*}
X_1 = \frac{\partial}{\partial t}, 
X_2 = \frac{\partial}{\partial x}, 
X_3 =  u \frac{\partial}{\partial u},
X_{\beta} = \beta(x,t)  \frac{\partial}{\partial u}.
\end{eqnarray*}
\noindent\textbf{2. Principal Case}\\
Symmetry algebra is two dimensional and spanned by 
\begin{eqnarray*}
X_1 = \frac{\partial}{\partial t}, 
X_2 = \frac{\partial}{\partial x} .
\end{eqnarray*}
\noindent\textbf{3. $\boldsymbol{\Phi = \epsilon (u+k_1)^{k_2} }$, where $\boldsymbol{\epsilon = \pm 1}$}\\
Symmetry algebra is three dimensional and spanned by 
\begin{eqnarray*}
X_1 = \frac{\partial}{\partial t}, 
X_2 = \frac{\partial}{\partial x}, 
X_3 = \frac{k_2}{2} t \frac{\partial}{\partial t} + u \frac{\partial}{\partial u}.
\end{eqnarray*}
\noindent\textbf{4. $\boldsymbol{\Phi = \epsilon \exp (ku) }$, where $\boldsymbol{\epsilon = \pm 1}$}\\
Symmetry algebra is three dimensional and spanned by 
\begin{eqnarray*}
X_1 = \frac{\partial}{\partial t}, 
X_2 = \frac{\partial}{\partial x}, 
X_3 = \frac{k}{2} t \frac{\partial}{\partial t} +  \frac{\partial}{\partial u}.
\end{eqnarray*}

\noindent\textbf{5. $\boldsymbol{\Phi = \epsilon (u)^{k_1} }$, where $\boldsymbol{\epsilon = \pm 1}$}\\
Symmetry algebra is three dimensional and spanned by 
\begin{eqnarray*}
X_1 = \frac{\partial}{\partial t}, 
X_2 = \frac{\partial}{\partial x}, 
X_3 = -k_1 t \frac{\partial}{\partial t}  - \frac{1}{2} k_1 x \frac{\partial}{\partial x} + u \frac{\partial}{\partial u}.
\end{eqnarray*}

\noindent\textbf{6. $\boldsymbol{\Phi = \epsilon \exp (u) }$, where $\boldsymbol{\epsilon = \pm 1}$}\\
Symmetry algebra is three dimensional and spanned by 
\begin{eqnarray*}
X_1 = \frac{\partial}{\partial t}, 
X_2 = \frac{\partial}{\partial x}, 
X_3 = -k t \frac{\partial}{\partial t}  - \frac{1}{2}  x \frac{\partial}{\partial x} +  \frac{\partial}{\partial u}.
\end{eqnarray*}
The adjoint representation is given by
\begin{equation} \label{E05}
 Ad(\exp(\epsilon X_i))(X_j) = X_j - \epsilon [{X_i}, X_j] + \frac{\epsilon^2}{2!} [{X_i}, [{X_i}, X_j]] - \cdots,
\end{equation}
where $\epsilon$ is a real number and $[X_i,X_j]$ denotes the Lie product defined by
\begin{equation}
[X_i, X_j] = X_iX_j - X_j X_i.
\end{equation}

\subsection{Optimal system: $\boldsymbol{\Phi^{\prime} = 0}$}

\noindent For $\Phi^{\prime}=0$, equation (\ref{Eq_3}) reduces to linear PDE given by  
\begin{eqnarray} \label{lin1op}
u_{tt} = cu_{xx} + h_1 u_{xxxx} + h_2 u_{xxtt}.
\end{eqnarray}
has infinite dimensional algebra spanned by  
\begin{eqnarray*}
X_1 = \frac{\partial}{\partial t}, 
X_2 = \frac{\partial}{\partial x}, 
X_3 =  u \frac{\partial}{\partial u},
X_{\beta} = \beta(x,t)  \frac{\partial}{\partial u}.
\end{eqnarray*}
The algebra is not closed as in commutator table coefficients turn out to be functions and its derivatives. The non-zero commutators for this case are 
\begin{eqnarray}
[X_1,X_{\beta}] = \beta_t  \frac{\partial}{\partial u}, \quad [X_2,X_{\beta}] = \beta_x  \frac{\partial}{\partial u}, \quad [X_3,X_{\beta}] =-\beta  \frac{\partial}{\partial u}.
\end{eqnarray}
The infinite symmetry algebra generated by $X_{\beta}$ is not essential \cite{s17} as if $X+X_{\beta}$ is part of the three-dimensional algebra of the linear PDE (\ref{lin1op}), then we can always find $X_{\alpha}={\alpha}(t,x) \frac{\partial}{\partial u} $, where $\alpha$ is a solution of the linear PDE, such that $Ad(\exp X_{\alpha})(X+X_{\beta})=X$. The $X_{\alpha}$ does not lead to group-invariant solutions. For instance, if $\mathbf{X}=\mathbf{X}_2=\frac{\partial}{\partial x}$, then
\begin{eqnarray}
\beta(x, t)=-\int_0^x \alpha(y, t) d y-\int_0^t \alpha_x(0, s) d s.
\end{eqnarray}
When dealing with the infinite dimensional algebra admitted by linear PDE, constructing optimal system require a careful and systematic analysis. One way to study optimal system for this case, is to consider the semi-direct product of a finite dimensional algebra $\mathfrak{g} = \mbox{span}\{ X_1, X_2, X_3 \}$ acting on an infinite dimensional linear subalgebra $\mathfrak{h} = \mbox{span} \{ X_{\beta}  |  {\forall \beta(x, t)} \mbox{ is an arbitrary function.} \}$ in the space $\mathfrak{g} \oplus \mathfrak{h}$.

\begin{theorem}
Consider the PDE (\ref{lin1op}) admits an infinite dimensional algebra spanned by
\begin{eqnarray}
\mathfrak{g} = \mbox{span}\{ X_1, X_2, X_3 \}, \ \ \mbox{and} \quad   \mathfrak{h} = \mbox{span} \{ X_{\beta}  |  {\forall \beta(x, t)} \mbox{ is an arbitrary function.} \}
\end{eqnarray}
The set of one-dimensional optimal system of sub-algebra, which can not be reduced further under any adjoint action is given by the following two lists. For the finite-dimensional part $\mathfrak{g}=\left\{X_1, X_2, X_3\right\}$
\[  
   \langle
                \begin{array}{llll}
                 X^1 =  X_1 \  , \ \   \  \  &X^2 = X_2  \  , \ \   \  \  &
                 X^3 = X_3  .
                \end{array}
               \rangle,
  \]
and for the infinite-dimensional part $\mathfrak{h}=\left\{X_\beta\right\}$
\[  
   \langle
                \begin{array}{llll}
                X^4 =   \frac{\partial}{\partial u}  \  , \ \   \  \ 
                 X^5 = x \frac{\partial}{\partial u} \  , \ \   \  \  &X^6 = t \frac{\partial}{\partial u}  \  , \ \   \  \  &
                 X^7 = e^{\lambda x} \frac{\partial}{\partial u}  \  , \ \   \  \  &X^8 = e^{\lambda t} \frac{\partial}{\partial u}. \\
                \end{array}
               \rangle.
  \]  
\end{theorem}
\noindent \textbf{Proof:} Since the finite dimensional part is an Abelian Lie algebra, i.e. 
\begin{eqnarray}
\{  [X_i, X_j ]=0, \forall X_i, X_j \in \mathfrak{g}, i,j=1,2,3 \}.
\end{eqnarray}
So, for three-dimensional Abelian algebra $\mathfrak{g}=\operatorname{span} \{X_1, X_2, X_3 \}$, the one dimensional optimal system can be represented by 
\begin{eqnarray}
 \{ \langle X_1 \rangle, \langle X_2 \rangle, \langle X_3 \rangle \}.
\end{eqnarray}
Any other one-dimensional subalgebra $ \langle c_1 X_1+ c_2 X_2+ c_3 X_3 \rangle$ is equivalent to one of the above for the finite dimensional part. The adjoint representations for this case are presented in Table-\ref{tab:title6-2a2inf1}.\\

\begin{minipage}{\linewidth}
\centering
\begin{tabularx}{\linewidth}{@{} C{1in} C{.85in} *4X @{}}\toprule[1.5pt]
\bf Ad & \bf $X_1$ & \bf $X_2$ & \bf $X_3$ & \bf $X_{\beta}$ \\
\bottomrule[1.25pt]
  $X_1$ & $X_1$ & $X_2$   & $X_3$ & $\beta(x, t-\epsilon) \frac{\partial}{\partial u}$  \\
\hline
$X_2$ & $ X_1$ & $X_2$ & $X_3$ & $\beta(x-\epsilon, t) \frac{\partial}{\partial u}$   \\
\hline
$X_3$ & $ X_1$ & $X_2$ & $X_3$ & $e^{\epsilon} X_\beta$   \\
\hline
$X_{\beta}$ & $ X_1+\epsilon \beta_t \frac{\partial}{\partial u} $ & $X_2+\epsilon \beta_x \frac{\partial}{\partial u}$ & $X_3-\epsilon X_\beta$ & $X_\beta$   \\
\bottomrule[1.25pt]
\end {tabularx}\par
\captionof{table}{Adjoint Table} \label{tab:title6-2a2inf1}
\bigskip
\end{minipage}
The adjoint action of $X_1$, $X_2$, and $X_3$ on $X_{\beta}$  is given by  
\begin{align*}
\text{Ad}(\exp{ \epsilon X_1})X_{\beta} &=  (\beta-\epsilon \beta_t+\frac{\epsilon^2}{2!} \beta_{t t}-\cdots ) \frac{\partial}{\partial u} = \beta(x, t-\epsilon) \frac{\partial}{\partial u},  \\
\text{Ad}(\exp{ \epsilon X_2})X_{\beta} &=  (\beta-\epsilon \beta_x+\frac{\epsilon^2}{2!} \beta_{x x}-\cdots ) \frac{\partial}{\partial u} = \beta(x-\epsilon, t) \frac{\partial}{\partial u},  \\
\text{Ad}(\exp{ \epsilon X_3})X_{\beta} &= \beta(x, t) \frac{\partial}{\partial u} (1+\epsilon+\frac{\epsilon^2}{2!}+\cdots ) = e^{\epsilon} X_\beta.
\end{align*}
The adjoint action of $X_{\beta}$ on $X_1$, $X_2$, and $X_3$ is given by  
\begin{align*}
\text{Ad}(\exp{ \epsilon X_\beta})X_{1} &=  X_1+\epsilon \beta_t \frac{\partial}{\partial u},  \\
\text{Ad}(\exp{ \epsilon X_\beta})X_{2} &= X_2+\epsilon \beta_x \frac{\partial}{\partial u},  \\
\text{Ad}(\exp{ \epsilon X_\beta})X_{3} &= X_3-\epsilon X_\beta.
\end{align*}
To construct optimal system of sub algebra of infinite dimensional algebra, consider a general element $X \in \mathfrak{g} \oplus \mathfrak{h}$, we have
\begin{eqnarray}
\tilde{X}=a_1 X_1 + a_2 X_2 + a_3 X_3 + X_\beta .
\end{eqnarray}
Coefficients $a_1, a_2$ and $a_3$ are already classified, the coefficient function of the infinite symmetry should be reduced to specific forms under adjoint action. This can be done by acting the finite dimensional algebra $\mathfrak{g} = \mbox{span}\{ X_1, X_2, X_3 \}$ on an infinite dimensional linear subalgebra $\mathfrak{h} = \mbox{span} \{ X_{\beta}  |  {\forall \beta(x, t)} \}$ given by
\begin{align*}
\operatorname{Adj}  ( \exp  (a X_1 )  ) \tilde{X} &= \beta(x, t-\epsilon) \frac{\partial}{\partial u}, \\
\operatorname{Adj}  ( \exp  (a X_2 )  ) \tilde{X} &= \beta(x-\epsilon, t) \frac{\partial}{\partial u}, \\
\operatorname{Adj}  ( \exp  (a X_3 )  ) \tilde{X} &=  e^{\epsilon} X_\beta.
\end{align*}
Following cases can be constructed from the above adjoint actions: \\ 
Under uniform shift in $u$:
\begin{eqnarray}
\beta(x, t) = 1, \implies X_\beta=\frac{\partial}{\partial u}. \nonumber 
\end{eqnarray}
Under spatial shifts along $t$ and $x$, respectively $\beta(x, t-\epsilon)$ and $\beta(x-\epsilon, t)$ in $u$:
\begin{eqnarray}
 \beta(x, t) = t, \implies X_\beta= t \frac{\partial}{\partial u}, \nonumber \\
  \beta(x, t) = x, \implies X_\beta= x \frac{\partial}{\partial u}, \nonumber 
\end{eqnarray}
Under scaling in $u$ ($\beta(x, t)$ scale by $e^\epsilon$.):
\begin{eqnarray}
  \beta(x, t) = e^{\lambda t}, \implies X_\beta= e^{\lambda t} \frac{\partial}{\partial u}, \nonumber \\
  \beta(x, t) = e^{\lambda x}, \implies X_\beta= e^{\lambda x} \frac{\partial}{\partial u}. \nonumber 
\end{eqnarray}

\subsection{Optimal system: Principal Case - For Arbitrary $\boldsymbol{\Phi}$}
\begin{theorem}
The set of one-dimensional optimal system of sub-algebra for the principal case, which can not be reduced further under any adjoint action, can be represented by
\[  
   \langle
                \begin{array}{ll}
                 X^1 =  X_1 \  , \ \   \  \  &X^2 = X_2 . 
                \end{array}
               \rangle.
  \]
\end{theorem}

\noindent \textbf{Proof:}  The algebra is two dimensional and the adjoint representations for this case is presented in Table-\ref{tab:title6-2a2}.\\

\begin{minipage}{\linewidth}
\centering
\begin{tabularx}{\linewidth}{@{} C{1in} C{.85in} *4X @{}}\toprule[1.5pt]
\bf Ad & \bf $X_1$ & \bf $X_2$ \\
\bottomrule[1.25pt]
  $X_1$ & $X_1$ & $X_2$    \\
\hline
$X_2$ & $ X_1$ & $X_2$  \\
\bottomrule[1.25pt]
\end {tabularx}\par
\captionof{table}{Adjoint Table} \label{tab:title6-2a2}
\bigskip
\end{minipage}
Since, all the commutators are zero, so, for two-dimensional Abelian algebra $\mathfrak{g}=\operatorname{span} \{X_1, X_2 \}$, the one dimensional optimal system can be represented by 
\begin{eqnarray}
\{  \langle X_1 \rangle, \langle X_2 \rangle \}.
\end{eqnarray}
Any other one-dimensional subalgebra $ \langle c_1 X_1+ c_2 X_2  \rangle$ is equivalent to one of the above. 

\subsection{Optimal system: $\boldsymbol{\Phi = \epsilon (u+k_1)^{k_2} }$, where $\boldsymbol{\epsilon = \pm 1}$}
\begin{theorem}
The set of one-dimensional optimal system of sub-algebra for this case, which can not be reduced further under any adjoint action, can be represented by
\[  
   \langle
                \begin{array}{ll}
                 X^1 =  X_2+ X_3 \  , \ \   \  \  &X^2 = X_3 , \\
                 X^3 = X_2 \pm X_1 \  , \ \   \  \ &X^4 =   X_1 ,\\
                  X^5 = X_1+ a_2 X_2.
                \end{array}
               \rangle.
  \]
\end{theorem}

\noindent  The algebra in this case is three-dimensional and spanned by the principal algebra and 
$$ X_3 =  \frac{k_2}{2}  t \frac{\partial}{\partial t} + u \frac{\partial}{\partial u}.$$
In this case the only non zero commutator is $ [ X_1 , X_3 ] = \frac{k_2}{2} X_1$. The adjoint representation is required to find the optimal system of algebras. The adjoint representation can be calculated by the expression (\ref{E05}) given by
\begin{align*}
\text{Ad}(\exp{ \epsilon X_1})X_1 &= X_1 , \\
\text{Ad}(\exp{ \epsilon X_1})X_2 &= X_2.
\end{align*}
The adjoint action of $X_1$ on $X_3$ is given by  
\begin{align*}
\text{Ad}(\exp{ \epsilon X_1})X_3 &= X_3-\epsilon [X_1, X_3 ]+\frac{\epsilon^2}{2 !} [X_1, [X_1, X_3 ] ]+\cdots, \\
\text{Ad}(\exp{ \epsilon X_1})X_3 &= X_3-\epsilon (\frac{k_2}{2} X_1  )+\frac{\epsilon^2}{2 !} [X_1, \frac{k_2}{2} X_1 ]+\cdots, \\
 &= X_3-\epsilon \frac{k^2}{2} X_1.
\end{align*}
The adjoint action of $X_2$ on $X_1$, $X_2$ and $X_3$ is given by  
\begin{align*}
\text{Ad}(\exp{ \epsilon X_2})X_1 &= X_1,  \\
\text{Ad}(\exp{ \epsilon X_2})X_2 &= X_2, \\
\text{Ad}(\exp{ \epsilon X_2})X_3 &= X_3.
\end{align*}
Similarly, the adjoint action of $X_3$ on $X_2$, $X_1$ is trivial but the action on $X_1$ is calculated below   
\begin{align*}
\text{Ad}(\exp{ \epsilon X_3})X_1 &= e^{\frac{\epsilon k_2}{2}} X_1 , \\
\text{Ad}(\exp{ \epsilon X_3})X_2 &= X_2, \\
\text{Ad}(\exp{ \epsilon X_3})X_3 &= X_3.
\end{align*}
Based on the values we calculated previously this lead us to the construction of adjoint Table-\ref{tab:title5aa3} given by \\

\begin{minipage}{\linewidth}
\centering
\begin{tabularx}{\linewidth}{@{} C{1in} C{.85in} *4X @{}}\toprule[1.5pt]
\bf Ad & \bf $X_1$ & \bf $X_2$ & \bf $X_3$  \\\midrule
 $X_1$ & $X_1$ & $X_2$ & $X_3 - \epsilon \frac{k_2}{2} X_1$  \\
\hline
 $X_2$ & $ X_1 $ & $X_2$ &  $X_3$  \\
 \hline
$X_3$ & $e^{\epsilon \frac{k_2}{2}} X_1$ & $ X_2 $ & $X_3$  \\
\bottomrule[1.25pt]
\end {tabularx}\par
\bigskip
\captionof{table}{Adjoint Table} \label{tab:title5aa3}
\end{minipage}\\

\noindent  \textbf{Proof}: Consider a general element $X \in \mathcal{L}_3$, we have
\begin{equation*}
\tilde{X}=a_1 X_1 + a_2 X_2 + a_3 X_3 .
\end{equation*}
By the adjoint action of $X_1$ on general element $X \in \mathcal{L}_3$ will lead us to the construction of optimal system given by 
\begin{align*}
\operatorname{Adj}  ( \exp  (a X_1 )  ) \tilde{X} &=a_1 X_1+a_2 X_2+a_3 (X_3-\frac{a k_2}{2} X_1 ), \\ 
&= (a_1-a_3 \frac{a k_2}{2} ) X_1+a_2 X_2+a_3 X_3.
\end{align*}
From here, we can start constructing the cases by letting $a_3 \neq 0$. When $a_3 \neq 0$, we have  
$a_1 = a_3 a \frac{ k_2}{2}$, since $k_2 \neq 0$ so, $k_2$ will not occur in constructing cases for the optimal system, and we get $a =\frac{2 a_1}{a_3 k_2}$ and $\tilde{X}$ becomes
\begin{align*}
\tilde{X} &= a_2 X_2+X_3,  \ &;  \  \   \ a_3 \neq 0 \\
\Rightarrow \tilde{X} &= X_2+X_3, \ &;  \  \   \ a_3 \neq 0, a_2 \neq 0 \\ 
\Rightarrow \tilde{X} &= X_3.  &;  \  \   \ a_3 \neq 0, a_2 = 0
\end{align*}
It cannot be reduced further under any adjoint action. So, now we can consider the case when $a_3=0$, for which we have 
\begin{align*}
\Rightarrow & \tilde{X}=a_1 X_1+a_2 X_2, \quad &; \quad a_3=0 \\
  \Rightarrow          & \tilde{X}=a_1 X_1+X_2. \quad &; \quad a_3=0, a_2 \neq 0
\end{align*}
From the adjoint table, $\tilde{X}$ can further be reduced by the action of $X_3$ given by 
\begin{align*}
\operatorname{Adj} (\operatorname{exp(aX_3)}  ) X &= e^{\frac{a k_2}{2}} a_1 X_1+X_2.
\end{align*}
By considering $e^{\frac{a k_2}{2}} a_1=\pm 1$, we end up with $a=\frac{2}{k_2} \ln  |\pm \frac{1}{a_1} |$ and this yields $\tilde{X}$ given by 
\begin{align*}
\tilde{X} &= \pm X_1+X_2,  &; a_3=0, a_2 \neq 0, \\
\tilde{X} &=  X_1. \quad &; a_3=0, a_2=0
\end{align*}
Consider the case when $a_3=0$ and $a_1 \neq 0$  for which we have 
\begin{align*}
\Rightarrow & \tilde{X}=a_1 X_1+a_2 X_2, \quad &; \quad a_3=0 \\
  \Rightarrow          & \tilde{X}= X_1+ a_2 X_2. \quad &; \quad a_3=0, a_1 \neq 0
\end{align*}
The tree diagram of all the cases formed to construct one-dimensional optimal system of all the  subalgebras is presented below
\begin{center}
	\Tree[ .a_3
		 [.a_3$\neq$0  [.a_2$\neq 0$ {Case-I \\ $  X^1 =  X_2+ X_3$ } ] 
		[ .a_2$=0$ {Case-II \\ $X^2 = X_3$  }   ]]
 	 	 [.a_3=0 [.a_2$\neq0$ {Case-III  \\ $X^3 = X_2 \pm X_1$ } ] 
 	 	 [.a_2=0    {Case-IV  \\ $X^4 =    X_1$} ] [.a_1$\neq0$ {Case-V  \\ $X^5 = X_1+ a_2 X_2$ } ]
 	 	 ]]
\end{center}

\subsection{Optimal system:  $\boldsymbol{\Phi = \epsilon \exp (ku) }$, where $\boldsymbol{\epsilon = \pm 1}$ }
\begin{theorem}
The set of one-dimensional optimal system of sub-algebra for this case, which can not be reduced further under any adjoint action, can be represented by
\[  
   \langle
                \begin{array}{ll}
                 X^1 =  X_2+ X_3 \  , \ \   \  \  &X^2 = X_3 , \\
                 X^3 = X_2 \pm X_1 \  , \ \   \  \ &X^4 =   X_1 ,\\
                  X^5 = X_1+ a_2 X_2.
                \end{array}
               \rangle.
  \]
\end{theorem}

\noindent  The algebra in this case is three-dimensional and spanned by the principal algebra and $X_3$ with the only non zero commutator  $ [ X_1 , X_3 ] = \frac{k}{2} X_1$. 
$$ X_3 =  \frac{k}{2}  t \frac{\partial}{\partial t} +  \frac{\partial}{\partial u}.$$
In this case the adjoint representation is required to find the optimal system of algebras. The adjoint representation can be calculated by the expression (\ref{E05}) given by
\begin{align*}
\text{Ad}(\exp{ \epsilon X_1})X_1 &= X_1 , \\
\text{Ad}(\exp{ \epsilon X_1})X_2 &= X_2, \\
\text{Ad}(\exp{ \epsilon X_1})X_3 &= X_3-\epsilon \frac{k}{2} X_1.
\end{align*}
The adjoint action of $X_2$ on $X_1$, $X_2$ and $X_3$ is trivial 
\begin{align*}
\text{Ad}(\exp{ \epsilon X_2})X_1 &= X_1,  \\
\text{Ad}(\exp{ \epsilon X_2})X_2 &= X_2, \\
\text{Ad}(\exp{ \epsilon X_2})X_3 &= X_3.
\end{align*}
Similarly, the adjoint action of $X_3$ on $X_2$, $X_3$ is trivial but the action on $X_1$ is   
\begin{align*}
\text{Ad}(\exp{ \epsilon X_3})X_1 &=  e^{\frac{\epsilon k}{2}} X_1, \\
\text{Ad}(\exp{ \epsilon X_3})X_2 &= X_2, \\
\text{Ad}(\exp{ \epsilon X_3})X_3 &= X_3.
\end{align*}
Based on the values we calculated previously lead us to the construction of adjoint table (\ref{tab:title5aa4}) given by \\
\begin{minipage}{\linewidth}
\centering
\begin{tabularx}{\linewidth}{@{} C{1in} C{.85in} *4X @{}}\toprule[1.5pt]
\bf Ad & \bf $X_1$ & \bf $X_2$ & \bf $X_3$  \\\midrule
 $X_1$ & $X_1$ & $X_2$ & $X_3 - \epsilon \frac{k}{2} X_1$  \\
\hline
 $X_2$ & $ X_1 $ & $X_2$ &  $X_3$  \\
 \hline
$X_3$ & $e^{\epsilon \frac{k}{2}} X_1$ & $ X_2 $ & $X_3$  \\
\bottomrule[1.25pt]
\end {tabularx}\par
\bigskip
\captionof{table}{Adjoint Table} \label{tab:title5aa4}
\end{minipage}\\

\noindent  \textbf{Proof}: Consider a general element $X \in \mathcal{L}_3$. We have
\begin{equation*}
\tilde{X}=a_1 X_1 + a_2 X_2 + a_3 X_3 .
\end{equation*}
$\begin{aligned} \operatorname{Adj}  ( \exp  (a X_1 )  ) \tilde{X} &=a_1 X_1+a_2 X_2+a_3 (X_3-\frac{a k}{2} X_1 ), \\ 
&= (a_1-a_3 \frac{a k}{2} ) X_1+a_2 X_2+a_3 X_3.   \end{aligned}$ \\
From here, we can start constructing the cases by letting $a_3 \neq 0$. When $a_3 \neq 0$, we have  
$a_1 = a_3 a \frac{ k}{2}$, since $k \neq 0$ so, we get $a =\frac{2 a_1}{a_3 k}$ and $\tilde{X}$ becomes
\begin{align*}
\tilde{X} &= a_2 X_2+X_3,  \ &;  \  \   \ a_3 \neq 0 \\
\Rightarrow \tilde{X} &= X_2+X_3, \ &;  \  \   \ a_3 \neq 0, a_2 \neq 0 \\ 
\Rightarrow \tilde{X} &= X_3.  &;  \  \   \ a_3 \neq 0, a_2 = 0
\end{align*}
It can not be reduced further under any adjoint action. So, now we can consider the case when $a_3=0$, we have 
\begin{align*}
\Rightarrow & \tilde{X}=a_1 X_1+a_2 X_2, \quad &; \quad a_3=0, \\
  \Rightarrow          & \tilde{X}=a_1 X_1+X_2. \quad &; \quad a_3=0, a_2 \neq 0.
\end{align*}
From the adjoint table, $\tilde{X}$ can further reduced by the action of $X_3$ given by 
\begin{align*}
\operatorname{Adj} (\operatorname{exp(aX_3)}  ) X &= e^{\frac{a k}{2}} a_1 X_1+X_2.
\end{align*}
By considering $e^{\frac{a k}{2}} a_1=\pm 1$, it provides us with $a=\frac{2}{k} \ln  |\pm \frac{1}{a_1} |$ and yield $\tilde{X}$ given by 
\begin{align*}
\tilde{X} &= \pm X_1+X_2,  &; a_3=0, a_2 \neq 0, \\
\tilde{X} &= X_1. \quad &; a_3=0, a_2=0.
\end{align*}
Consider the case when $a_3=0$ and $a_1 \neq 0$  for which we have 
\begin{align*}
\Rightarrow & \tilde{X}=a_1 X_1+a_2 X_2, \quad &; \quad a_3=0 \\
  \Rightarrow          & \tilde{X}= X_1+ a_2 X_2. \quad &; \quad a_3=0, a_1 \neq 0
\end{align*}
The tree diagram of all the cases formed to construct one-dimensional optimal system of all the  subalgebras is presented below
\begin{center}
	\Tree[ .a_3
		 [.a_3$\neq$0  [.a_2$\neq 0$ {Case-I \\ $  X^1 =  X_2+ X_3$ } ] 
		[ .a_2$=0$ {Case-II \\ $X^2 = X_3$  }   ]]
 	 	 [.a_3=0 [.a_2$\neq0$ {Case-III  \\ $X^3 = X_2 \pm X_1$ } ] 
 	 	 [.a_2=0    {Case-IV  \\ $X^4 =    X_1$} ] [.a_1$\neq0$ {Case-V  \\ $X^5 = X_1+ a_2 X_2$ } ]
 	 	 ]]
\end{center}

\subsection{Optimal system: $\boldsymbol{\Phi = \epsilon (u)^{k_1} }$, where $\boldsymbol{\epsilon = \pm 1}$ }
\begin{theorem}
The set of one-dimensional optimal system of sub-algebra for this case, which can not be reduced further under any adjoint action, can be represented by
\[  
   \langle
                \begin{array}{ll}
                 X^1 = X_3,  \\
                 X^2 = c_2 X_2 \pm X_1, \\
                 X^3 = X_1 .
                \end{array}
               \rangle.
  \]
\end{theorem}

\noindent The algebra in this case is three-dimensional and spanned by the principal algebra and $X_3$ with the non zero commutators $ [ X_1 , X_3 ] = k_1 X_1$ and $ [ X_2 , X_3 ] = -\frac{k_1}{2} X_2$. 
$$ X_3 =  - k_1  t \frac{\partial}{\partial t} - \frac{1}{2} k_1 x \frac{\partial}{\partial x}  + u \frac{\partial}{\partial u}.$$
In this case the adjoint representation is required to find the optimal system of algebras. The adjoint representation can be calculated by the expression (\ref{E05}). The adjoint action of $X_1$ on $X_1$, $X_2$ and $X_3$ are given by
\begin{align*}
\text{Ad}(\exp{ \epsilon X_1})X_1 &= X_1,  \\
\text{Ad}(\exp{ \epsilon X_1})X_2 &= X_2, \\
\text{Ad}(\exp{ \epsilon X_1})X_3 &= X_3 - \epsilon k_1 X_1  .
\end{align*}
The adjoint action of $X_2$ on $X_1$, $X_2$ and $X_3$ is given by  
\begin{align*}
\text{Ad}(\exp{ \epsilon X_2})X_1 &= X_1,  \\
\text{Ad}(\exp{ \epsilon X_2})X_2 &= X_2, \\
\text{Ad}(\exp{ \epsilon X_2})X_3 &= X_3 + \frac{\epsilon k_2}{2} X_2.
\end{align*}
Similarly, the adjoint action of $X_3$ on $X_1$, $X_2$ and $X_3$ are given by    
\begin{align*}
\text{Ad}(\exp{ \epsilon X_3})X_1 &=  e^{ \epsilon k_1} X_1, \\
\text{Ad}(\exp{ \epsilon X_3})X_2 &=  e^{-\frac{ \epsilon k_1}{2}} X_2 , \\
\text{Ad}(\exp{ \epsilon X_3})X_3 &= X_3.
\end{align*}
Based on the values we calculated, lead us to the construction of adjoint table (\ref{tab:title5aa5}) given by \\

\begin{minipage}{\linewidth}
\centering
\begin{tabularx}{\linewidth}{@{} C{1in} C{.85in} *4X @{}}\toprule[1.5pt]
\bf Ad & \bf $X_1$ & \bf $X_2$ & \bf $X_3$  \\\midrule
 $X_1$ & $X_1$ & $X_2$ & $X_3 - \epsilon k_1 X_1$  \\
\hline
 $X_2$ & $ X_1 $ & $X_2$ &  $X_3 + \frac{\epsilon k_1}{2} X_2 $  \\
 \hline
$X_3$ & $e^{\epsilon k_1} X_1$ & $ e^{- \epsilon \frac{k_1}{2}} X_2 $ & $X_3$  \\
\bottomrule[1.25pt]
\end {tabularx}\par
\bigskip
\captionof{table}{Adjoint Table} \label{tab:title5aa5}
\end{minipage}\\

\noindent  \textbf{Proof}: Consider a general element $X \in \mathcal{L}_3$. We have
\begin{equation*}
\tilde{X}=a_1 X_1 + a_2 X_2 + a_3 X_3 .
\end{equation*}
Apply the action of $X_1$ on $\tilde{X}$ we have 
\begin{align*}
\operatorname{Adj}  ( \exp  (a X_1 )  ) \tilde{X} &=  a_1 X_1+a_2 X_2+a_3 (X_3-a k_1 X_1 ), \\
                            &=  (a_1-a_3 a k_1  ) X_1+a_2 X_2+a_3 X_3.
\end{align*}
From here, we can start constructing the cases by letting $a_3 \neq 0$. When $a_3 \neq 0$, we have  
$a_1 = a_3 a k_1$, since $k_1 \neq 0$ so, we get $a =\frac{ a_1}{a_3 k_1}$ and $\tilde{X}$ becomes
\begin{align*}
\tilde{X} &= a_2 X_2+ a_3 X_3.  \ &;  \  \   \ a_3 \neq 0 
\end{align*}
By the action of $X_2$, we get 
\begin{align*}
\operatorname{Adj}  ( \exp  (a X_2  )  ) \tilde{X} &= a_2 X_2 + a_3 (X_3+\frac{a}{2} k X_2 ),  \ &;  \  \   \ a_3 \neq 0 \\
                     &=  (a_2+\frac{a_3 a}{2} k ) X_2+ a_3 X_3.  \ &;  \  \   \ a_3 \neq 0
\end{align*}
By letting $a_2+\frac{a}{2} a_3 k = 0$, we get $\frac{2 a_2}{a_3 k}=a, a_3 \neq 0$ which provides us 
\begin{align*}
 \tilde{X} =   a_3 X_3 = X_3.  \ &;  \  \   \ a_3 \neq 0
\end{align*}
Now check for $a_3=0$, we have 
\begin{align*}
\Rightarrow & \tilde{X}=a_1 X_1+a_2 X_2. \quad &; \quad a_3=0 
\end{align*}
From the adjoint table, by the action of $X_3$ given by 
\begin{align*}
\operatorname{Adj} (\operatorname{exp(aX_3)}  ) \tilde{X} &= e^{\frac{3 a k}{2}} a_1 X_1+ e^{- \epsilon \frac{k_1}{2}} X_2. &; a_3=0, a_2 \neq 0, a^{\prime}_2 = e^{- \epsilon \frac{k_1}{2}}
\end{align*}
Here, by letting $e^{\frac{3}{2} a k} a_1=\pm 1$, we obtain   
\begin{align*}
\frac{3}{2} a k &=\ln  |\frac{\pm 1}{a_1} | \quad \ \Rightarrow a=\frac{2}{3 k} \ln  |\pm \frac{1}{a_1} | .
\end{align*}
The value of $a=\frac{2}{3 k} \ln  |\pm \frac{1}{a_1} | $ results in the $\tilde{X}$ given by  
\begin{align*}
\tilde{X} &= \pm X_1 + a^{\prime}_2 X_2, &; a_3=0, a_2 \neq 0, a^{\prime}_2 = e^{- \epsilon \frac{k_1}{2}}, \\
\tilde{X} &= X_1.  &; a_3=0, a_2=0.
\end{align*}
The tree diagram of all the cases formed to construct one-dimensional optimal system of all the  subalgebras is presented below
\begin{center}
	\Tree[ .a_3
		 [.a_3$\neq$0  {Case-I \\ $  X^1 =  X_3$ } ] 
 	 	 [.a_3=0 [.a_2$\neq0$ {Case-II  \\ $X^2 = c_2 X_2 \pm X_1 $ } ]
 	 	 [.a_2=0    {Case-III  \\ $X^3 =   X_1$} ]
 	 	 ]]
\end{center}


\subsection{Optimal system: $\boldsymbol{\Phi = \epsilon \exp (u) }$, where $\boldsymbol{\epsilon = \pm 1}$ }
\begin{theorem}
The set of one-dimensional optimal system of sub-algebra for this case, which can not be reduced further under any adjoint action, can be represented by
\[  
   \langle
                \begin{array}{ll}
                 X^1 = X_3,  \\
                 X^2 = c_2 X_2 \pm X_1 , \\
                 X^3 = X_1 .
                \end{array}
               \rangle
  \]
\end{theorem}

\noindent The algebra in this case is three-dimensional and spanned by the principal algebra and $X_3$ with the non zero commutators $ [ X_1 , X_3 ] = - k_1 X_1$ and $ [ X_2 , X_3 ] = -\frac{1}{2} X_2$. 
$$ X_3 =  - k  t \frac{\partial}{\partial t} - \frac{1}{2} x \frac{\partial}{\partial x}  +  \frac{\partial}{\partial u}.$$
In this case the adjoint representation is required to find the optimal system of algebras. The adjoint representation can be calculated by the expression (\ref{E05}). The adjoint action of $X_1$, $X_2$ and $X_3$ given by
\begin{align*}
\text{Ad}(\exp{ \epsilon X_1})X_1 &= X_1 , \\
\text{Ad}(\exp{ \epsilon X_1})X_2 &= X_2 , \\
\text{Ad}(\exp{ \epsilon X_1})X_3 &= X_3 + \epsilon  k X_1 . 
\end{align*}
The adjoint action of $X_2$ on $X_1$, $X_2$ and $X_3$ is given by  
\begin{align*}
\text{Ad}(\exp{ \epsilon X_2})X_1 &= X_1,  \\
\text{Ad}(\exp{ \epsilon X_2})X_2 &= X_2 ,\\
\text{Ad}(\exp{ \epsilon X_2})X_3  &= X_3 + \frac{\epsilon }{2} X_2.
\end{align*}
Similarly, the adjoint action of $X_3$ on $X_1$, $X_2$ and $X_3$ is calculated below   
\begin{align*}
\text{Ad}(\exp{ \epsilon X_3})X_1 &= e^{ - \epsilon k} X_1, \\
\text{Ad}(\exp{ \epsilon X_3})X_2 &= e^{\frac{ \epsilon }{2}} X_2 , \\
\text{Ad}(\exp{ \epsilon X_3})X_3 &= X_3.
\end{align*}
Based on the values we calculated, lead us to the construction of adjoint table (\ref{tab:title5aa5}) given by \\

\begin{minipage}{\linewidth}
\centering
\begin{tabularx}{\linewidth}{@{} C{1in} C{.85in} *4X @{}}\toprule[1.5pt]
\bf Ad & \bf $X_1$ & \bf $X_2$ & \bf $X_3$  \\\midrule
 $X_1$ & $X_1$ & $X_2$ & $X_3 + \epsilon k X_1$  \\
\hline
 $X_2$ & $ X_1 $ & $X_2$ &  $X_3 + \frac{\epsilon }{2} X_2 $  \\
 \hline
$X_3$ & $e^{-\epsilon k} X_1$ & $ e^{  \frac{\epsilon }{2}} X_2 $ & $X_3$  \\
\bottomrule[1.25pt]
\end {tabularx}\par
\bigskip
\captionof{table}{Adjoint Table} \label{tab:title5aa5}
\end{minipage}\\

\noindent  \textbf{Proof}: Consider a general element $X \in \mathcal{L}_3$. We have
\begin{equation*}
\tilde{X}=a_1 X_1 + a_2 X_2 + a_3 X_3 .
\end{equation*}
Apply the action of $X_1$ on $\tilde{X}$ we have 
\begin{align*}
\operatorname{Adj}  ( \exp  (a X_1 )  ) \tilde{X} &=  a_1 X_1+a_2 X_2+a_3 (X_3-a k X_1 ), \\
                            &=  (a_1-a_3 a k  ) X_1+a_2 X_2+a_3 X_3.
\end{align*}
By letting $a_3 \neq 0$, we have $a_1 + a_3 a k = 0 $, so, we get $a =\frac{ - a_1}{a_3 k}$ and $\tilde{X}$ becomes
\begin{align*}
\tilde{X} &= a_2 X_2+ a_3 X_3.  \ &;  \  \   \ a_3 \neq 0 
\end{align*}
By the action of $X_2$, we get 
\begin{align*}
\operatorname{Adj}  ( \exp  (a X_2  )  ) \tilde{X} &= a_2 X_2 + a_3 (X_3+\frac{a}{2}  X_2 ),  \ &;  \  \   \ a_3 \neq 0 \\
                     &=  (a_2+\frac{a_3 a}{2}  ) X_2+ a_3 X_3.  \ &;  \  \   \ a_3 \neq 0
\end{align*}
By letting $a_2+\frac{a}{2} a_3  = 0$, we get $\frac{ - 2 a_2}{a_3 }=a, a_3 \neq 0$ which provides us with  
\begin{align*}
 \tilde{X} = X_3.  \ &;  \  \   \ a_3 \neq 0
\end{align*}
Now check for $a_3=0$, we have 
\begin{align*}
\Rightarrow & \tilde{X}=a_1 X_1+a_2 X_2. \quad &; \quad a_3=0 
\end{align*}
From the adjoint table, by the action of $X_3$ given by 
\begin{align*}
\operatorname{Adj} (\operatorname{exp(aX_3)}  ) \tilde{X} &= e^{\frac{3 a k}{2}} a_1 X_1+ e^{- \epsilon \frac{k_1}{2}} X_2. &; a_3=0, a_2 \neq 0, a^{\prime}_2 = e^{- \epsilon \frac{k_1}{2}}
\end{align*}
Here, by letting $e^{\frac{3}{2} a k} a_1=\pm 1$, we obtain   
\begin{align*}
\frac{3}{2} a k &=\ln  |\frac{\pm 1}{a_1} | \quad \ \Rightarrow a=\frac{2}{3 k} \ln  |\pm \frac{1}{a_1} | .
\end{align*}
The value of $a=\frac{2}{3 k} \ln  |\pm \frac{1}{a_1} | $ results in the $\tilde{X}$ given by  
\begin{align*}
\tilde{X} &= \pm X_1 + a^{\prime}_2 X_2, &; a_3=0, a_2 \neq 0, a^{\prime}_2 = e^{- \epsilon \frac{k_1}{2}}, \\
\tilde{X} &= X_1.  &; a_3=0, a_2=0.
\end{align*}
The tree diagram of all the cases formed to construct one-dimensional optimal system of all the  subalgebras is presented below
\begin{center}
	\Tree[ .a_3
		 [.a_3$\neq$0  {Case-I \\ $  X^1 =  X_3$ } ] 
 	 	 [.a_3=0 [.a_2$\neq0$ {Case-II  \\ $X^2 = c_2 X_2 \pm X_1 $ } ]
 	 	 [.a_2=0    {Case-III  \\ $X^3 =   X_1$} ]
 	 	 ]]
\end{center}

\section{Symmetry Reduction under Optimal System} \label{sec_a3}
We presented the complete reduction of the PDE (\ref{Eq_3}) under optimal system that further lead us to the construction of exact solutions, explained in standard textbooks on the subject \cite{s8,s13,s17}. In this section, we focus on the invariant form method and the corresponding characteristic equation is given by
\begin{equation}
\frac{dt}{\xi^1(t,x,u)} = \frac{dx}{\xi^2(t,x,u)} = \frac{du}{\eta^1(t,x,u)}.
\end{equation}
Symmetry reduction for case-$(1)$ is excluded as the algebra is infinite dimensional. Consider the symmetry generator $X_1= \frac{\partial}{\partial t}$ from principal case, the respective invariant variables are 
\begin{eqnarray} \label{R_eq1}
 u = F(\alpha),  \ \ \alpha=x.
\end{eqnarray}
Differentiating (\ref{R_eq1}), we obtain the derivatives of reductions given by 
\begin{eqnarray}
u_t=u_{tt}=u_{xxtt}=0 \ \ \mbox{and} \ \ u_{x}=F^{\prime}, u_{xx}=F^{\prime \prime}, u_{xxxx}=F^{\prime \prime \prime \prime}.
\end{eqnarray} 
Therefore, (\ref{R_eq1}) is the solution of the equation (\ref{Eq_3}), If $F$ satisfies the ODE given by
\begin{eqnarray}
\Phi^{\prime} {F^{\prime}}^2 + \Phi F^{\prime \prime} + h_1 F^{\prime \prime \prime \prime}  =0.
\end{eqnarray} 
Similarly, symmetry generator $X_1= \frac{\partial}{\partial x}$ has invariant solution given by  
\begin{eqnarray}
u= F(\alpha)=c_1 t + c_2, \ \ \alpha = t, 
\end{eqnarray}
which is the invariant solution of (\ref{Eq_3}) provided it satisfies the ODE given by
\begin{eqnarray}
F^{\prime \prime} = 0, 
\end{eqnarray}
where the derivatives of reductions are given as
\begin{eqnarray}
u_x=u_{xx}=u_{xxxx}=u_{xxtt}=0 \ \ \mbox{and} \ \ u_{tt}=F^{\prime \prime}.
\end{eqnarray}

\subsection{Symmetry Reduction : $\boldsymbol{\Phi = \epsilon (u+k_1)^{k_2} }$}
The symmetry reduction of equation (\ref{Eq_3}) for $\Phi = \epsilon (u+k_1)^{k_2}$ is presented in Table \ref{tabler7-1}.
\begin{center}
\begin{tabular}{ l l }\toprule[1.5pt]
 \multicolumn{2}{c}{\bf Symmetry Reduction : $\boldsymbol{\Phi = \epsilon (u+k_1)^{k_2} }$} \\
 \hline
 \bf  Symmetries & \bf Reduced ODE \\
\midrule[1.5pt]
  $ X^{1} = X_2+ X_3$          & $h_2 k_2^2 [4 F^{\prime \prime \prime \prime} \alpha^{4}+20 F^{\prime \prime \prime} \alpha^3+16 F^{\prime \prime} \alpha^2 ]-16 F^{\prime \prime} \alpha^2=0$ \\
    &  Invariant solution: $ u = F(t e^{- \frac{k_2 x}{2}})$, $\alpha = t e^{- \frac{k_2 x}{2}} $.  \\
  \hline
    $ X^{2} = X_3$          & $h_1 F^{\prime \prime \prime \prime}+e^2 (F+k_1 )^{k_2} F^{\prime \prime}+e k_2 (F)+k_1 )^{k_2-1} [F^{\prime} ]^3=0.$ \\
     &  Invariant solution: $ u = F(\alpha)$, $ \alpha = x$.                  \\
  \hline
 $ X^{3}=X_2 + X_1$ &
       $ -(h_1 + h_2) F^{\prime \prime \prime \prime} + e k_2 (F + k_1)^{k_2-1} \left(F^{\prime}\right)^3 + F^{\prime \prime} \left(1 - e^2 (F + k_1)^{k_2}\right) = 0. $  \\
        &  Invariant solution: $ u =  F(t-x)$, $ \alpha = t-x$.               \\
         \hline
 $ X^{4}=X_1 $ &
       $  -e (F+k_1 )^{k_2-1} k_2 (F^{\prime} )^3-e^2 (F+k_1 )^{k_2} F^{\prime \prime}-h_1 F^{\prime \prime \prime \prime}=0 .$  \\
        &  Invariant solution: $ u =  F(\alpha)$, $ \alpha = x$.                 \\
\midrule[1.5pt]
\end{tabular}
\captionof{table}{Symmetry Reduction : $\Phi = \epsilon (u+k_1)^{k_2} $} \label{tabler7-1}
\end{center}

\begin{figure}[h]
    \centering
    \begin{minipage}{0.4\textwidth}
        \includegraphics[width=\linewidth]{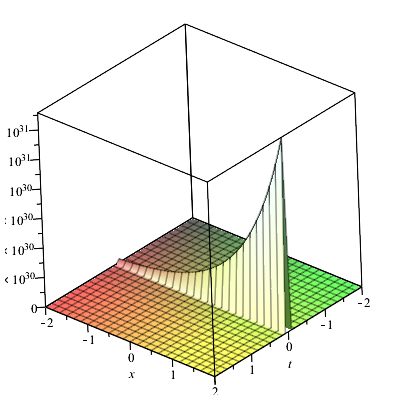}
    \end{minipage}
    \hfill
    \begin{minipage}{0.4\textwidth}
        \includegraphics[width=\linewidth]{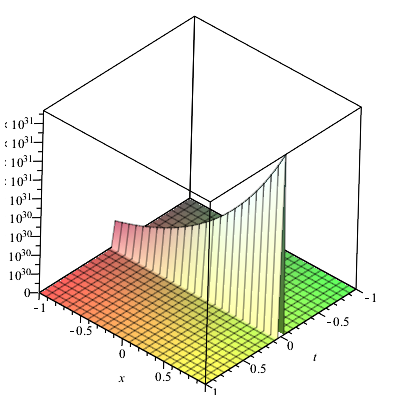}
    \end{minipage}
    \caption{ Solution $u(x,t)$ : Visualised for $t,x=-2,..,2$ and $t,x=-1,..,1$.}
    \label{invfig1k}
\end{figure}

\noindent \textbf{Reduced ODE-1:} The first reduced ODE corresponding to the first symmetry generator  $ X^{1} = X_2+ X_3$  
\begin{eqnarray}
h_2 k_2^2 [4 F^{\prime \prime \prime \prime} \alpha^{4}+20 F^{\prime \prime \prime} \alpha^3+16 F^{\prime \prime} \alpha^2 ]-16 F^{\prime \prime} \alpha^2=0,
\end{eqnarray}
can be solved further and has the following solution given by 
\begin{eqnarray}
F(\alpha)=c_1+c_2 \alpha+c_3 \alpha^{\frac{2}{\sqrt{h_2 k_2}}}+c_4 \alpha^{-\frac{2}{\sqrt{h_2} k_2}},
\end{eqnarray}
and finally the solution of PDE (\ref{Eq_3}) can be obtained by substituting similarity variables, we get  
\begin{eqnarray}
u(x,t) = c_1+c_2 t e^{- \frac{k_2 x}{2}} +c_3 (t e^{- \frac{k_2 x}{2}})^{\frac{2}{\sqrt{h_2 k_2}}}+c_4 (t e^{- \frac{k_2 x}{2}})^{-\frac{2}{\sqrt{h_2} k_2}}.
\end{eqnarray}
Solution $u(x,t)$ is presented in Figure-\ref{invfig1k}.

\noindent \textbf{Reduced ODE-2:} The second reduced ODE is a non-linear fourth order equation      
\begin{eqnarray} \label{red1}
h_1 F^{\prime \prime \prime \prime}(\alpha)+e^2 (F(\alpha)+k_1 )^{k_2} F^{\prime \prime}(\alpha)+e k_2 (F(\alpha)+k_1 )^{k_2-1} [F^{\prime}(\alpha) ]^3=0,
\end{eqnarray}
For above ODE (\ref{red1}) the solution of the form $F  ( \alpha  ) =k{\alpha}^{n}$ can be studied for $k_2=1$. The equation (\ref{red1}) reduced to the expression
\begin{eqnarray}
 [e k^2 (\alpha^n )^2 n^2 \alpha+e^2 k \alpha^2(n-1) \alpha^n+(n-1) (\alpha^2 e^2 k_1+n^2 h_1-5 n h_1+6 h_1 ) ] \alpha^{n-4} =0,
\end{eqnarray}
that provide us the value $n=4$ and further provide the value of $k$ given by
\begin{eqnarray}
k=- \frac{1}{32} \frac{3 e^2 \alpha^2+\sqrt{9 \alpha^4 e^4-192 \alpha^3 e^3 k_1-384 \alpha e h_1}}{e \alpha^5}.
\end{eqnarray}
Thus, solution for (\ref{red1}) can be presented by 
\begin{eqnarray}
F  ( \alpha  ) = - \frac{1}{32} \frac{3 e^2 \alpha^2+\sqrt{9 \alpha^4 e^4-192 \alpha^3 e^3 k_1-384 \alpha e h_1}}{e \alpha},
\end{eqnarray}
and the solution of PDE (\ref{Eq_3}) can be obtained by substituting the corresponding similarity variables, we have  
\begin{eqnarray}
u(x,t) = \frac{1}{32} \frac{3 \mathrm{e}^2 x^2+\sqrt{9 \mathrm{e}^4 x^4-192 \mathrm{e}^3 x^3-384 \mathrm{e} x}}{\mathrm{e} x}.
\end{eqnarray}

\noindent  \textbf{Reduced ODE-3:} The third reduced ODE is also a non-linear fourth order equation
\begin{eqnarray} \label{red2}
 - (h_1+h_2 ) F^{\prime \prime \prime \prime}+e k_2 (F+k_1 )^{k_2-1} (F^{\prime} )^3-e^2 (F+k_1 )^{k_2} F^{\prime \prime}+F^{\prime \prime}=0.
\end{eqnarray}
For above ODE (\ref{red2}) the solution of the form $F  ( \alpha  ) =k{\alpha}^{n}$ can be studied for $k_2=1$. The equation (\ref{red2}) reduced to the expression
\begin{eqnarray}
\alpha^{n-4} k n (-e k^2 (\alpha^n )^2 n^2 \alpha+k e^2 \alpha^2(n-1) \alpha^n+  \nonumber \\  (n-1) ( (h_1+h_2 ) n^2  + (-5 h_1-5 h_2 ) n+ (e^2 k_1-1 ) \alpha^2+6 h_1+6 h_2 )) =0,
\end{eqnarray}
that provide us the value $n=4$ and further provide the value of $k$ given by
\begin{eqnarray}
k=\frac{1}{32} \frac{3 \alpha^2 e^2+\sqrt{9 \alpha^4 e^4+192 \alpha^3 e^3 k_1-192 \alpha^3 e+384 \alpha e h_1+384 \alpha e h_2}}{e \alpha^5}.
\end{eqnarray}
Thus, solution for (\ref{red1}) can be presented by 
\begin{eqnarray}
F  ( \alpha  ) = \frac{1}{32} \frac{3 \alpha^2 e^2+\sqrt{9 \alpha^4 e^4+192 \alpha^3 e^3 k_1-192 \alpha^3 e+384 \alpha e h_1+384 \alpha e h_2}}{e \alpha},
\end{eqnarray}
and the solution of PDE (\ref{Eq_3}) can be obtained by substituting the corresponding similarity variables, we have  
\begin{eqnarray*}
u(x,t) = \frac{3 (t-x)^2 e^2+\sqrt{9 (t-x)^4 e^4+ (t-x)^3 [192e^3 k_1-192e]+(t-x) [384  e h_1+384 e h_2]}}{32e (t-x)}.
\end{eqnarray*}

\noindent  \textbf{Reduced ODE-4:} Similarly ODE-4 has the solution of the form $F  ( \alpha  ) =k{\alpha}^{n}$ for $k_2=1$ given by
\begin{eqnarray}
F  ( \alpha  ) = -\frac{1}{32} \frac{3 e^2 \alpha^2+\sqrt{9 \alpha^4 e^4-192 \alpha^3 e^3 k_1-384 \alpha e h_1}}{e \alpha},
\end{eqnarray}
and the solution of PDE (\ref{Eq_3}) can be obtained by substituting the corresponding similarity variables, we get   
\begin{eqnarray}
u(x,t) = \frac{1}{32} \frac{-3(\mathrm{e})^2 x^2-\sqrt{9 x^4(\mathrm{e})^4-192 x^3(\mathrm{e})^3 k_1-384 x \mathrm{e} h_1}}{x \mathrm{e}}.
\end{eqnarray}

\subsection{Symmetry Reduction : $\boldsymbol{ \Phi = \epsilon \exp (ku) }$}
The symmetry reduction of equation (\ref{Eq_3}) for $\Phi = \epsilon \exp (ku)$ is presented in Table \ref{tabler7-2}.
\begin{center}
\begin{tabular}{ l l }\toprule[1.5pt]
 \multicolumn{2}{c}{\bf Symmetry Reduction: $\Phi = \epsilon \exp (ku)$} \\
 \hline
 \bf  Symmetries & \bf Reduced ODE \\
\midrule[1.5pt]
  $ X^{1} = X_2+X_3$          & $F^{\prime \prime} \alpha^2 \left(16 - 16 h_2 k^2\right) - 20 k^2 \alpha^3 F^{\prime \prime \prime} h_2 - 4 k^2 \alpha^4 F^{\prime \prime \prime \prime} h_2 = 0.$ \\
    &  Invariant solution: $ u = F(\alpha)$, $\alpha = t e^{- \frac{kx}{2}} x^c$. \\
  \hline
    $ X^{2} = X_3$          & $-h_1 F^{\prime \prime \prime \prime}-\epsilon e^{k F}  [k F^{\prime 3}+F^{\prime \prime} ]=0.$ \\
     &   Invariant solution: $ u = F(\alpha)$, $ \alpha = x$.                \\
      \hline
    $ X^{3} = X_2 + X_1$          & $-(h_1 + h_2) F^{\prime \prime \prime \prime} + F^{\prime \prime \prime} e^{k F} \epsilon k + F^{\prime \prime} \left(1 - e^{k F} \epsilon\right) = 0.$ \\
     &   Invariant solution: $ u = F(\alpha)$, $ \alpha = t- x$.                \\
         \hline
    $ X^{4} =  X_1$          & $h_1 F^{\prime \prime \prime \prime}-\epsilon e^{k F} [k F^{\prime 3}+F^{\prime \prime} ]=0$ \\
     &  Invariant solution: $ u = F(\alpha)$, $ \alpha = x$.                  \\
\midrule[1.5pt]
\end{tabular}
\captionof{table}{Symmetry Reduction : $\Phi = \epsilon \exp (ku)$} \label{tabler7-2}
\end{center}

\begin{figure}[h]
    \centering
    \begin{minipage}{0.42\textwidth}
        \includegraphics[width=\linewidth]{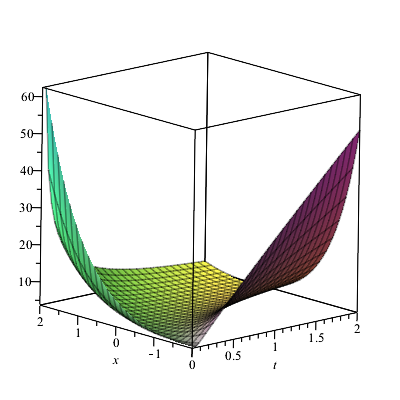}
    \end{minipage}
    \hfill
    \begin{minipage}{0.42\textwidth}
        \includegraphics[width=\linewidth]{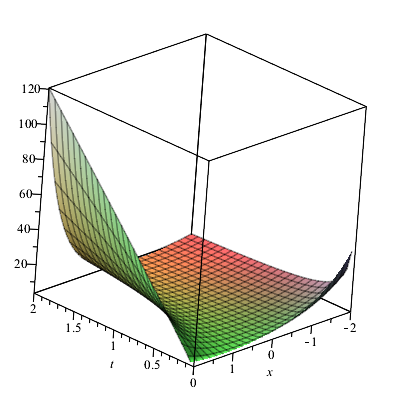}
    \end{minipage}
    \caption{  Solution $u(x,t)$ : Visualised for $t,x=0,..,2$ and $t,x=0,..,2$.}
    \label{invfig2k2}
\end{figure}

\noindent  \textbf{Reduced ODE-1:} Reduced ODE corresponding to symmetry generator $X^{1} = X_2+X_3$ has the exact solution given by
\begin{eqnarray}
F(\alpha) = c_1 + c_2 \alpha + c_3 \alpha^{\frac{2}{k}} + c_4 \alpha^{-\frac{2}{k}} .
\end{eqnarray}
Now by substituting the values of $F(\alpha)$ and $\alpha=  t e^{-\frac{1}{2} k x}$ back in the similarity variable $u(x, t) = F(\alpha)$  provide us the similarity solutions for equation (\ref{Eq_3}) given by 
\begin{eqnarray}
u(x, t) = c_1 + c_2 (t e^{-\frac{1}{2} k x}) + c_3 (t e^{-\frac{1}{2} k x})^{\frac{2}{k}} + c_4 (t e^{-\frac{1}{2} k x})^{-\frac{2}{k}} .
\end{eqnarray}
The geometrical plot of solution $u(x, t)$ is presented in Figure-\ref{invfig2k2}.

\subsection{Symmetry Reduction : $\boldsymbol{ \Phi = \epsilon (u)^{k_1} }$}
The symmetry reduction of equation (\ref{Eq_3}) for $\Phi = \epsilon (u)^{k_1}$ is presented in Table \ref{tabler7-3}.
\begin{center}
\begin{tabular}{ l l }\toprule[1.5pt]
 \multicolumn{2}{c}{\bf Symmetry Reduction : $\Phi = \epsilon (u)^{k_1}$} \\
 \hline
 \bf  Symmetries & \bf Reduced ODE \\
\midrule[1.5pt]
  $ X^{1} = X_3$          & $-4 F^{\prime \prime \prime \prime} h_2 \alpha^2 - 22 F^{\prime \prime \prime} h_2 \alpha - 20 F^{\prime \prime} h_2 =0.$ \\
    &  Invariant solution: $ u = F(\alpha)$, $\alpha = \frac{t}{x^2}$. \\
  \hline
    $ X^{2} = X_1+X_2$          & $F^{\prime \prime} \left(1 - \epsilon F^{k_1}\right) - F^{\prime \prime \prime \prime} \left(h_1 + h_2\right) + \epsilon F^{k_1-1} k_1 \left(F^{\prime}\right)^3 = 0.$ \\
     &   Invariant solution: $ u = F(\alpha)$, $ \alpha = t-x$.            \\
      \hline
    $ X^{3} =  X_1$          & $- \epsilon F^{k_1-1} k (F^{\prime})^3 - \epsilon F^{k_1} F^{\prime \prime}-h_1 F^{\prime \prime \prime \prime}=0$ \\
     &   Invariant solution: $ u = F(\alpha)$, $ \alpha = x$.                  \\
\midrule[1.5pt]
\end{tabular}
\captionof{table}{Symmetry Reduction : $\Phi = \epsilon (u)^{k_1}$} \label{tabler7-3}
\end{center}

\begin{figure}[h]
    \centering
    \begin{minipage}{0.42\textwidth}
        \includegraphics[width=\linewidth]{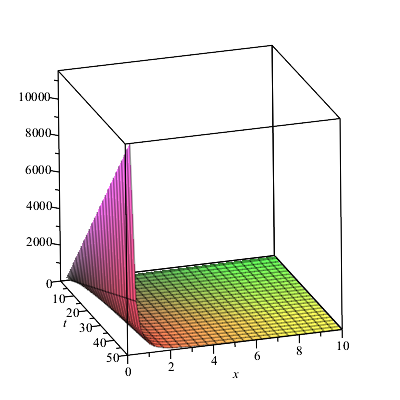}
    \end{minipage}
    \hfill
    \begin{minipage}{0.42\textwidth}
        \includegraphics[width=\linewidth]{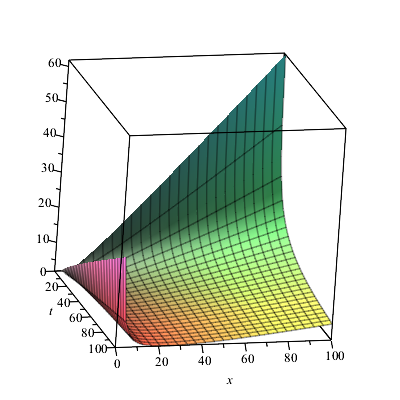}
    \end{minipage}
    \caption{ Solution $u(x,t)$ : Visualised for $t,x=0,..,10$ and $t,x=0,..,100$.}
    \label{invfig3k}
\end{figure}

\noindent \textbf{Reduced ODE-1:} The reduced ODE under $ X^{1} = X_3$ has the exact solution given by 
\begin{eqnarray}
F(\alpha)=c_1+c_2 \ln (\alpha)+\frac{c_3}{\sqrt{\alpha}}+c_4 \alpha.
\end{eqnarray}
The invariant solutions of equation (\ref{Eq_3}) is given by 
\begin{eqnarray}
u(x, t) = c_1+c_2 \ln  (\frac{t}{x^2} )+\frac{c_3}{\sqrt{\frac{t}{x^2}}}+\frac{c_4 t}{x^2} .
\end{eqnarray}
The 3D plot of solution $u(x, t)$ is presented in Figure-\ref{invfig3k}.

\noindent \textbf{Reduced ODE-2:} ODE under symmetry generator $ X^{2} = X_1+X_2$  has the solution of the form $F  ( \alpha  ) =k{\alpha}^{n}$ for $k_1=1$ given by
\begin{eqnarray}
F  ( \alpha  ) = \frac{[3 e \alpha^2+\sqrt{9 \alpha^4 e^2-192 \alpha^3 e+384 \alpha e h_1+384 \alpha e h_2}]}{32\alpha e},
\end{eqnarray}
and the solution of PDE (\ref{Eq_3}) can be obtained by substituting the corresponding similarity variables, we get   
\begin{eqnarray*}
u(x,t) = \frac{[3 e(t-x)^2+\sqrt{9 e^2(t-x)^4-192 e(t-x)^3+384(t-x) e h_1+384(t-x) e h_2}]}{32(t-x) e}.
\end{eqnarray*}

\noindent \textbf{Reduced ODE-3:} ODE under symmetry generator $ X^{3} = X_1$  has the solution of the form $F  ( \alpha  ) =k{\alpha}^{n}$ for $k_1=1$ given by
\begin{eqnarray}
F  ( \alpha  ) = -[{3 e \alpha^2-\sqrt{9 \alpha^4 e^2-384 \alpha e h_1}}]/({32 \alpha e}),
\end{eqnarray}
and the solution of PDE (\ref{Eq_3}) can be obtained by substituting the corresponding similarity variables, we get   
\begin{eqnarray}
u(x,t) = - [{3 e x^2-\sqrt{9 e^2 x^4-384 e x h_1}}]/({32x e}).
\end{eqnarray}

\subsection{Symmetry Reduction : $\boldsymbol{ \Phi = \epsilon \exp (u) }$}
The symmetry reduction of equation (\ref{Eq_3}) for $\Phi = \epsilon \exp (u)$ is presented in Table \ref{tabler7-4}.
\begin{center}
\begin{tabular}{ l l }\toprule[1.5pt]
 \multicolumn{2}{c}{\bf Symmetry Reduction : $\Phi = \epsilon \exp (u)$} \\
 \hline
 \bf  Symmetries & \bf Reduced ODE \\
\midrule[1.5pt]
  $ X^{1} = X_3$          & $-\alpha^2 F^{\prime \prime}  (16 h_2 k^2 + 4 h_2 k  ) - \alpha^3 F^{\prime \prime \prime}  (20 h_2 k^2 + 2 h_2 k  ) - 4 h_2 k^2 \alpha^4 F^{\prime \prime \prime \prime}   = 0$ \\
    & Invariant solution: $ u = F(\alpha)$, $\alpha = t x^{-2k}$. \\
  \hline
    $ X^{2} = a_1 X_1+X_2$          & $ F^{\prime \prime} \left(1 - e^{F} a_1^2 \epsilon \right) - F^{\prime \prime \prime \prime} \left(a_1^4 h_1 + a_1^2 h_2 \right) + (F^{\prime})^3 e^{F} a_1^3 \epsilon = 0.$ \\
     &   Invariant solution: $ u = F(\alpha)$, $ \alpha = \-a_1 x + t$.                  \\
      \hline
    $ X^{3} =  X_1$          & $-h_1 F^{\prime \prime \prime \prime}-\epsilon e^F  [(F^{\prime})^3+F^{\prime \prime} ]=0$ \\
     &  Invariant solution: $ u = F(\alpha)$, $ \alpha = x$.                  \\
\midrule[1.5pt]
\end{tabular}
\captionof{table}{Symmetry Reduction : $\Phi = \epsilon \exp (u)$} \label{tabler7-4}
\end{center}

\begin{eqnarray}
 , 
\end{eqnarray}

\noindent \textbf{Reduced ODE-1:} The reduced ODE under $ X^{1} = X_3$ has the exact solution given by 
\begin{eqnarray}
F(\alpha)=c_1+c_2 \ln (\alpha)+c_3 \alpha+c_4 \alpha^{-\frac{1}{2 k}}.
\end{eqnarray}
The invariant solutions of equation (\ref{Eq_3}) is given by 
\begin{eqnarray}
u(x, t) = c_1+c_2 \ln  (t x^{-2 k} )+c_3 t x^{-2 k}+c_4 (t x^{-2 k} )^{-\frac{1}{2 k}} .
\end{eqnarray}

\subsection{Invariant Solutions of PDE with Power Non-linearity }
\begin{figure}[h]
    \centering
    \begin{minipage}{0.49\textwidth}
        \includegraphics[width=\linewidth]{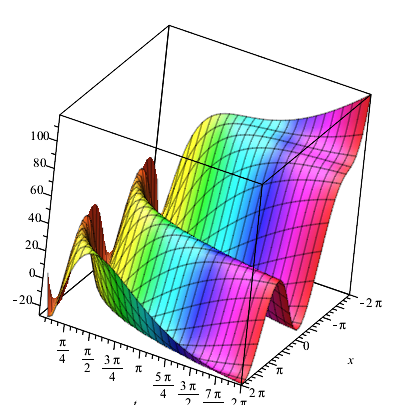}
    \end{minipage}
    \hfill
    \begin{minipage}{0.49\textwidth}
        \includegraphics[width=\linewidth]{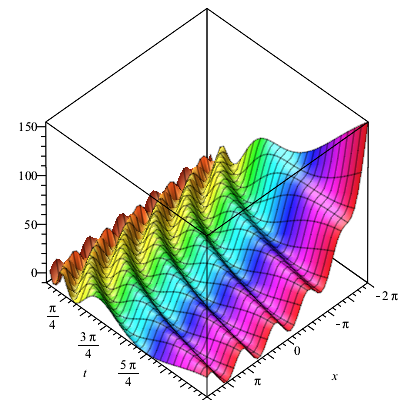}
    \end{minipage}
    \caption{Similarity solution is depicted along the axes $x = -2\pi \ldots 2\pi, t = 0 \ldots 2\pi $.}
    \label{invfig1}
\end{figure}

With the power law non-linearity, the equation (\ref{Eq_3}) takes the form given by
\begin{eqnarray}
 u_{tt} - c {u^{k-1} k u_x^3} - c u^k u_{xx} - h_1 u_{xxxx} - h_2 u_{xxtt} = 0. \nonumber
\end{eqnarray} 
For instance, characteristic equation corresponding to symmetry generator $$X^{1} = X_2+X_3= \frac{\partial}{\partial x} +  \frac{k_2}{2} t \frac{\partial}{\partial t} + u \frac{\partial}{\partial u}$$ is given by
\begin{equation} \label{chac1}
\frac{dt}{ \frac{k_2}{2} t} = \frac{dx}{1} = \frac{du}{u}.
\end{equation}
Under the similarity variables $\xi = t \exp  ( -\frac{1}{2} k_2 x  )$ and $u(x, t) = F(\xi)$. The set of ODEs for some finite values of $h_1, h_2, k_1$ and $k_2$ are given by 
\begin{eqnarray}
4 F^{(4)} \xi^4+20 F^{\prime \prime \prime} \xi^3+32 F^{\prime \prime} \xi^2=0 .  \label{ode1o}  
\end{eqnarray}
ODE (\ref{ode1o}) can easily be solved and its solution is 
\begin{eqnarray}
F(\xi) = c_1 + c_2 \xi + c_3 \sin(2 \ln(\xi)) + c_4 \cos(2 \ln(\xi)) .
\end{eqnarray}
Now by substituting the values of $F(\xi)$ and $\xi = t \exp  ( -\frac{1}{2} k_2 x  ) = te^{-\frac{1}{2} k_2 x}$ back in the similarity variable $u(x, t) = F(\xi)$  provide us the similarity solutions for equation (\ref{Eq_3}) given by 
\begin{eqnarray}
u_1(x, t) = c_1 + c_2 (te^{-\frac{1}{2} k_2 x}) + c_3 \sin(2 \ln(te^{-\frac{1}{2} k_2 x})) + c_4 \cos(2 \ln(te^{-\frac{1}{2} k_2 x})) .
\end{eqnarray}
The geometrical plot of the similarity solution $u_1(x, t)$ for finite values of coefficients is presented in Figure-\ref{invfig1}. 

\vspace{0.4cm}

\section{Multipliers and Corresponding Conservation Laws} \label{sec_a4}
Conservation laws are important in studying the invariance property of the surface generated by the solution of PDEs. In recent work, Stephen C. Anco \cite{s19} presented the general method to find conserved integrals based on multipliers. The approach is direct, comprehensive and systematic in finding conservation laws. The Euler operator is defined as
\begin{align}
\mathcal{E}  &= \frac{\partial }{\partial u} - D_x \frac{\partial }{\partial u_x} - D_t  \frac{\partial }{\partial u_t} + D^2_x  \frac{\partial }{\partial u_{xx}} + D_x D_t  \frac{\partial }{\partial u_{xt}} + \cdots
\end{align}
where $D_x$ and $D_t$ are the total derivative operators that can be obtained from
\begin{align*}
D_i = \frac{\partial}{\partial x^i} + u_i \frac{\partial}{\partial u} + \cdots, \, (x^1,x^2)=(t,x).
\end{align*}
A conserved vector of (\ref{Eq_3}) is a $2$-tuple $T=(T^t,T^x)$ such that
\begin{equation} \label{11aa}
D_t T^t + D_x T^x = 0 ,
\end{equation}
holds for all solutions of (\ref{Eq_3}). Equation (\ref{11aa}) is called the local conservation law. The multipliers $\mathcal{M}(x,t,u)$ of the PDE (\ref{Eq_3}) has the property
\begin{equation} \label{E12aa}
D_t T^t + D_x T^x = \mathcal{M} [  (u_{t t}-\phi_u u_x^2-\phi u_{x x} - h_1 u_{x x x x} - h_2 u_{x x t t})  ]  .
\end{equation}
The multipliers $\mathcal{M}(x,t,u)$ are obtained by taking the variational derivative of (\ref{E12aa}), which provides us the invariance criterion for finding multipliers $\mathcal{M}(x,t,u)$ given by
\begin{equation} \label{E13aa}
\mathcal{E} [\mathcal{M}   (u_{t t}-\phi_u u_x^2-\phi u_{x x} - h_1 u_{x x x x} - h_2 u_{x x t t} )  ]=0,
\end{equation}
where $\mathcal{E} = \frac{\delta}{\delta u} $ is the standard Euler operator. Consider the multiplier $\mathcal{M}(x,t,u)$ and substitute it in expression (\ref{E13aa}). Its simplification yields the following system of determining equations given by 
\begin{eqnarray}
\frac{\delta}{\delta u} [ \mathcal{M}  (u_{t t}-\phi_u u_x^2-\phi u_{x x} - h_1 u_{x x x x} - h_2 u_{x x t t} ) ]=0
\end{eqnarray}
We obtained the following multipliers $\{ \mathcal{M}_1, \mathcal{M}_2, \mathcal{M}_3, \mathcal{M}_4 \} = \{ xt, x, t, 1 \}$ by solving above system. The multipliers of wave equation (\ref{Eq_3}) for an arbitrary function $\Phi(u)$ indicating that the wave equation (\ref{Eq_3}) possesses four conservation laws.

\subsection{Conservation Laws for $\boldsymbol{\Phi = \epsilon (u)^{k_2} }$, where $\boldsymbol{\epsilon = \pm 1}$}
Consider the function $\Phi(u) = \epsilon (u)^{k_2}$, we obtain the the complete set of local conservation laws of (\ref{Eq_3}) for $\Phi(u)= \epsilon (u)^{k_2} $  which are presented in the form of conserved vectors:
\begin{eqnarray}
T_1 &= (T^t , \  \ T^x ) ,
\end{eqnarray}
where $T^t$ and $T^x$ are given by 
\begin{align*}
T^t &= -xt \epsilon u^3 u_x + \frac{1}{4} \epsilon t u^4 + \frac{1}{2} x t u_{xtt} + xt u_{xxx} + \frac{1}{6} t u_{tt} + \frac{1}{3} x u_{xt} + t u_{xx} - \frac{1}{3}  u_t, \\
T^x &= \frac{1}{2} x t u_{xxt} - \frac{1}{3} t u_{tx} - \frac{1}{6} x u_{xx} + x t u_t + \frac{1}{3} u_x - xu.
\end{align*}
\begin{equation}
T_2 = (T^t , \  \ T^x)
\end{equation}
where $T^t$ and $T^x$ are given by 
\begin{align*}
T^t &= -x \epsilon u^3 u_x + \frac{1}{4} \epsilon  u^4 + \frac{1}{2} x  u_{xtt} + x u_{xxx} - \frac{1}{6}  u_{tt} -  u_{xx} , \\
T^x &= \frac{1}{2} x  u_{xxt} - \frac{1}{3}  u_{tx} + x u_t .
\end{align*}
\begin{equation}
T_3 = (T^t, \  \ T^x ) = ( \frac{1}{2} t u_{xtt} + t u_{xxx} - \frac{1}{3} u_{xt} - t \epsilon u_x u^{k_2}   , \  \  \frac{1}{2} t u_{xxt} - \frac{1}{6} u_{xx} + t u_x - u ),
\end{equation}
\begin{equation}
T_4 = (T^t, \  \ T^x ) = ( \frac{u_{xtt}}{2} + u_{xxx} - \epsilon u^3 u_x  , \  \  \frac{u_{xxt}}{2} +u_t ).
\end{equation}
which corresponds to the following four multipliers  $\{ \mathcal{M}_1, \mathcal{M}_2, \mathcal{M}_3, \mathcal{M}_4 \}$ = $ \{ xt, x, $ $ t, 1 \}$ respectively.

\subsection{Conservation Laws for $\boldsymbol{\Phi = \epsilon \exp (ku) }$, where $\boldsymbol{\epsilon = \pm 1}$}
The conserved vectors for the function $\Phi(u) =  \epsilon \exp (ku) $ are obtained. For the multiplier $\mathcal{M}(x,t,u)=xt$, we deduce
\begin{align}
T_1 = (T^t , \  T^x) =    (-\frac{1}{2}x t (2 \epsilon \exp (k u) u_x-2 u_{x x x}-u_{x t t} ), \frac{1}{2} x t (u_{x x t}+2 u_t ) ).
\end{align}
For the multiplier $\mathcal{M}(x,t,u)=x$, we find
\begin{align}
T_2 &= (T^t , \  T^x), \nonumber \\
    = &  \ \ (-\frac{1}{6 k} (6 x k \epsilon \mathrm{e}^{k u} u_x-6 x u_{x x x} k-3 x u_{x t t} k-6 \epsilon \mathrm{e}^{k u}+6 u_{x x} k+u_{t t} k+6 \epsilon), \\
    &  \ \  \frac{1}{2} x u_{x x t}-\frac{1}{3} u_{x t}+x u_t ).  \nonumber
\end{align}
For the multiplier $\mathcal{M}(x,t,u)=t$, we obtain
\begin{align}
T_3 &= (T^t , \  T^x) \nonumber \\
    &=    (\frac{t u_{x t t}}{2}+t u_{x x x}-\frac{u_{x t}}{3}-t \epsilon \mathrm{e}^{k u} u_x, \frac{1}{2} t u_{x x t}-\frac{1}{6} u_{x x}+t u_t-u ).
\end{align}
For the multiplier $\mathcal{M}(x,t,u)=1$, we have
\begin{align}
T_4 = (T^t , \  T^x) =    (\frac{u_{x t t}}{2}+u_{x x x}-\epsilon \mathrm{e}^{k u^2} u_x, \frac{u_{x x t}}{2}+u_t ).
\end{align}

\subsection{Conservation Laws for $\boldsymbol{\Phi = \epsilon \exp (u) }$, where $\boldsymbol{\epsilon = \pm 1}$}
The conserved vectors for the function $\Phi(u) =  \epsilon \exp (u) $ are obtained.
For the multiplier $\mathcal{M}(x,t,u)=xt$, we deduce
\begin{align}
T_1 = (T^t , \  T^x) =    (-\frac{1}{2}x t (2 \epsilon \exp ( u) u_x-2 u_{x x x}-u_{x t t} ), \frac{1}{2} x t (u_{x x t}+2 u_t ) ).
\end{align}
For the multiplier $\mathcal{M}(x,t,u)=x$, we find
\begin{align}
T_2 &= (T^t , \  T^x), \nonumber \\
    = &  \ \ (-\frac{1}{6} (6 x  \epsilon \mathrm{e}^{ u} u_x-6 x u_{x x x} -3 x u_{x t t} -6 \epsilon \mathrm{e}^{ u}+6 u_{x x} +u_{t t} +6 \epsilon), \\
    &  \ \  \frac{1}{2} x u_{x x t}-\frac{1}{3} u_{x t}+x u_t ).  \nonumber
\end{align}
For the multiplier $\mathcal{M}(x,t,u)=t$, we obtain
\begin{align}
T_3 &= (T^t , \  T^x) \nonumber \\
    &=    (\frac{t u_{x t t}}{2}+t u_{x x x}-\frac{u_{x t}}{3}-t \epsilon \mathrm{e}^{ u} u_x, \frac{1}{2} t u_{x x t}-\frac{1}{6} u_{x x}+t u_t-u ).
\end{align}
For the multiplier $\mathcal{M}(x,t,u)=1$, we have
\begin{align}
T_4 = (T^t , \  T^x) =    (\frac{u_{x t t}}{2}+u_{x x x}-\epsilon \mathrm{e}^{ u} u_x, \frac{u_{x x t}}{2}+u_t ).
\end{align}

In a recent work, Anco \cite{s19} proposed a study that establishes a connection between multipliers and the variational symmetry that underlies Noether's theorem. In another study, Anco and Gandarias \cite{anga1} presented the application of conservation laws in multi-reductions. All of the conservation laws obtained by the multiplier approach have a clear application in mathematical analysis. Although some conservation laws describe physical quantities and explain natural phenomena like as momentum, energy, and object motion, the rest of the conservation laws describe the geometry of the surface formed by the partial differential equation.

\section{The Nerve Equation } \label{sec_a0}

In this section, we'll discuss the particular physical model which represent the phenomenon of liquid flows in the bio-membranes by considering the particular valued of $\Phi(u)$ in equation (\ref{Eq_3}). We perform a complete analysis of the equations that govern the models of the nerve membranes, i.e., the second (\ref{bio1})  and fourth order (\ref{bio3}) cases.

\subsection{Analysis of the class of Second Order Nerve Equation} \label{sec_a1}
The invariance criterion for the second order case is  
\begin{eqnarray}
X^{[2]} [u_{tt} - (2pu + q) u^2_x - (p u^2 + qu + r) u_{xx}]_{\LARGE|_{u_{tt} = ( (p u^2 + qu + r)\cdot u_{x}  )_x}}=0, 
\end{eqnarray}

for which we construct cases based on the forms of $p,q$ and $r$. 
\subsubsection{Invariance Analysis and Optimal System: $p \neq 0 $, $p \neq \frac{1}{4} \cdot \frac{q^2}{r}, q \neq 0$  }
For this case, we have obtained the following set of infinitesimals
\begin{align*}
\xi^1 =  \frac{3}{2} c_1 t + c_4 , \ \ \
\xi^2     =   \frac{3}{2} c_1 x + \tilde{c_2} , \ \ \ \
\eta   = 0 .
\end{align*}
The algebra in this case is spanned by $X_1, \ X_2$ and $X_3$.  
\begin{align*}
 X_1 = \frac{\partial}{\partial t}, \ \ \ \ \  
  X_2 = \frac{\partial}{\partial x}, \ \ \ \ \ \ 
 X_3 = t \frac{\partial}{\partial t} +  x \frac{\partial}{\partial x}. \ \ \ \ \ 
\end{align*}
Where the nonzero commutators given by 
\begin{equation}
[X_1,X_3]=X_1, \, [X_2,X_3]=X_2.
\end{equation}
An optimal system of non equivalent symmetry generators for the general class of wave equations are studied in \cite{s2}. Non equivalent symmetry generators are given by 
\[  
   \langle
                \begin{array}{ll}
                {X^1}  =   X_3 \pm X_1  \  , \ \   \  \  &{X^2} =  X_3\\
               {X^3} =  X_2 + c_1 X_1 \  , \ \   \  \ &{X^4} =  X_1 .
                \end{array}
               \rangle
  \]
Reduction under optimal system has performed for the equation-(\ref{bio1}). The procedure is outlined and demonstrated in \cite{s2}.\\

\begin{figure}[h]
    \centering
    \begin{minipage}{0.49\textwidth}
        \includegraphics[width=\linewidth]{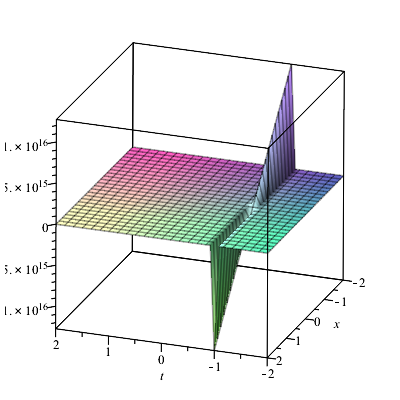}
    \end{minipage}
    \hfill
    \begin{minipage}{0.49\textwidth}
        \includegraphics[width=\linewidth]{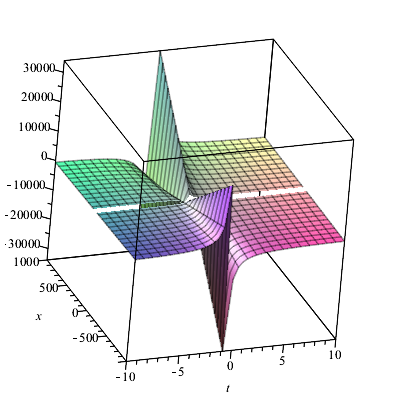}
    \end{minipage}
    \caption{ Solution $u(x,t)$ : Visualised for two sets of values $t,x$}
    \label{invfig1za12}
\end{figure}

\noindent \textbf{Reduction under ${X^1}$:} Reduced equation corresponding to ${X^1} $ is 
\begin{eqnarray} \label{redode_a1}
\begin{gathered}
F^{\prime \prime}(\alpha) (-F(\alpha)^2 p \alpha^2-F(\alpha) q \alpha^2-r \alpha^2+1 ) +F^{\prime}(\alpha)^2 (-2 F(\alpha) p \alpha^2-q \alpha^2 ) \\ +F^{\prime}(\alpha) (-2 F(\alpha)^2 p \alpha-2 F(\alpha) q \alpha-2 r \alpha ) = 0
\end{gathered}
\end{eqnarray}
where the similarity variables are $u(x, t)=F (\frac{1+t}{x} )$ and $\alpha=\frac{1+t}{x}$. Integrating equation (\ref{redode_a1}) simplifies to equation  
\begin{eqnarray}
-F^{\prime}(\alpha) (F(\alpha)^2 p \alpha^2+F(\alpha) q \alpha^2+r \alpha^2-1 )=0,
\end{eqnarray}
has the solution $F(\alpha)$ given by
\begin{eqnarray}
F(\alpha)=\frac{-\alpha q+\sqrt{-4 \alpha^2 p r+\alpha^2 q^2+4 p}}{2p \alpha}.
\end{eqnarray}
The solution of PDE (\ref{bio1}) via similarity variables $u(x, t)=F (\frac{1+t}{x} )$ and $\alpha=\frac{1+t}{x}$ can be represented by
\begin{eqnarray}
u(x,t) = \frac{1}{2} \frac{ (-\frac{(1+t) q}{x}+\sqrt{-\frac{4(1+t)^2 p r}{x^2}+\frac{(1+t)^2 q^2}{x^2}+4 p} ) x}{p(1+t)}.
\end{eqnarray}
The solution of second order nerve equation (\ref{bio1}) in bio-membranes can be visualised in Figure-(\ref{invfig1za12}). The graph of the solution in this case is presented for some finite values of parameters $p,q$ and $r$. \\
 
\noindent \textbf{Reduction under ${X^2}$:} Reduced equation corresponding to ${X^2} $ is 
\begin{eqnarray} \label{redode_a2}
\begin{gathered}
 -F(\alpha)^2 p \alpha^2 F^{\prime \prime}(\alpha)-2 F(\alpha) p \alpha^2 F^{\prime}(\alpha)^2 -F(\alpha) q \alpha^2 F^{\prime \prime}(\alpha)-q \alpha^2 F^{\prime}(\alpha)^2 \\
 -2 F(\alpha)^2 p \alpha F^{\prime}(\alpha)-r \alpha^2 F^{\prime \prime}(\alpha)  -2 F(\alpha) q \alpha F^{\prime}(\alpha)-2 r \alpha F^{\prime}(\alpha)  +F^{\prime \prime}(\alpha)=0
\end{gathered}
\end{eqnarray}
where the new variables are $u(x, t)=F (\frac{t}{x} )$ and $\alpha=\frac{t}{x}$. ODE (\ref{red_a2}) has the solution 
\begin{eqnarray}
F(\alpha)= [{-q \alpha+\sqrt{-4 \alpha^2 p r+\alpha^2 q^2+4 p}}]/({2p \alpha}),
\end{eqnarray}
that provide the solution of nerve equation (\ref{bio1}) under similarity variables given by 
\begin{eqnarray}
u(x,t) = [{ (-\frac{q t}{x}+\sqrt{-\frac{4 t^2 p r}{x^2}+\frac{t^2 q^2}{x^2}+4 p} ) x}]/({2p t}).
\end{eqnarray}

\noindent \textbf{Reduction under ${X^3}$:} Reduced equation corresponding to ${X^3} $ is
\begin{eqnarray} \label{redode_a3}
\begin{gathered}
F''(\alpha)  (1 - c^2 p F(\alpha)^2 - c^2 q F(\alpha) - c^2 r ) - F'(\alpha)^2 c^2  (2p F(\alpha) + q ) = 0,
\end{gathered}
\end{eqnarray}
where the new variables are $u(x, t)=F(\alpha)$ and $\alpha=-cx+t$. ODE (\ref{red_a3}) has the solution free from $\alpha$ given by 
\begin{eqnarray}
F(\alpha)=-[{q c+\sqrt{-4 c^2 p r+c^2 q^2+4 p}}]/({2p c}).
\end{eqnarray} 
which is not interesting case to study.

\noindent \textbf{Reduction under ${X^4}$:} Reduced equation corresponding to ${X^4} $ is 
\begin{eqnarray} \label{redode_a4}
\begin{gathered}
-2 F'(\alpha)^2 p F(\alpha) - F'(\alpha)^2 q - F''(\alpha)  ( p F(\alpha)^2 + q F(\alpha) + r  ) = 0,
\end{gathered}
\end{eqnarray}
where the new variables are $u(x, t)=F(\alpha)$ and $\alpha=x$. Above ODE has the solution given by
\begin{eqnarray}
F(\alpha)=- [{q+\sqrt{-4 p r+q^2}}]/{2p},
\end{eqnarray}
which is also independent of $\alpha$ and not interesting case to study further.

\subsubsection{Invariance Analysis and Optimal System: $q,r \neq 0, p=0$}
The PDE in this case reduces to 
\begin{eqnarray} \label{pde_red1}
u_{tt} = ( (qu + r) u_x)_x. 
\end{eqnarray}
The infinitesimal in this case are 
\begin{align*}
\xi^1 = c_5 + c_6 t , \quad 
\xi^2 =  \frac{1}{3}c_6 x  + c_1 x + \tilde{c_2}  \ \ \ \ , \ \ \tilde{c_2} = c_2 - \frac{2}{3}c_7  ,  \quad
\eta = \frac{2}{k} \tilde{c_1} .
\end{align*}
The algebra is spanned by $X_1, \ X_2 , \ X_3$ and by one extra symmetry $X_4$. 
\begin{align*}
X_1 = \frac{\partial}{\partial t} ,   \quad
X_2 = \frac{\partial}{\partial x}  ,   \quad
X_3 =  t \frac{\partial}{\partial t} + x \frac{\partial}{\partial x}  ,  \quad
X_4 =  x \frac{\partial}{\partial x} + \frac{2}{k} \frac{\partial}{\partial u} ,
\end{align*}
where $X_4$ equivalently can be written as 
\begin{align*}
X_4 &=  xq \frac{\partial}{\partial x} + 2(qu+r) \frac{\partial}{\partial u} .
\end{align*}
The non-zero commutators in this algebra are 
\begin{equation}
[X_1,X_3]=X_1 ,  [X_2,X_3]=X_2 \ , [X_2,X_4]=X_2.
\end{equation}
The optimal system including non equivalent symmetry generators is presented by   
\[  
   \langle
                \begin{array}{ll}
                X^1 =  X_3+c_4 X_4  \  , \ \   \  \  &X^2 = \pm X_2 + X_3 - X_4 \\
                X^3 = X_3 - X_4 \  , \ \   \  \ &X^4 =   \pm X_1 + X_4 \\
                X^5 =  X_4 \  , \ \   \  \ &X^6 =  X_1 \pm  X_2  \\
                X^7 = X_1 \  , \ \   \  \ &X^8 = X_2.
                \end{array}
               \rangle
  \]

\noindent \textbf{Reduction under ${X^1}$:} Reduced equation corresponding to ${X^1} $ is 
\begin{eqnarray} \label{red_a1}
\begin{gathered}
-  ( q F(\alpha) + r  ) F''(\alpha) \alpha^2 - q F'(\alpha)^2 \alpha^2 - 3  ( q F(\alpha) + r  ) F'(\alpha) \alpha=0,
\end{gathered}
\end{eqnarray}
where the new variables are $u(x, t)=F(\alpha)$ and $\alpha=\frac{t}{\sqrt{x}}$. The solution of the reduced ODE is 
\begin{eqnarray}
F(\alpha)=-[{r \alpha+\sqrt{2 \alpha^2 c_2 q+\alpha^2 r^2-c_1 q}}]/({q \alpha}),
\end{eqnarray}
The solution of PDE (\ref{pde_red1}) under similarity variables are given by
\begin{eqnarray}
u(x,t) = -[{ (\frac{r t}{\sqrt{x}}+\sqrt{\frac{2 t^2 c_2 q}{x}+\frac{t^2 r^2}{x}-c_1 q} ) \sqrt{x}}]/({q t}).
\end{eqnarray}

\noindent \textbf{Reduction under ${X^2}$:} Reduced equation corresponding to ${X^2} $ is 
\begin{eqnarray} \label{red_a2}
\begin{gathered}
F''(\alpha) \alpha^2 ( F(\alpha) q + r) + q F'(\alpha)^2 \alpha^2 + \alpha F'(\alpha) (F(\alpha) q + r) = 0
\end{gathered}
\end{eqnarray}
where the new variables are $u(x, t)=F(\alpha)$ and $\alpha= t e^{-x}$. The solution of reduced ODE is give by 
\begin{eqnarray}
F(\alpha)=[{-r+\sqrt{2 \ln (\alpha) c_1 q+2 c_2 q+r^2}}]/{q}.
\end{eqnarray}
Finally the solution of PDE (\ref{pde_red1}) under similarity variables has the form
\begin{eqnarray}
u(x,t) = [{-r+\sqrt{2 \ln  (t \mathrm{e}^{-x} ) c_1 q+2 c_2 q+r^2}}]/{q}.
\end{eqnarray}

\noindent \textbf{Reduction under ${X^3}$:} Reduced equation corresponding to ${X^3} $ has the exact solution which is presented below 
\begin{eqnarray} \label{red_a3}
\begin{gathered}
-q F'(\alpha)^2 - F''(\alpha) (q F(\alpha) + r) = 0,
\end{gathered}
\end{eqnarray}
Since, the new variables are $u(x, t)=F(\alpha)$ and $\alpha= x$. Thus, the exact solution $u(x, t)$ for ODE (\ref{red_a3}) and the PDE in this case are given by 
\begin{eqnarray}
u(x,t) = -[{r+\sqrt{2 c_1 q x+2 c_2 q+r^2}}]/{q}.
\end{eqnarray}

\noindent \textbf{Reduction under ${X^4}$:} Reduced equation corresponding to ${X^4} $ is 
\begin{eqnarray} \label{red_a4}
\begin{gathered}
-qF(\alpha)F''(\alpha) - q(F'(\alpha))^2 - qF(\alpha)F'(\alpha) - rF''(\alpha) - rF'(\alpha) = 0
\end{gathered}
\end{eqnarray}
where the new variables are $u(x, t)=F(\alpha)$ and $\alpha= t - \ln(x) $.	 The exact solution $F(\alpha)$ for ODE (\ref{red_a4}) are given by
\begin{eqnarray}
F(\alpha)=[{-r+\sqrt{-2 \mathrm{e}^{-\alpha} c_1 q+2 c_2 q+r^2}}]/{q}.
\end{eqnarray}
Since, $\alpha= - \ln(x) + t$. Thus, general class of exact solutions of PDE in this case are  
\begin{eqnarray}
u(x,t) = [{-r+\sqrt{-2 \mathrm{e}^{\ln (x)-t} c_1 q+2 c_2 q+r^2}}]/{q}.
\end{eqnarray}

\noindent \textbf{Reduction under ${X^5}$:} Reduced equation corresponding to ${X^5} $ is 
\begin{eqnarray}
F''(\alpha) = 0,
\end{eqnarray}
where the new variables are $u(x, t)=F(\alpha)$ and $\alpha= t$.\\

\noindent \textbf{Reduction under ${X^6}$:} Reduced equation corresponding to ${X^6} $ is 
\begin{eqnarray} \label{red_a6}
\begin{gathered}
F''(\alpha) - q(F'(\alpha))^2 - qF(\alpha)F''(\alpha) - rF''(\alpha) = 0
\end{gathered}
\end{eqnarray}
where the new variables are $u(x, t)=F(\alpha)$ and $\alpha= t-x$. The exact solution $F(\alpha)$ for ODE (\ref{red_a6}) are given by
\begin{eqnarray}
F(\alpha) = \pm [\sqrt{c_{2}+2c_{1}\alpha+1}-1].
\end{eqnarray}
Since, $\alpha= t-x$. Thus, general class of exact solutions of PDE in this case are  
\begin{eqnarray}
u(x, t) =  \pm [\sqrt{c_{2}+2c_{1}(t-x)+1}-1].
\end{eqnarray}

\noindent \textbf{Reduction under ${X^7}$:} Reduced equation corresponding to ${X^7} $ is 
 \begin{eqnarray} \label{red_a7}
\begin{gathered}
-q(F'(\alpha))^2 - qF(\alpha)F''(\alpha) - rF''(\alpha) = 0
\end{gathered}
\end{eqnarray}
where the new variables are $u(x, t)=F(\alpha)$ and $\alpha= x$.
Since, $\alpha= x$, thus general class of exact solutions of PDE in this case are  
\begin{eqnarray}
u(x, t) = \pm [\frac{\sqrt{c_{2}+4c_{1}x+9}}{2}-\frac{3}{2}].
\end{eqnarray}

\noindent \textbf{Reduction under ${X^8}$:} Reduced equation corresponding to ${X^8} $ is 
\begin{eqnarray}
F''(\alpha) = 0,
\end{eqnarray}
where the new variables are $u(x, t)=F(\alpha)$ and $\alpha= t$.

\subsubsection{Invariance Analysis and Optimal system: $q,r =0, p \neq 0$}
The PDE in this case can take the form 
\begin{eqnarray} \label{pde_red3}
u_{tt} = ( (p u^2 ) u_x)_x.
\end{eqnarray}
The algebra in this case is spanned by $X_1, \ X_2 , \ X_3$ and by one extra symmetry $X_4$. 
\begin{align*}
X_1 = \frac{\partial}{\partial t} ,  \quad
X_2 = \frac{\partial}{\partial x} ,  \quad
X_3 =  t \frac{\partial}{\partial t} + x \frac{\partial}{\partial x}  ,  \quad
X_4 =  x \frac{\partial}{\partial x} + u \frac{\partial}{\partial u}  .
\end{align*}
The non-zero commutators are 
\begin{equation*}
[X_1,X_3]=X_1 ,  [X_2,X_3]=X_2 \ , [X_2,X_4]=X_2.
\end{equation*}
The optimal system including non equivalent symmetry generators is presented by   
\[  
   \langle
                \begin{array}{ll}
                X^1 =  X^3+c_4 X_4  \  , \ \   \  \  &X^2 = \pm X_2 + X_3 - X_4 \\
                X^3 = X_3 - X_4 \  , \ \   \  \ &X^4 =   \pm X_1 + X_4 \\
                X^5 =  X_4 \  , \ \   \  \ &X^6 =  X_1 \pm  X_2  \\
                X^7 = X_1 \  , \ \   \  \ &X^8 = X_2.
                \end{array}
               \rangle
  \]

\noindent \textbf{Reduction under ${X^1}$:} Reduced equation corresponding to ${X^1} $ is 
\begin{eqnarray}
- pF(\alpha)^2 F''(\alpha) \alpha^2 - 2pF(\alpha) (F'(\alpha))^2 \alpha^2 - 3pF(\alpha)^2 F'(\alpha) \alpha = 0,
\end{eqnarray}
where the new variables are $u(x, t)=F(\alpha)$ and $\alpha=\frac{t}{\sqrt{x}}$. The solution of the reduced ODE is given by
\begin{eqnarray}
F(\alpha)=\frac{1}{2} \frac{ ( (24 \alpha^2 c_2-12 c_1 ) \alpha )^{1 / 3}}{\alpha}.
\end{eqnarray} 
Further provide us the solution of PDE (\ref{pde_red3}) is given by
\begin{eqnarray}
u(x,t) = \frac{1}{2} \frac{ ( (24(\frac{t}{\sqrt{x}})^2 c_2-12 c_1 )(\frac{t}{\sqrt{x}}) )^{1 / 3}}{[\frac{t}{\sqrt{x}}]}.
\end{eqnarray}

\noindent \textbf{Reduction under ${X^2}$:} Reduced equation corresponding to ${X^2} $ is 
\begin{eqnarray}
- pF(\alpha)^2 F''(\alpha) \alpha^2 - 2pF(\alpha) (F'(\alpha))^2 \alpha^2 - pF(\alpha)^2 F'(\alpha) \alpha = 0,
\end{eqnarray}
has the solution 
\begin{eqnarray}
F(\alpha)= (3 c_1 \ln (\alpha)+3 c_2 )^{1 / 3},
\end{eqnarray}
where the new variables are $u(x, t)=F(\alpha)$ and $\alpha= t e^{-x}$. The solution of the PDE (\ref{pde_red3}) under similarity variables is given by 
\begin{eqnarray}
u(x,t) =  (3 c_1 \ln  (t \mathrm{e}^{-x} )+3 c_2 )^{1 / 3}.
\end{eqnarray}

\noindent \textbf{Reduction under ${X^3}$:} Reduced equation corresponding to ${X^3} $ is 
 \begin{eqnarray} \label{red_b1}
\begin{gathered}
-p F(\alpha)  ( F(\alpha) F''(\alpha) + 2 (F'(\alpha))^2  )=0,
\end{gathered}
\end{eqnarray}
where the new variables are $u(x, t)=F(\alpha)$ and $\alpha= x$. The solution of PDE (\ref{pde_red3}) is 
\begin{eqnarray}
u(x, t) &= 3^{1/3}(c_2+c_1x)^{1/3}.
\end{eqnarray}

\noindent \textbf{Reduction under ${X^4}$:} Reduced equation corresponding to ${X^4} $ is 
 \begin{eqnarray} \label{red_b2}
\begin{gathered}
-p F(\alpha)^2 F''(\alpha) - p F(\alpha)^2 F'(\alpha) - 2p F(\alpha) (F'(\alpha))^2 = 0 ,
\end{gathered}
\end{eqnarray}
where the new variables are $u(x, t)=F(\alpha)$ and $\alpha= t - \ln(x) $. The exact solution $F(\alpha)$ for ODE is given by
\begin{eqnarray}
F(\alpha) =   (c_{1}+\mathrm{e}^{c_{2}-\alpha} )^{1/3}.
\end{eqnarray}
Since, $\alpha= t - \ln(x) $. Thus, general class of exact solutions of PDE (\ref{pde_red3}) is 
\begin{eqnarray}
u(x, t) &=   (c_{1}+\mathrm{e}^{c_{2}-(t - \ln(x))} )^{1/3}.
\end{eqnarray}
  
\noindent \textbf{Reduction under ${X^5}$:} Reduced equation corresponding to ${X^5} $ is 
\begin{eqnarray}
F''(\alpha) = 0,
\end{eqnarray}
where the new variables are $u(x, t)=F(\alpha)$ and $\alpha= t$.\\

\noindent \textbf{Reduction under ${X^6}$:} Reduced equation corresponding to ${X^6} $ is 
\begin{eqnarray}
F''(\alpha) - 2pF(\alpha) (F'(\alpha))^2 - pF(\alpha)^2 F''(\alpha) = 0,
\end{eqnarray}
where the new variables are $u(x, t)=F(\alpha)$ and $\alpha= t-x$. The solution of the reduced ODE is presented by 
\begin{eqnarray}
F(\alpha)=\frac{1}{8} \frac{ ( ( (12 c_1 \alpha+4 \sqrt{\frac{-4+9 (\alpha c_1+c_2 )^2 p}{p}}+12 c_2 ) p^2 )^{2 / 3}+4 p ) 4^{2 / 3}}{p ( (3 c_1 \alpha+3 c_2+\sqrt{\frac{-4+9 (\alpha c_1+c_2 )^2 p}{p}} ) p^2 )^{1 / 3}}.
\end{eqnarray}
Solution of PDE (\ref{pde_red3}) is 
\begin{eqnarray}
u(x,t)=\frac{1}{8} \frac{ (4^{2 / 3} ( (3 c_1( t-x)+3 c_2+\sqrt{\frac{-4+9 (c_1( t-x)+c_2 )^2 p}{p}} ) p^2 )^{2 / 3}+4 p ) 4^{2 / 3}}{p ( (3 c_1( t-x)+3 c_2+\sqrt{\frac{-4+9 (c_1( t-x)+c_2 )^2 p}{p}} ) p^2 )^{1 / 3}}.
\end{eqnarray}

\noindent \textbf{Reduction under ${X^7}$:} Reduced equation corresponding to ${X^7} $ is 
\begin{eqnarray}
-p F(\alpha)  ( F(\alpha) F''(\alpha) + 2 (F'(\alpha))^2  ) = 0,
\end{eqnarray}
where the new variables are $u(x, t)=F(\alpha)$ and $\alpha= x$. The exact solution of PDE (\ref{pde_red3}) can be obtained by simple substituting similarity variables to get solution in original variables.  
\begin{eqnarray}
u(x, t) &= 3^{1/3}(c_2+c_1x)^{1/3}.
\end{eqnarray}

\noindent \textbf{Reduction under ${X^8}$:} Reduced equation corresponding to ${X^8} $ is 
\begin{eqnarray}
F''(\alpha) = 0,
\end{eqnarray}
where the new variables are $u(x, t)=F(\alpha)$ and $\alpha= t$.

\subsubsection{Invariance Analysis and Optimal system: $p,r \neq 0, q=0$ }
The PDE in this case tak the form 
\begin{eqnarray} \label{PDE_red4}
u_{tt} = ( (p u^2 + r) u_x)_x .
\end{eqnarray}
The algebra is spanned by $X_1, \ X_2 , \ X_3$ given by 
\begin{align*}
X_1 = \frac{\partial}{\partial t} ,  \quad
X_2 = \frac{\partial}{\partial x}  ,  \quad
X_3 =  t \frac{\partial}{\partial t} + x \frac{\partial}{\partial x}  .
\end{align*}
with nonzero commutators  
\begin{equation}
[X_1,X_3]=X_1, \, [X_2,X_3]=X_2.
\end{equation}
The optimal system including non equivalent symmetry generators is listed by 
\[  
   \langle
                \begin{array}{ll}
                {X^1}  =   X_3 \pm X_1  \  , \ \   \  \  &{X^2} =  X_3  \\
                {X^3} =  X_2 + c_1 X_1 \  , \ \   \  \ &{X^4} =  X_1 .
                \end{array}
               \rangle
  \]

\noindent \textbf{Reduction under ${X^1}$:} Reduced equation corresponding to ${X^1} $ is 
\begin{eqnarray}
F^{\prime \prime} \left(1 - F^2 p \alpha^2 - r \alpha^2 \right) - 2 F \left(F^{\prime}\right)^2 p \alpha^2 - 2 F^2 F^{\prime} p \alpha - 2 F^{\prime} r \alpha = 0, 
\end{eqnarray}
where the new variables are $u(x, t)=F(\alpha)$ and $\alpha=\frac{1+t}{x}$. The reduced ODE has the solution 
\begin{eqnarray}
F(\alpha)=[{\sqrt{-p (\alpha^2 r-1 )}}]/({p \alpha}).
\end{eqnarray} 
that further provide us the solution of PDE (\ref{PDE_red4}) given by
\begin{eqnarray}
u(x,t) = [{\sqrt{-p (\frac{r(1+t)^2}{x^2}-1 )} x}]/(p(1+t)).
\end{eqnarray}

\noindent \textbf{Reduction under ${X^2}$:} Reduced equation corresponding to ${X^2} $ is 
\begin{eqnarray}
F^{\prime \prime} \left(1 - p F^2 \alpha^2 - r \alpha^2 \right) - 2 p F \left(F^{\prime}\right)^2 \alpha^2 - 2 p F^2 F^{\prime} \alpha - 2 r F^{\prime} \alpha = 0, 
\end{eqnarray}
where the new variables are $u(x, t)=F(\alpha)$ and $\alpha=\frac{t}{x}$. The reduced ODE has the solution given by
\begin{eqnarray}
F(\alpha)=[{\sqrt{-p (\alpha^2 r-1 )}}]/(p \alpha).
\end{eqnarray} 
The solution of PDE (\ref{PDE_red4}) is given by
\begin{eqnarray}
u(x,t) = [{\sqrt{-p (\frac{r t^2}{x^2}-1 )} x}]/(p t).
\end{eqnarray}

\noindent \textbf{Reduction under ${X^3}$:} Reduced equation corresponding to ${X^3} $ is 
\begin{eqnarray}
F^{\prime \prime} \left(1 - c^2 p F^2 - c^2 r \right) - 2 p F \left(F^{\prime}\right)^2 c^2 = 0, 
\end{eqnarray}
where the new variables are $u(x, t)=F(\alpha)$ and $\alpha= -cx+t$. The reduced ODE has the solution given by  
\begin{eqnarray}
F(\alpha)=\frac{1}{24} \frac{ (-4 c^2 p r+ ( (12 c_1 \alpha c+12 c_2 c+4 \sqrt{\frac{A}{p}} ) p^2 )^{2 / 3}+4 p ) 12^{2 / 3}}{c p (p^2 (\frac{1}{3} \sqrt{\frac{A}{p}}+c (\alpha c_1+c_2 ) ) )^{1 / 3}},
\end{eqnarray}
where $-4+4\,{c}^{6}{r}^{3}-12\,{c}^{4}{r}^{2}+  ( 9\,  ( \alpha\,c_{1}+c_{2}  ) ^{2}p+12\,r  ) {c}^{2}=A
$. The solution of ODE (\ref{PDE_red4}) is given by
\begin{eqnarray}
u(x,t) = \frac{1}{24} \frac{ (-4 c^2 p r+ ( (12 c_1(-c x+t) c+12 c_2 c+4 \sqrt{\frac{B}{p}} ) p^2 )^{2 / 3}+4 p ) 12^{2 / 3}}{c p (p^2 (\frac{1}{3} \sqrt{\frac{B}{p}}+c ((-c x+t) c_1+c_2 ) ) )^{1 / 3}}.
\end{eqnarray}
where $-4+4\,{c}^{6}{r}^{3}-12\,{c}^{4}{r}^{2}+  ( 9\,  (   ( -cx+t  ) c_{1}+c_{2}  ) ^{2}p+12\,r  ) {c}^{2}=B$.\\

\noindent \textbf{Reduction under ${X^4}$:} Reduced equation corresponding to ${X^4} $ is 
\begin{eqnarray}
F^{\prime \prime} \left(-p F^2 - r \right) - 2 p F F^{\prime 2} = 0,
\end{eqnarray}
where the new variables are $u(x, t)=F(\alpha)$ and $\alpha= x$. The solution of the reduced ODE is 
\begin{eqnarray}
F(\alpha)=\frac{1}{8} \frac{ ( ( (12 c_1 \alpha+4 \sqrt{\frac{9 (\alpha c_1+c_2 )^2 p+4 r^3}{p}}+12 c_2 ) p^2 )^{2 / 3}-4 r p ) 4^{2 / 3}}{p ( (3 c_1 \alpha+3 c_2+\sqrt{\frac{9 (\alpha c_1+c_2 )^2 p+4 r^3}{p}} ) p^2 )^{1 / 3}}.
\end{eqnarray}
The solution of ODE (\ref{PDE_red4}) is
\begin{eqnarray}
u(x,t)=\frac{1}{8} \frac{ ( ( (12 c_1 x+4 \sqrt{\frac{9 (c_1 x+c_2 )^2 p+4 r^3}{p}}+12 c_2 ) p^2 )^{2 / 3}-4 r p ) 4^{2 / 3}}{p ( (3 c_1 x+3 c_2+\sqrt{\frac{9 (c_1 x+c_2 )^2 p+4 r^3}{p}} ) p^2 )^{1 / 3}}.
\end{eqnarray}

\subsection{Analysis of the Class of Fourth Order Nerve Equation} \label{sec_a2}

The analysis of fourth order nerve equation in bio-membranes is given below.

\subsubsection{Invariance Analysis, Optimal System and Reduction : $ h_1 =0$}
The PDE in this case reduces to 
\begin{eqnarray} \label{4thPDE1}
u_{tt} = ( (p u^2 + qu + r) u_x  )_x + h_2 u_{xxtt}.
\end{eqnarray}
The infinitesimals are
\begin{eqnarray}
\xi^1 = c_1 + \frac{k_2}{2} c_3 t ,  \quad
\xi^2 = c_5 ,  \quad
\eta = c_3 u + k_1 c_3.
\end{eqnarray}
From the infinitesimals, we obtain three symmetries which form a three dimensional algebra spanned by 
\begin{eqnarray*}
X_1 = \frac{\partial}{\partial t},  \quad
X_2 = \frac{\partial}{\partial x} ,  \quad
X_3 = \frac{k_2}{2} t \frac{\partial}{\partial t} + (k_1 + u )  \frac{\partial}{\partial u}.
\end{eqnarray*}
Equivalence transformations are given by 
\begin{eqnarray}
\bar{t}=a_{1} t+a_{2}, \ \  \ \bar{x}=b_{1} x+b_{2}, \ \  \ \bar{u}=c_{1} u +c_{2}, \\
\bar{h}_{1}=\frac{b_{1}^{2}}{a_{1}^{2}} h_{1}, \ \  \  \bar{h}_{2} =b_{1}^{2} h_{2}, \ \  \ \bar{( p u^2 + qu + r)}=\frac{b_{1}^{2}}{a^{2}} ( p u^2 + qu + r).
\end{eqnarray}
Using these transformations, we obtain 
\begin{eqnarray*}
X_1 = \frac{\partial}{\partial t},  \quad 
X_2 = \frac{\partial}{\partial x} ,  \quad
X_3 = \frac{k_2}{2} t \frac{\partial}{\partial t} + u \frac{\partial}{\partial u}.
\end{eqnarray*}

\noindent The set of optimal sytem of subalgebras for this case is given by 
\[  
   \langle
                \begin{array}{ll}
                 X^1 =  X_2+ X_3 \  , \ \   \  \  &X^2 = X_3 \\
                 X^3 = X_2 \pm X_1 \  , \ \   \  \ &X^4 =   \pm X_1 .
                \end{array}
               \rangle
  \]
\noindent \textbf{Reduction under ${X^1}$:} Reduced equation corresponding to ${X^1} $ is 
\begin{eqnarray}
-h_2 F''''(\alpha) k_2^2 \alpha^4 - 5 h_2 F'''(\alpha) k_2^2 \alpha^3 - 4 h_2 F''(\alpha) \alpha^2 k_2^2 + 4 F''(\alpha) \alpha^2 = 0,
\end{eqnarray}
where the new variables are $u(x, t)=F(\alpha)$ and $\alpha= t e^{-(1/2) k_2 x} $. The reduced ODE has the solution
\begin{eqnarray}
F  ( \alpha  ) =c_{{1}}+c_{{2}}\alpha+c_{{3}}{\alpha}^{\,{
\frac {2}{\sqrt {h_{2}}k_2}}}+c_{{4}}{\alpha}^{\,{\frac {-2}{\sqrt {h_{
2}}k_2}}}.
\end{eqnarray}
Similarity variables $u(x, t)=F(\alpha)$ and $\alpha= t e^{-(1/2) k x} $ provide us the solution of PDE (\ref{4thPDE1}) given by
\begin{eqnarray}
u(x,t) = c_{{1}}+c_{{2}}t{{\rm e}^{-1/2\,kx}}+c_{{3}}  ( t{{\rm e}^{-1/2\,k
x}}  ) ^{2\,{\frac {1}{\sqrt {h_{2}}k_2}}}+c_{{4}}  ( t{{\rm e}
^{-1/2\,kx}}  ) ^{-2\,{\frac {1}{\sqrt {h_{2}}k_2}}}.
\end{eqnarray}

\noindent \textbf{Reduction under ${X^2}$:} Reduced equation corresponding to ${X^2} $ is 
\begin{eqnarray}
 (-p F^2-q F-r ) F^{\prime \prime}-2 (p F+\frac{1}{2} q ) (F^{\prime} )^2=0, 
\end{eqnarray}
where the new variables are $u(x, t)=F(\alpha)$ and $\alpha=x$. The above reduced ODE has the solution 
\begin{eqnarray}
F(\alpha)=\frac{1}{2} \frac{ (12 c_1 \alpha p^2+12 c_2 p^2+6 r q p-q^3+2 \sqrt{B} p )^{2 / 3}-4 r p+q^2}{p (2 \sqrt{B} p+ (12 \alpha c_1+12 c_2 ) p^2+6 r q p-q^3 )^{1 / 3}},
\end{eqnarray}
where $36\,  ( \alpha\,c_{{1}}+c_{{2}}  ) ^{2}{p}^{2}+36\,  ( 
  ( \alpha\,c_{{1}}+c_{{2}}  ) q+4/9\,{r}^{2}  ) rp+
  ( -6\,\alpha\,c_{{1}}-6\,c_{{2}}  ) {q}^{3}-3\,{q}^{2}{r}^{
2}=B$. From here we can easily find the solution of PDE (\ref{4thPDE1}) by substituting similarity variable $\alpha=x$.\\

\begin{figure}[h]
    \centering
    \begin{minipage}{1\textwidth}
        \includegraphics[width=\linewidth]{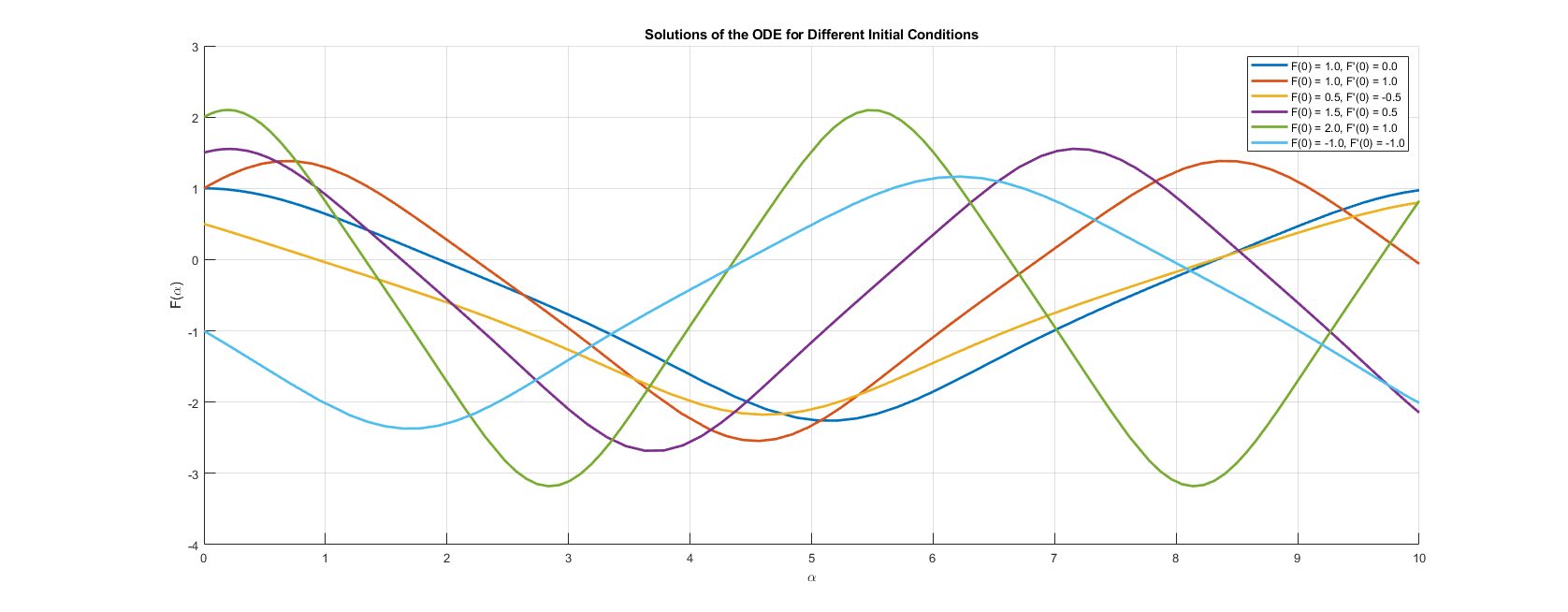}
    \end{minipage}
    \caption{ Numerical solutions $F(\alpha)$: Visualised for several pairs of IVCs }
    \label{invfig1za1}
\end{figure}

\noindent \textbf{Reduction under ${X^3}$:} Reduced equation corresponding to ${X^3} $ is 
\begin{eqnarray}
-h_2 F^{\prime \prime \prime \prime}+ (-p F^2-q F-r+1 ) F^{\prime \prime}-2 (F^{\prime} )^2 (p F+\frac{1}{2} q )=0,
\end{eqnarray}
where the new variables are $u(x, t)=F(\alpha)$ and $\alpha=  t-x$. The reduced ODE after twice integration, can be reduced to second order ODE 
\begin{eqnarray} \label{4thode_z1}
-\frac{1}{3} p F^3-\frac{1}{2} q F^2 + (1-r)F-h_2 F^{\prime \prime}=0, 
\end{eqnarray} 
which is still complicated to solve using analytical techniques. The numerical technique is used to solve ODE (\ref{4thode_z1}). The numerical solutions of ODE for different values of initial value conditions are presented in Figure-\ref{invfig1za1}.

\noindent \textbf{Reduction under ${X^4}$:} Reduced equation corresponding to ${X^4} $ is 
\begin{eqnarray}
 (-p F^2-q F-r ) F^{\prime \prime}-2 (p F+\frac{1}{2} q ) (F^{\prime} )^2=0,
\end{eqnarray}
and has solution.
\begin{eqnarray}
F(\alpha)= [{-q+\sqrt{-4 p r+q^2}}]/(2p).
\end{eqnarray}
where the new variables are $u(x, t)=F(\alpha)$ and $\alpha= x$.

\subsubsection{Invariance Analysis, Optimal System and Reduction : $h_2=0$}
The PDE in this case reduces to 
\begin{eqnarray} \label{4thPDE2}
u_{tt} = ( (p u^2 + qu + r) u_x  )_x  - h_1 u_{xxxx} .
\end{eqnarray}
The set of infinitesimals given by 
\begin{eqnarray}
\xi^1 = -k_{1} c_{4} t+c_{2} , \quad
\xi^{2} = -\frac{1}{2} k_{1} c_{4} x + c_{3},  \quad
\eta = c_{4} u + k_{2}.
\end{eqnarray}
Further, we obtain the set of symmetries given by 
\begin{eqnarray*}
X_1 = \frac{\partial}{\partial t},  \quad
X_2 =\frac{\partial}{\partial x},  \quad
X_3 = -k_1 t \frac{\partial}{\partial t}  - \frac{1}{2} k_1 x \frac{\partial}{\partial x} + (u+k_2) \frac{\partial}{\partial u}.
\end{eqnarray*}
By equivalence transformation, symmetries simplify to  
\begin{eqnarray*}
X_1 = \frac{\partial}{\partial t},  \quad
X_2 = \frac{\partial}{\partial x} ,  \quad
X_3 = -k_1 t \frac{\partial}{\partial t}  - \frac{1}{2} k_1 x \frac{\partial}{\partial x} + u \frac{\partial}{\partial u}.
\end{eqnarray*}
The set of optimal sytem of subalgebras for this case is given by 
\[  
   \langle
                \begin{array}{ll}
                 X^1 = X_3  \\
                 X^2 = \pm X_1 + X_2 \\
                 X^3 = X_1 .
                \end{array}
               \rangle
  \]

\noindent \textbf{Reduction under ${X^1}$:} Reduced equation corresponding to ${X^1} $ is 
\begin{eqnarray}
-4 (\alpha (p F^2+q F+r ) F^{\prime \prime}+2 F^{\prime} (\alpha (F p+\frac{1}{2} q ) F^{\prime}+\frac{3}{4} p F^2+\frac{3}{4} q F+\frac{3}{4} r ) ) \alpha=0,
\end{eqnarray}
has the solution  
\begin{eqnarray}
F(\alpha)=\frac{1}{2} \frac{ (\frac{A}{\sqrt{\alpha}} )^{1 / 3}}{p}+\frac{1}{2} \frac{-4 p r+q^2}{p (\frac{A}{\sqrt{\alpha}} )^{1 / 3}}-\frac{1}{2} \frac{q}{p},
\end{eqnarray}
under new variables are $u(x, t)=F(\alpha)$ and $\alpha= \frac{t}{x^2} $ and
\begin{eqnarray}
A=  ( 12\,c_{{2}}{p}^{2}+  ( 6\,rq+2\,\sqrt {-{{B/
\alpha}}}  ) p-{q}^{3}  ) \sqrt {\alpha}-24\,c_{{1}}{p}^{2}
,  \\
B=144\,  ( c_{{2}}{p}^{2}+1/2\,rqp-1/12\,{q}^{3}  ) c_{{1}}
\sqrt {\alpha}+  ( -36\,{c_{{2}}}^{2}\alpha-144\,{c_{{1}}}^{2}
  ) {p}^{2} \nonumber \\ +  ( -36\,qrc_{{2}}-16\,{r}^{3}  ) \alpha\,p+
  ( 6\,{q}^{3}c_{{2}}+3\,{q}^{2}{r}^{2}  ) \alpha.
\end{eqnarray}
The solution of PDE (\ref{4thPDE2}) can be obtained by simple substituting similarity variable $u(x, t)=F(\alpha)$ and $\alpha= \frac{t}{x^2} $ to get solution in original variables. \\

\noindent \textbf{Reduction under ${X^2}$:} Reduced equation corresponding to ${X^2} $ is 
\begin{eqnarray}
-h_1 F^{\prime \prime \prime \prime}+ (-p F^2-q F-r+1 ) F^{\prime \prime}-2 (F^{\prime} )^2 (p F+\frac{1}{2} q )=0,
\end{eqnarray}
where the new variables are $u(x, t)=F(\alpha)$ and $\alpha=t-x$. The reduced equation can be further simplified to the following form by integrating twice given by
\begin{eqnarray} \label{4thode_z2}
-\frac{1}{3} F^3 p-\frac{1}{2} F^2 q-r F+F-F^{\prime \prime} h_1=0,
\end{eqnarray} 
which is also still complicated to solve for exact solution.\\

\noindent \textbf{Reduction under ${X^3}$:} Reduced equation corresponding to ${X^3} $ is 
\begin{eqnarray}
-h_1 F^{\prime \prime \prime \prime}+ (-p F^2-q F-r ) F^{\prime \prime}-2 (F^{\prime} )^2 (p F+\frac{1}{2} q )=0,
\end{eqnarray}
where the new variables are $u(x, t)=F(\alpha)$ and $\alpha= x$.

\subsubsection{Invariance Analysis, Optimal System and Reduction : Arbitrary $p,q,r$}
The PDE in this case reduces to 
\begin{eqnarray} \label{4thPDE3}
u_{tt} = ( (p u^2 + qu + r) u_x  )_x  - h_1 u_{xxxx} + h_2 u_{xxtt} .
\end{eqnarray}
For arbitrary values of $p$, $q$ and $r$, the infinitesimals are given by 
\begin{eqnarray}
\xi^1 = c_2,  \quad
\xi^2 = c_3,  \quad
\eta = 0.
\end{eqnarray}
and the corresponding symmetries are 
\begin{eqnarray*}
X_1 = \frac{\partial}{\partial t},  \quad  X_2 = \frac{\partial}{\partial x} .
\end{eqnarray*}
The set of optimal sytem of subalgebras for this case is trivial given by 
\[  
   \langle
                \begin{array}{ll}
                 X^1 = X_1  \\
                 X^2 = X_2 
                \end{array}
               \rangle
  \]

\noindent \textbf{Reduction under ${X^1}$:} Reduced equation corresponding to ${X^1} $ is 
\begin{eqnarray}
-h_1 F^{\prime \prime \prime \prime}+ (-p F^2-q F-r ) F^{\prime \prime}-2 (p F+\frac{1}{2} q ) (F^{\prime} )^2=0,
\end{eqnarray}
where the new variables are $u(x, t)=F(\alpha)$ and $\alpha= x $.\\ 

\noindent \textbf{Reduction under ${X^2}$:} Reduced equation corresponding to ${X^2} $ is 
\begin{eqnarray}
F''(\alpha) =0,
\end{eqnarray}
where the new variables are $u(x, t)=F(\alpha)$ and $\alpha=t$.

\subsubsection{ Invariance Analysis, Optimal System and Reduction : $h_1,p=0$ }
The PDE in this case reduces to 
\begin{eqnarray} \label{4thPDE4}
u_{tt} = ( ( qu + r) u_x  )_x   + h_2 u_{xxtt} .
\end{eqnarray}
The algebra is spanned by $X_1, \ X_2 , \ X_3$.
\begin{align*}
X_1 = \frac{\partial}{\partial t} ,  \quad
X_2 = \frac{\partial}{\partial x} ,  \quad
X_3 = tq \frac{\partial}{\partial t} - 2(qu+r) \frac{\partial}{\partial u} . \ \ \ \ \ 
\end{align*}
The set of optimal sytem of subalgebras for this case is given by 
\[  
   \langle
                \begin{array}{ll}
                 X^1 =  X_2+ X_3 \  , \ \   \  \  &X^2 = X_3 \\
                 X^3 = X_2 \pm X_1 \  , \ \   \  \ &X^4 =   \pm X_1 .
                \end{array}
               \rangle
  \]

\noindent \textbf{Reduction under ${X^1}$:} Reduced equation corresponding to ${X^1} $ is 
\begin{eqnarray}
-\alpha^2  ( F^{\prime \prime \prime \prime} \alpha^2 h_2 + 5 F^{\prime \prime \prime} \alpha h_2 + F^{\prime \prime} (4 h_2 - 1)  ) = 0,
\end{eqnarray}
has the solution 
\begin{eqnarray}
F(\alpha)=[{-r+\sqrt{2 \ln (\alpha) c_1 q+2 c_2 q+r^2}}]/q,
\end{eqnarray}
where the new variables are $u(x, t)=F(\alpha)$ and $\alpha= t e^{-x} $. The solution of PDE (\ref{4thPDE4}) can be obtained by simple substituting similarity variables to get solution in original variables. \\

\noindent \textbf{Reduction under ${X^2}$:} Reduced equation corresponding to ${X^2} $ is 
\begin{eqnarray}
(q F + r) F^{\prime \prime} + q  (F^{\prime} )^2 = 0,
\end{eqnarray}
has the solution 
\begin{eqnarray}
F(\alpha)=[{-r+\sqrt{2 \alpha q c_1+2 q c_2+r^2}}]/q,
\end{eqnarray}
where the new variables are $u(x, t)=F(\alpha)$ and $\alpha= x$. \\

\noindent \textbf{Reduction under ${X^3}$:} Reduced equation corresponding to ${X^3} $ is 
\begin{eqnarray}
-h_2 F^{\prime \prime \prime \prime} + (-F q - r + 1) F^{\prime \prime} - q (F^{\prime} )^2 = 0,
\end{eqnarray}
where the new variables are $u(x, t)=F(\alpha)$ and $\alpha=  t-x$. \\

\noindent \textbf{Reduction under ${X^4}$:} Reduced equation corresponding to ${X^4} $ is 
\begin{eqnarray}
(-q F - r) F^{\prime \prime} - q  (F^{\prime} )^2 = 0,
\end{eqnarray}
has the solution 
\begin{eqnarray}
F(\alpha)=-[{r+\sqrt{2 \alpha q c_1+2 q c_2+r^2}}]/q,
\end{eqnarray}
where the new variables are $u(x, t)=F(\alpha)$ and $\alpha= x$.

\subsubsection{ Invariance Analysis, Optimal System and Reduction : $p = \frac{q^2}{4r}$}
The PDE in this case reduces to 
\begin{eqnarray} \label{4thPDE5}
u_{tt} = ( (  \frac{q^2}{4r} u^2 + qu + r) u_x  )_x + h_2 u_{xxtt}.
\end{eqnarray}
The algebra is spanned by $X_1, \ X_2 , \ X_3$.
\begin{align*}
X_1 = \frac{\partial}{\partial t} ,  \quad
X_2 = \frac{\partial}{\partial x} ,  \quad
X_3 = tq \frac{\partial}{\partial x} - (qu+2r) \frac{\partial}{\partial u} . \ \ \ \ \ 
\end{align*}
The set of optimal sytem of subalgebras for this case is given by 
\[  
   \langle
                \begin{array}{ll}
                 X^1 =  X_2+ X_3 \  , \ \   \  \  &X^2 = X_3 \\
                 X^3 = X_2 \pm X_1 \  , \ \   \  \ &X^4 =   \pm X_1 .
                \end{array}
               \rangle
  \]

\noindent \textbf{Reduction under ${X^1}$:} Reduced equation corresponding to ${X^1} $ is 
\begin{eqnarray}
- ( \alpha (q F + 2 r) F^{\prime \prime} +  ( 2  (q \alpha F^{\prime} + \frac{1}{2} q F + r )  ) F^{\prime}  ) (q F + 2 r) \alpha = 0,
\end{eqnarray}
has the solution
\begin{eqnarray}
F(\alpha)= [{ ( (3 c_1 \ln (\alpha)+3 c_2 ) q )^{1 / 3}-2 r}]/q,
\end{eqnarray}
where the new variables are $u(x, t)=F(\alpha)$ and $\alpha= t e^{-x} $. The solution of PDE (\ref{4thPDE5}) can be obtained by simple substituting similarity variables to get solution in original variables. \\ 

\noindent \textbf{Reduction under ${X^2}$:} Reduced equation corresponding to ${X^2} $ is 
 \begin{eqnarray} \label{red_b2}
- ( (q F + 2 r) F^{\prime \prime} + 2  (F^{\prime})^2 q ) (q F + 2 r) = 0,
\end{eqnarray}
has the solution
\begin{eqnarray}
F(\alpha)= [{ (-3 c_1 q (c_2+\alpha ) )^{1 / 3}-2 r}]/q,
\end{eqnarray}
where the new variables are $u(x, t)=F(\alpha)$ and $\alpha= x$. The solution of PDE (\ref{4thPDE5}) can be obtained by simple substituting similarity variables to get solution in original variables. \\ 

\noindent \textbf{Reduction under ${X^3}$:} Reduced equation corresponding to ${X^3} $ is 
\begin{eqnarray}
-4 h_2 F^{\prime \prime \prime \prime} r +  (-q^2 F^2 - 4 q r F - 4 r^2 + 4 r ) F^{\prime \prime} - 2 q  (F^{\prime} )^2  (q F + 2 r ) = 0,
\end{eqnarray}
where the new variables are $u(x, t)=F(\alpha)$ and $\alpha=  t-x$. \\

\noindent \textbf{Reduction under ${X^4}$:} Reduced equation corresponding to ${X^4} $ is 
 \begin{eqnarray} \label{red_b4}
-(q F + 2 r)  ( (q F + 2 r) F^{\prime \prime} + 2  (F^{\prime} )^2 q  ) = 0, 
\end{eqnarray}
has the solution
\begin{eqnarray}
F(\alpha)=[{ (-3 c_1 q (c_2+\alpha))^{1 / 3}-2 r}]/q,
\end{eqnarray}
where the new variables are $u(x, t)=F(\alpha)$ and $\alpha= x$. The solution of PDE (\ref{4thPDE5}) can be obtained by simple substituting similarity variables to get solution in original variables.

\subsubsection{ Invariance Analysis, Optimal System and Reduction : $h_1,q,r=0$ }
The PDE in this case reduces to 
\begin{eqnarray} \label{4thPDE6}
u_{tt} = ( (p u^2  ) u_x  )_x  + h_2 u_{xxtt}.
\end{eqnarray}
The algebra is spanned by $X_1, \ X_2, \ X_3 $ given by  
\begin{align*}
X_1 = \frac{\partial}{\partial t} ,  \quad
X_2 = \frac{\partial}{\partial x} ,  \quad
X_3 = t \frac{\partial}{\partial t} - u \frac{\partial}{\partial u} . \ \ \ \ \
\end{align*}
The set of optimal sytem of subalgebras for this case is given by 
\[  
   \langle
                \begin{array}{ll}
                 X^1 =  X_2+ X_3 \  , \ \   \  \  &X^2 = X_3 \\
                 X^3 = X_2 \pm X_1 \  , \ \   \  \ &X^4 =   \pm X_1 .
                \end{array}
               \rangle
  \]

\noindent \textbf{Reduction under ${X^1}$:} Reduced equation corresponding to ${X^1} $ is 
 \begin{eqnarray} \label{red_c1}
-F \alpha p ( F^{\prime \prime} F \alpha + 2 (F^{\prime})^2 \alpha + F F^{\prime} ) = 0, 
\end{eqnarray}
has the solution 
\begin{eqnarray}
F(\alpha)= (3 c_1 \ln (\alpha)+3 c_2 )^{1 / 3},
\end{eqnarray}
where the new variables are $u(x, t)=F(\alpha)$ and $\alpha= t e^{-x} $. The solution of PDE (\ref{4thPDE6}) can be obtained by simple substituting similarity variables to get solution in original variables. \\ 

\noindent \textbf{Reduction under ${X^2}$:} Reduced equation corresponding to ${X^2} $ is 
\begin{eqnarray}
-F p (F F'' + 2 F'^2) = 0,
\end{eqnarray}
has the solution 
\begin{eqnarray}
F(\alpha)=(-3 \alpha c_1+c_2)^{1 / 3},
\end{eqnarray}
where the new variables are $u(x, t)=F(\alpha)$ and $\alpha= x$. The solution of PDE (\ref{4thPDE6}) can be obtained by simple substituting similarity variables to get solution in original variables. \\ 

\noindent \textbf{Reduction under ${X^3}$:} Reduced equation corresponding to ${X^3} $ is 
\begin{eqnarray}
F''- 2 p F F'^2 - p F^2 F'' - h_2 F'''' = 0,
\end{eqnarray}
where the new variables are $u(x, t)=F(\alpha)$ and $\alpha=  t-x$.\\ 

\noindent \textbf{Reduction under ${X^4}$:} Reduced equation corresponding to ${X^4} $ is 
\begin{eqnarray}
-F(\alpha) p ( F(\alpha) F''(\alpha) + 2 F'(\alpha)^2 ) = 0,
\end{eqnarray}
has the solution 
\begin{eqnarray}
F(\alpha)=(-3 \alpha c_1+c_2)^{1 / 3},
\end{eqnarray}
where the new variables are $u(x, t)=F(\alpha)$ and $\alpha= x$. The solution of PDE (\ref{4thPDE6}) can be obtained by simple substituting similarity variables to get solution in original variables. \\

\subsection{Conservation Laws for the class of Second Order Nerve Equation} \label{sec_a3}
We study the conservation laws for the second order nerve equation (\ref{bio1}) in this section using the direct approach \cite{s11,s12,s13,s14,s15,s16,s17}  that explain the geometrical properties of the invariant solutions of PDEs. Anco and Bluman \cite{s11,s12} proposed the direct approach to investigate conservation laws. 

In this work, all the multipliers of the form $\mathcal{M}(x,t,u)$ are constructed and their corresponding conservation laws are presented. By the direct approach, multiplier $\mathcal{M}(x,t,u)$ satisfy the given expression
\begin{equation}\label{euler}   
\mathcal{E} (\mathcal{M}(x,t,u) [u_{tt} - ( (p u^2 + qu + r) u_{x}  )_x])=0,
\end{equation}
where $\mathcal{E}$ is Euler operator. The multipliers $ \mathcal{M} \{ x,t,u \} $ of the PDE (\ref{bio1}) has the property
\begin{equation} \label{E12}
D_tT^t(x,t,u)+D_xT^x(x,t,u)= \mathcal{M}(x,t,u) [u_{tt} - ( (p u^2 + qu + r) u_{x}  )_x] . 
\end{equation}
where multipliers $\mathcal{M}$ are depending independent variables on $x, \ t$ and dependent variable $  u $. Consider the multipliers by $\mathcal{M}(x,t,u)$ and expanding the expression (\ref{E12}) which is the invariance criteria to find conservation laws, leads us to the expression helps in finding the set of determining equations that further leads us to the multipliers given by 
$$
\mathcal{M}(x, t, u)= ( c_1 x+ c_3  ) t+x c_2 + c_4
$$
Above equation provides four multipliers $\{ xt,x,t,1 \}$ when $c_1, c_2, c_3$ and $c_4$ are non zero respectively, and their corresponding conserved flows a for the PDE $u_{tt} = ( (p u^2 + qu + r) u_x)_x $, when $p \neq 0 $, $p \neq \frac{1}{4} \cdot \frac{q^2}{r}  $ and $q \neq 0$ are given by 
\begin{center}
$\begin{gathered}
 T_1 = (T^x,T^t) = ( x (t u_t-u_{} ) , - p t x u_{}^2 u_x + \frac{1}{3} p t u_{}^3 - q t x u_{} u_x + \frac{1}{2} q t u_{}^2 - r t x u_x + r t u_{} ) \\
 T_2 = (T^x,T^t) = ( x u_t , - p x u_{}^2 u_x + \frac{1}{3} p u_{}^3 - q x u_{]} u_x + \frac{1}{2} q u_{}^2 - r x u_x + r u_{} ) \\
  T_3 = (T^x,T^t) =  (t u_t-u_{} , -p t u_{}^2 u_x - q t u_{} u_x - r t u_x ) \\
  T_4 = (T^x,T^t) = ( u_t , - p u_{}^2 u_x - q u_{} u_x - r u_x )
\end{gathered}$
\end{center}
Similar multipliers $\{ xt,x,t,1 \}$ are obtained for the PDE $u_{tt} = ( (qu + r) u_x)_x $ and their corresponding conserved flows are presented by
\begin{center}
$\begin{gathered}
 T_1 = (T^x,T^t) = (x(t u_t -u_{}) , - q t x u_{} u_x + \frac{1}{2} q t u_{}^2 - r t x u_x + r t u_{} ) \\
 T_2 = (T^x,T^t) = (x u_t , - q x u_{} u_x + \frac{1}{2} q u_{}^2 - r x u_x + r u_{} )  \\
  T_3 = (T^x,T^t) =  (t u_t - u_{} , - t q u_x u_{}  - r t u_x )  \\
  T_4 = (T^x,T^t) = ( u_t , - p u u_x - r u_x )
\end{gathered}$
\end{center}
Multipliers $\{ xt,x,t,1 \}$ and corresponding conservation laws for the PDE $u_{tt} = ( (p u^2 ) u_x)_x $ are obtained
\begin{center}
$\begin{gathered}
 T_1 = (T^x,T^t) = (x(t u_t-u_{}), - (x u_x - \frac{u_{}}{3} ) t p u_{}^2 )  \\
 T_2 = (T^x,T^t) = ( x u_t , -u_{}^2 p (x u_x -\frac{u_{}}{3} ) ) \\
  T_3 = (T^x,T^t) =  ( t u_t -u_{} , -t p u_{}^2 u_x ) \\
  T_4 = (T^x,T^t) = ( u_t , - p u^2 u_x )
\end{gathered}$
\end{center}
Similar multipliers $\{ xt,x,t,1 \}$ are obtained for the PDE $u_{tt} = ( (p u^2 + r) u_x)_x $  and their corresponding conserved flows are listed by
\begin{center}
$\begin{gathered}
 T_1 = (T^x,T^t) = (x(t u_t-u_{}), -(p x u_{}^2 u_x-\frac{1}{3} p u_{}^3+r x u_x-r u_{})t )  \\
 T_2 = (T^x,T^t) = ( x u_t , - p x u_{}^2 u_x + \frac{1}{3} p u_{}^3 - r x u_x + r u_{} ) \\
  T_3 = (T^x,T^t) =  ( t u_t -u_{} , -t ( p u_{}^2 +r) u_x ) \\
  T_4 = (T^x,T^t) = ( u_t , - (p u^2+r) u_x )
\end{gathered}$
\end{center}

\subsection{Conservation Laws for the Class of Fourth Order Nerve Equation} \label{sec_a4}

In this section, we study the conservation laws for the fourth order nerve equation (\ref{bio2}) and (\ref{bio3}) using the direct approach \cite{s11,s12,s13,s14,s15,s16,s17}. Consider the multipliers by $\mathcal{M}(x,t,u)$ and expanding the expression (\ref{E12}) which is the invariance criteria to find conservation laws, leads us to the expression helps in finding the set of determining equations. By solving the set of determine equations, we will get the multipliers
$$
\mathcal{M}(x, t, u)= ( c_1 x+ c_3  ) t+x c_2 + c_4
$$
Above equation provides four multipliers $\{ xt,x,t,1 \}$ when $c_1, c_2, c_3$ and $c_4$ are non zero respectively, and their corresponding conserved flows for the PDE $u_{tt} = ( (p u^2 + qu + r) u_x  )_x + h_2 u_{xxtt}$ are given by 
\[
\begin{aligned}
T_{1}&= \begin{cases}
T^{x} =\frac{1}{2} x t h_2 u_{xxt}-\frac{1}{3} t h_2 u_{xt}-\frac{1}{6} h_2 x u_{xx}-x t u_t+\frac{1}{3} h_2 u_x+x u_{} , \\
T^{t} =  p t x u_{}^2 u_x-\frac{1}{3} p t u_{}^3+q t x u_{} u_x-\frac{1}{2} q t u_{}^2+\frac{1}{2} x t h_2 u_{xtt}-\frac{1}{6} t h_2 u_{tt} \\
 \quad \quad -\frac{1}{3} h_2 x u_{xt}+\frac{1}{3} h_2 u_t+r t x u_x-r t u_{}. 
\end{cases}\\
T_{2}&= \begin{cases}
T^{x} =\frac{1}{2} x h_2 u_{xxt}-\frac{1}{3} h_2 u_{xt}-x u_t , \\
T^{t} = p x u_{}^2 u_x-\frac{1}{3} p u_{}^3+q x u_{} u_x-\frac{1}{2} q u_{}^2+\frac{1}{2} x h_2 u_{xtt}-\frac{1}{6} h_2 u_{tt}+r x u_x-r u_{} 
\end{cases}\\
T_{3}&= \begin{cases}
T^{x} =\frac{1}{2} t h_2 u_{xxt}-\frac{1}{6} h_2 u_{xx}-t u_t+u_{}, \\
T^{t} = \frac{1}{2} t h_2 u_{xtt}-\frac{1}{3} h_2 u_{xt}+p t u_{}^2 u_x+q t u_{} u_x+r t u_x
\end{cases}\\
T_{4}&= \begin{cases}
T^{x} =\frac{1}{2}h_2 u_{xxt}-u_t, \\
T^{t} = \frac{1}{2} h_2 u_{xtt}+p u_{}^2 u_x+q u_{} u_x+r u_x
\end{cases}
\end{aligned}
\]
Similar multipliers $\{ xt,x,t,1 \}$ are obtained for the PDE $u_{tt} = ( (p u^2 + qu + r) u_x  )_x  - h_1 u_{xxxx} $ and their corresponding conserved flows $(T^x,T^t)$ are listed by
\begin{center}
$\begin{gathered}
 T_1 = (T^x,T^t) = (x(t u_t-u_{}) , tx(p u_{}^2+q u_{}+r) u_x - \frac{p u_{x}^3}{3}-\frac{q u_{}^2}{2} - r u_{}+h_1(x u_{xxx}-u_{xx} ))\\
 T_2 = (T^x,T^t) = (x u_t , p x u_{}^2 u_x - \frac{1}{3} p u_{}^3 + q x u_{} u_x - \frac{1}{2} q u_{}^2 + r x u_x + x h_1 u_{xxx} - r u_{} - h_1 u_{xx} )  
\end{gathered}$
\end{center}
\begin{center}
$\begin{gathered}
  T_3 = (T^x,T^t) = (t u_t-u_{} , t(p u_{}^2+q u_{}+r) u_x + h_1t u_{xxx} ) \\
  T_4 = (T^x,T^t) = (u_t , h_1 u_{xxx}+(p u_{}^2+q u_{}+r) u_x ) 
\end{gathered}$
\end{center}
Multipliers $\{ xt,x,t,1 \}$ are obtained for the PDE $u_{tt} = ( ( (\frac{1}{4} \frac{q^2}{r}) u^2 + qu + r) u_x  )_x + h_2 u_{xxtt}$ and their corresponding conserved flows are listed by
\[
\begin{aligned}
T_{1}&= \begin{cases}
T^{x} =\frac{1}{2} x t h_2 u_{xxt}-\frac{1}{3} t h_2 u_{xt}-\frac{1}{6} x h_2 u_{xx}-x t u_t+\frac{1}{3} h_2 u_x+x u_{}, \\
T^{t} =  \frac{1}{12 r} (3 q^2 t x u_{}^2 u_x-q^2 t u_{}^3+12 q r t x u_{} u_x-6 q r t u_{}^2+12 r^2 t x u_x+6 x t h_2 u_{xtt} r \\
 \quad \quad -12 r^2 t u_{}-2 t h_2 u_{tt} r-4 x h_2 u_{xt} r+4 h_2 u_t r).
\end{cases}\\
T_{2}&= \begin{cases}
T^{x} =\frac{1}{2} x h_2 u_{xxt}-\frac{1}{3} h_2 u_{xt}-x u_t , \\
T^{t} =\frac{1}{12 r}(3 q^2 x u_{}^2 u_x-q^2 u_{}^3+12 q r x u_{} u_x-6 q r u_{}^2+12 r^2 x u_x  +6 x h_2 u_{xtt} r-12 r^2 u_{}-2 h_2 u_{tt} r).
\end{cases}\\
T_{3}&= \begin{cases}
T^{x} =\frac{1}{2} t h_2 u_{xxt}-\frac{1}{6} h_2 u_{xx}-t u_t+u_{}, \\
T^{t} =  \frac{1}{12 r}(3 q^2 t u_{}^2 u_x+12 q r t u_{} u_x+12 r^2 t u_x+6 t h_2 u_{xtt} r-4 h_2 u_{xt} r).
\end{cases}\\
T_{4}&= \begin{cases}
T^{x} =\frac{1}{2}h_2 u_{xxt}-u_t, \\
T^{t} = \frac{1}{4 r}(q^2 u_{}^2 u_x+4 q r u_{} u_x+4 r^2 u_x+2 r h_2 u_{xtt}).
\end{cases}
\end{aligned}
\]
Multipliers $\{ xt,x,t,1 \}$ and their corresponding conserved flows $(T^x,T^t)$ are presented for the PDE $u_{tt} = ( (p u^2  ) u_x  )_x  + h_2 u_{xxtt}$ given by 

\begin{center}
$\begin{gathered}
 T_1 = (T^x,T^t) , \ \ \mbox{where,} \\
 \begin{aligned}
& \begin{cases}
T^x =\frac{1}{2} x t h_2 u_{xxt}-\frac{1}{3} t h_2 u_{xt}-\frac{1}{6} h_2 x u_{xx}-x t u_t+\frac{1}{3} h_2 u_x+x u_{},  \\
T^t =x t p u_{}^2 u_x-\frac{1}{3} p t u_{}^3+\frac{1}{2} x t h_2 u_{xtt}-\frac{1}{6} t h_2 u_{tt}-\frac{1}{3} h_2 x u_{xt}+\frac{1}{3} h_2 u_t.
\end{cases}
\end{aligned}\\
  T_2 = (T^x,T^t) , \ \ \mbox{where,} \\
   \begin{aligned}
& \begin{cases}
T^x = \frac{1}{2} h_2 x u_{xxt}-\frac{1}{3} h_2 u_{xt}-x u_t,  \\
T^t = x p u_{}^2 u_x-\frac{1}{3} p u_{}^3+\frac{1}{2} h_2 x u_{xtt}-\frac{1}{6} h_2 u_{tt}.
\end{cases}
\end{aligned}\\
 T_3 = (T^x,T^t) , \ \ \mbox{where,} \\
 \begin{aligned}
& \begin{cases}
T^x =\frac{1}{2} t h_2 u_{xxt}-\frac{1}{6} h_2 u_{xx}-t u_t+u_{},\\
 T^t = \frac{1}{2} t h_2 u_{xtt}-\frac{1}{3} h_2 u_{xt}+t p u_{}^2 u_x. \\
\end{cases}
\end{aligned}\\
 T_4 = (T^x,T^t) = (\frac{1}{2}h_2 u_{xxt}-u_t, \frac{1}{2}h_2 u_{xtt}+p u_{}^2 u_x) .
\end{gathered}$
\end{center}

\subsection{Soliton Solutions of Second Order Nonlinear Nerve Equation in Biomembranes with Extension to Fourth Order Dispersion}

The soliton solutions \cite{g20,g21} are discussed for a partial differential equations (PDE) that are nonlinear in the unknowns and their derivatives . A travelling wave solutions \cite{g22,g23} is an exact, closed-form solution, expressed as a finite power series of one of a particular set of functions \cite{g22} (we consider $\tanh$, $\sinh$ and $\cosh$) that has a linear combination of the independent variables. Concretely, given such a PDE system with $i$ unknown functions $ Z_i (x_1, \ldots, x_j ) $ of $j$ variables, solutions of the form
\begin{eqnarray}
Z_i(\tau)=\sum_{k=0}^{n_i} A_{i, k} \tau^k,
\end{eqnarray}
\noindent where the $n_i$ are finite, the $A_{i, k}$ are constants with respect to the $x_j$ and
\begin{eqnarray} \label{sol1s}
\tau=\tanh  (\sum_{k=0}^j c_k x_k ),\quad \tau= \mbox{sech} (\sum_{k=0}^j c_k x_k ) ,
\end{eqnarray}
\noindent where the $c_k$ are constants with respect to the $x_j$. Travelling wave solutions can be found by transforming the PDE system into a nonlinear ODE system by introducing $\tau$ as new variable, implying
\begin{eqnarray}
\frac{\partial}{\partial x_j} Z_i=c_j (-\tau^2+1 ) (\frac{\partial}{\partial \tau} Z_i ).
\end{eqnarray}
\noindent for all $x_j$. The resulting ODE system, polynomial in $Z_i(\tau)$ and its derivatives, has all the coefficients polynomial in $\tau$ and the unknowns $c_j$. To search for solutions $Z_i=\sum_{k=0}^{n_i} A_{i, k} \tau^k$ for that ODE system, an upper bound $n_i$ for each unknown $Z_i$ is computed. The complete finite expansion in $\tau$ for each $Z_{i}$, containing the unknowns $A_{i, k}$, is introduced in the ODE system obtained previously, and, by taking coefficients of different powers of $\tau$, a new algebraic nonlinear system for the expansion coefficients $ \{A_{i, k} , c_j \}$ is built and leads us to the solutions for the $A_{i, k}$ in terms of the $c_j$. In next sections we present the soliton solutions of second order nonlinear nerve equation in biomembranes with extension to fourth order dispersion. Further, one can also discuss the solutions of the forms not limited to
\begin{eqnarray} \label{sol2s}
 \tau=\sinh  (\sum_{k=0}^j c_k x_k ), \quad \tau=\cosh  (\sum_{k=0}^j c_k x_k ).
\end{eqnarray} 

\subsubsection{Soliton Solutions of Second Order Nerve Equation}
Now we present the soliton solution of equation (\ref{bio1}) using 
\begin{eqnarray}
\tau = \tanh(t c_2 + x c_1 + c_0),
\end{eqnarray}
\noindent which provide the equivalent ODE given by 
\begin{eqnarray}
  ( c_2^2 (\tau^2 - 1)^2 - (p u(\tau)^2 + q u(\tau) + r) c_1^2 (\tau^2 - 1)^2  ) \frac{d^2 u(\tau)}{d\tau^2} \nonumber \\
 - (2 p u(\tau) + q) c_1^2 (\tau^2 - 1)^2  ( \frac{d u(\tau)}{d\tau}  )^2  +  ( 2 c_2^2 (\tau^2 - 1) \tau  \nonumber \\ 
- 2 (p u(\tau)^2 + q u(\tau) + r) c_1^2 (\tau^2 - 1) \tau  ) \frac{d u(\tau)}{d\tau}=0 .
\end{eqnarray}
\noindent Thus, solution of the form 
$$u(\tau) = \tau^3 A_{1,3} + \tau^2 A_{1,2} + \tau A_{1,1} + A_{1,0} $$ 
for $n_i = 3$ and for the parameters $p = 0, q = 0, r = \frac{c_3^2}{c_2^2}$ are given by 
\begin{eqnarray}
u_1(x, t) = c_7 \tanh(c_2 x + c_3 t + c_1)^3 + c_5 \tanh(c_2 x + c_3 t + c_1) + c_4 ,\\
 u_2(x, t) = c_7 \tanh(c_2 x + c_3 t + c_1)^3 + c_6 \tanh(c_2 x + c_3 t + c_1)^2 \nonumber \\
 + c_5 \tanh(c_2 x + c_3 t + c_1) + c_4.
\end{eqnarray}
\noindent For the graphical analysis, we take the values of parameters $c_1=c_2=c_3=\cdots=c_7=1$ for the simplicity. For some cases, we also take non-identity values to study the behaviour over different values of parameters. The 3D illustration of soliton solutions $u_1(x, t)$ and $u_2(x, t)$ are presented in Figures (\ref{fig2za}) and (\ref{fig2zb}).

\subsubsection{Soliton Solutions of Fourth Order Nerve Equation}
Now we present the closed form solution of equation (\ref{bio3}) after updating the parameters, modified to
\begin{eqnarray} \label{Eq_3closed}
u_{tt} =  ( (p u^2 + qu + r) u_x  )_x  - h u_{xxxx}++ k u_{xxtt}.
\end{eqnarray}
By using the substitution
\begin{eqnarray}
\tau = \tanh(t c_2 + x c_1 + c_0)
\end{eqnarray}
the equivalent ODE obtained is given by 
\begin{eqnarray}
&  ( h c_1^4 (\tau^2 - 1) (\tau^6 - 3\tau^4 + 3\tau^2 - 1) - k c_2^2 (\tau^2 - 1) c_1^2 (\tau^6 - 3\tau^4 + 3\tau^2 - 1)  ) \frac{d^4 u(\tau)}{d\tau^4} \nonumber \\
& +  ( h c_1^4 (\tau^2 - 1) (12\tau^5 - 24\tau^3 + 12\tau) - k c_2^2 (\tau^2 - 1) c_1^2 (12\tau^5 - 24\tau^3 + 12\tau)  ) \frac{d^3 u(\tau)}{d\tau^3} \nonumber \\
& +  ( c_2^2 (\tau^2 - 1)^2 - (p u(\tau)^2 + q u(\tau) + r) c_1^2 (\tau^2 - 1)^2 + h c_1^4 (\tau^2 - 1) (36\tau^4 - 44\tau^2 + 8)  \nonumber \\
& \quad - k c_2^2 (\tau^2 - 1) c_1^2 (36\tau^4 - 44\tau^2 + 8)  ) \frac{d^2 u(\tau)}{d\tau^2} - (2p u(\tau) + q) c_1^2 (\tau^2 - 1)^2  ( \frac{d u(\tau)}{d\tau}  )^2 \nonumber  \\
& +  ( 2c_2^2 (\tau^2 - 1) \tau - 2 (p u(\tau)^2 + q u(\tau) + r) c_1^2 (\tau^2 - 1) \tau + h c_1^4 (\tau^2 - 1) (24\tau^3 - 16\tau) \nonumber  \\
& \quad - k c_2^2 (\tau^2 - 1) c_1^2 (24\tau^3 - 16\tau)  ) \frac{d u(\tau)}{d\tau} = 0.
\end{eqnarray}
Set of above ODE lead us to the form of soliton solutions of the form (\ref{sol1s}). Now we find soliton solutions by splitting into cases with respect to the parameters ${h, k, p, q}$ and $r$ listed here: \\

\noindent \textbf{Principal Case: For arbitrary values of parameters ${h, k, p, q}$ and $r$}\\

\noindent Thus, solution of the form 
\begin{eqnarray}
u(\tau) = \tau A_{1,1}+A_{1,0},
\end{eqnarray}
\noindent for $,n_1 = 1$ are given by 
\[
\begin{aligned}
 \begin{cases}
u_1(x, t) = \frac{\tanh  ( -\frac{1}{2} \sqrt{p  (2 c_2^2 k + 1 )  (8 c_2^2 h p + 4 p r - q^2 )} \frac{c_2 t}{p  (2 c_2^2 k + 1 )} + c_2 x + c_1  ) (\sqrt{6} c_2) \sqrt{ (2 c_2^2 k + 1 )  (-4 k p r + k q^2 + 4 h p )}  - 2 c_2^2 k q - q}{4 c_2^2 k p + 2 p} \\
u_2(x, t) = \frac{-\tanh  ( -\frac{1}{2} \sqrt{p  (2 c_2^2 k + 1 )  (8 c_2^2 h p + 4 p r - q^2 )} \frac{c_2 t}{p  (2 c_2^2 k + 1 )} + c_2 x + c_1  ) (\sqrt{6} c_2) \sqrt{ (2 c_2^2 k + 1 )  (-4 k p r + k q^2 + 4 h p )}  - 2 c_2^2 k q - q}{4 c_2^2 k p + 2 p} 
\end{cases}
\end{aligned}
\]

\[
\begin{aligned}
 \begin{cases}
u_3(x, t) = \frac{-\tanh  ( \frac{1}{2} \sqrt{p  (2 c_2^2 k + 1 )  (8 c_2^2 h p + 4 p r - q^2 )} \frac{c_2 t}{p  (2 c_2^2 k + 1 )} + c_2 x + c_1  ) (\sqrt{6} c_2) \sqrt{ (2 c_2^2 k + 1 )  (-4 k p r + k q^2 + 4 h p )}  - 2 c_2^2 k q - q}{4 c_2^2 k p + 2 p}\\
u_4(x, t) = \frac{\tanh  ( \frac{1}{2} \sqrt{p  (2 c_2^2 k + 1 )  (8 c_2^2 h p + 4 p r - q^2 )} \frac{c_2 t}{p  (2 c_2^2 k + 1 )} + c_2 x + c_1  ) (\sqrt{6} c_2) \sqrt{ (2 c_2^2 k + 1 )  (-4 k p r + k q^2 + 4 h p )}  - 2 c_2^2 k q - q}{4 c_2^2 k p + 2 p}.
\end{cases}
\end{aligned}
\]
Four soliton solutions $u_1(x, t), u_2(x, t), u_3(x, t)$ and $u_4(x, t)$ of equation (\ref{bio3}) is presented. For the better physical illustration and also for understanding the phenomenon of dynamics of the materials through the biomembranes, we discuss the geometrical explanation of the soliton solution of the nerve equation (\ref{bio3}). The 3D illustration  of Kink-Type solutions $u_1(x, t), u_2(x, t), u_3(x, t)$ and $u_4(x, t)$ are given in Figure-(\ref{fig2b}).\\

 Now we'll discuss the cases for $\tanh$-soliton solutions of the form (\ref{sol1s}) based on the values of parameters arises in this analysis.\\

\textbf{Case-1:} $h = \frac{1}{6}  (\frac{c_2^2 c_5^2 p - 3 c_3^2}{c_2^4} ),  k = -\frac{1}{2 c_2^2},  p \neq 0,  q \neq 0, \mbox{ and } r = -\frac{1}{12}  (\frac{4 c_2^2 c_5^2 p^2 - 3 c_2^2 q^2 - 12 c_3^2 p}{c_2^2 p} )$
\begin{eqnarray}
u_5(x, t) = c_5 \tanh(c_2 x + c_3 t + c_1) - \frac{1}{2} \frac{q}{p}
\end{eqnarray}

\textbf{Case-2:} $h = \frac{1}{6}  (\frac{6 c_3^2 k + c_5^2 p}{c_2^2} ), 
k \neq 0, p \neq 0,  q \neq 0, \mbox{ and } r = -\frac{1}{3}  (\frac{c_2^2 c_5^2 p - 3 c_3^2}{c_2^2} )$
\begin{eqnarray}
u_6(x, t) = c_5 \tanh(c_2 x + c_3 t + c_1)
\end{eqnarray}

\textbf{Case-3:} $ h = \frac{1}{6}  (\frac{6 c_3^2 k + c_5^2 p}{c_2^2} ), k \neq 0, p \neq 0,  q = -2 p c_4, \mbox{ and } r = \frac{1}{3}  (\frac{3 c_2^2 c_4^2 p - c_2^2 c_5^2 p + 3 c_3^2}{c_2^2} )$
\begin{eqnarray}
u_7(x, t) = c_5 \tanh(c_2 x + c_3 t + c_1) + c_4
\end{eqnarray}

\noindent The 3D graphs of Kink-Type solutions $u_5(x, t), u_6(x, t)$ and $u_7(x, t)$ are illustrated in Figure (\ref{fig2c}). The type of sech-Kink-Type solutions of the form (\ref{sol1s}) based on the values of parameters arises in this analysis.\\

\textbf{Case-4:} $ h = \frac{1}{6}  (\frac{6 c_3^2 k - c_5^2 p}{c_2^2} ), k \neq 0, p \neq 0, q = -2 p c_4, \mbox{ and } r = \frac{1}{6}  (\frac{6 c_2^2 c_4^2 p - c_2^2 c_5^2 p + 6 c_3^2}{c_2^2} )$
\begin{eqnarray}
u_8(x, t) = \operatorname{sech}(c_2 x + c_3 t + c_1) c_5 + c_4
\end{eqnarray}

\noindent The 3D graphs of soliton solutions $u_8(x, t)$ is illustrated in Figure-(\ref{fig2d1}),(\ref{fig2d2}).

\section{Conclusions} \label{sec_a5}
We studied the nonlinear wave equation with fourth order dissipation terms and the particular Nerve-Membrane equations. The complete symmetry classification of fourth order wave equation was presented for all the forms of smooth function $\Phi(u)$ that are studied. The corresponding optimal system of subalgebras of the Lie algebras was presented and complete reduction performed under optimal system of subalgebras. Furthermore, the conservation laws via multiplier approach were discussed and presented in the form of conserved vectors. Invariant solutions and their physical illustration is discussed.  

\vspace{1cm}

\noindent {\bf Acknowledgements} AR and AHK thank the African Institute of Mathematics (AIMS) in Muizenberg for providing research facilities.

\vspace{1cm}

\noindent {\bf Statement of Authors Participation} All authors included above have contributed to the manuscript in some but significant way.

\addcontentsline{toc}{chapter}{References}

\bibliographystyle{amsplain}

\addcontentsline{toc}{chapter}{Appendices}

\section{Appendix} \label{sec_App}

\subsection{ Explanation for general Lie symmetry classification}
The detailed explanation for solving fourth order PDE (\ref{bio3}) and finding the Lie symmetries is presented here. The fourth order wave equation with arbitrary fundtion $\Phi(u)$ and with higher order dissipation is given by
\begin{equation} \label{Eq2}
u_{tt} = ( (p u^2 + qu + r) u_x)_x - h_1 u_{xxxx} + h_2 u_{xxtt}.
\end{equation}
By Lie symmetry criterion, we require a fourth order prolongation $X^{[4]}$ given by  
\begin{equation*}
X^{[4]}= \xi^1 \frac{\partial}{\partial t} + \xi^2 \frac{\partial}{\partial x} + \eta \frac{\partial}{\partial u} + \zeta_{x} \frac{\partial}{\partial u_{x}}+  \zeta_{xx} \frac{\partial}{\partial u_{xx}} + \zeta_{tt} \frac{\partial}{\partial u_{tt}} + \zeta_{xxxx} \frac{\partial}{\partial u_{xxxx}} + \zeta_{xxtt} \frac{\partial}{\partial u_{xxtt}} +     \cdots,
\end{equation*}
where the corresponding extended transformations are given by  
\begin{equation}
\zeta_t = D_t \eta - u_t D_t \xi^1 - u_x D_t \xi^2 ,
\end{equation}
\begin{equation}
\zeta_x = D_x \eta - u_t D_x \xi^1 - u_x D_x \xi^2 ,
\end{equation}
The second order prolonged operators $\zeta_{xx} $ and $\zeta_{tt}$ are required   
\begin{eqnarray}
& \zeta_{xx} = D_x \zeta_x - u_{xt} D_x \xi^1 - u_{xx} D_x \xi^2  \nonumber  \\
&= \eta_{xx} + 2 \eta_{xu} u_x + \eta_{uu} u^2_x + \eta_u u_{xx} - 2u_{tx}(\xi^1_x + \xi^1_u \cdot  u_{x}) \nonumber  \\
&  -2u_{xx}(\xi^2_x + \xi^2_u \cdot u_x)  - u_t(\xi^1_{xx} + 2 \xi^1_{xu} \cdot  u_x + \xi^1_{uu} \cdot  u^2_x + \xi^1_u \cdot u_{xx} )   \\
& -u_x (\xi^2_{xx} + 2 \xi^2_{xu}u_x + \xi^2_{uu} u^2_x + \xi^2_u u_{xx}). \nonumber 
\end{eqnarray}
\begin{eqnarray}
& \zeta_{tt} = D_t \zeta_t - u_{tt} D_t \xi^1 - u_{tx} D_t \xi^2  \nonumber  \\
&= \eta_{tt} + 2 \eta_{tu} u_t + \eta_{uu} u^2_t + \eta_u u_{tt} - 2u_{tt}(\xi^1_t + \xi^1_u \cdot  u_{t}) \nonumber  \\
&  -2u_{tx}(\xi^2_t + \xi^2_u \cdot u_t)  - u_t(\xi^1_{tt} + 2 \xi^1_{tu} \cdot  u_t + \xi^1_{uu} \cdot  u^2_t + \xi^1_u \cdot u_{tt} )   \\
& -u_x (\xi^2_{tt} + 2 \xi^2_{tu}u_t + \xi^2_{uu} u^2_t + \xi^2_u u_{tt}). \nonumber 
\end{eqnarray}
Expressions for fourth order extended infinitesimals are given by 
\begin{eqnarray*}
& \zeta_{xxxx}   = \eta_{xxxx} + 4 \eta_{xxxu} u_x + 6 \eta_{xxuu} u^2_x +  6 \eta_{xxu} u_{xx} + 4\eta_{xuuu} u^3_x  + 12 \eta_{xuu} u_x u_{xx} \\
			 & +4 \eta_{xu} u_{xxx} + 6 \eta_{uuu} u^2_x u_{xx} + 3\eta_{uu} u^2_{xx} + 4\eta_{uu} u_x u_{xxx} + \eta_{uuuu} u^4_x  +  \eta_{u} u_{xxxx} \\
			   &  - \xi^1_{xxxx} u_{t}  -  4   \xi^1_{xxxu} u_{t} u_x - 6  \xi^1_{xxuu} u_{t} u^2_x -  6  \xi^1_{xxu} u_{t} u_{xx} - 4  \xi^1_{xuuu} u_{t} u^3_x  - 12  \xi^1_{xuu} u_{t} u_x u_{xx} \\
			 & - 4 \xi^1_{xu} u_{t}  u_{xxx} - 6 \xi^1_{uuu}  u_{t} u^2_x u_{xx} - 3  \xi^1_{uu} u_{t} u^2_{xx} - 4 \xi^1_{uu} u_{t}  u_x u_{xxx} -  \xi^1_{uuuu} u_{t} u^4_x  -  \xi^1_{u} u_{t}  u_{xxxx} \\
			   &  -  \xi^2_{xxxx} u_x - 4 \xi^2_{xxxu} u_x^2 - 6 \xi^2_{xxuu} u^3_x -  18   \xi^2_{xxu} u_x u_{xx} - 4  \xi^2_{xuuu} u^4_x  -24 \xi^2_{xuu} u^2_x u_{xx} \\
			 & -16 \xi^2_{xu} u_x u_{xxx} - 10 \xi^2_{uuu} u^3_x u_{xx} - 15 \xi^2_{uu} u_x u^2_{xx} - 10 \xi^2_{uu} u^2_x u_{xxx} - \xi^2_{uuuu} u^5_x  - 5  \xi^2_{u} u_x u_{xxxx} \\
			   & - 4 \xi^1_{xxx} u_{tx} -12 \xi^1_{xxu} u_{tx} u_x - 12 \xi^1_{xuu} u_{tx} u^2_x -12 \xi^1_{xu} u_{tx} u_{xx} - 12 \xi^1_{uu} u_{tx} u_x u_{xx} - 4 \xi^1_{uuu} u_{tx} u^3_x \\ 
			   &  - 4 \xi^1_{u} u_{tx} u_{xxx} - 4 \xi^2_{xxx} u_{xx} - 12  \xi^2_{xu} u^2_{xx} - 10 \xi^2_{u} u_{xx} u_{xxx}   - 6  \xi^1_{xx} u_{txx} - 12 \xi^1_{xu} u_{txx} u_x  \\ 
			   & - 6 \xi^1_{uu} u_{txx} u^2_x - 6 \xi^1_u u_{xx} u_{txx}  - 6  \xi^2_{xx} u_{xxx}  - 4  \xi^1_x u_{xxxt} - 4 \xi^1_u u_{xxxt} u_x - 4  \xi^2_x u_{xxxx} .
\end{eqnarray*}
Following the same procedure, we obtain 
\begin{align*}
\zeta_x       &= D_x \eta - u_t D_x \xi^1 - u_x D_x \xi^2 \\
\zeta_{xx}   &= D_x \zeta_x - u_{xt} D_x \xi^1 - u_{xx} D_x \xi^2 \\
\zeta_{xxt}      &= D_t \zeta_{xx} - u_t D_t \xi^1 - u_x D_t \xi^2 \\
\zeta_{xxtt}   &= D_t \zeta_{xxt}   - u_{tt} D_t \xi^1 - u_{tx} D_t \xi^2 
\end{align*}
\begin{eqnarray*}
& \zeta_{xxtt} =  \eta_{xxtt} + \eta_{uuuu}  u^2_{t} u^2_{x} + \eta_{uuu}  u_{tt} u^2_x 
             + \eta_{uuu}  u^2_{t} u_{xx}  + \eta_{uu}  u_{tt} u_{xx} 
              + 2 \eta_{uut}  u_{t} u_{xx} + 2 \eta_{uu}  u_{t} u_{txx} \\
              &  + 4 \eta_{uutx}  u_{t} u_x + 4 \eta_{uux}  u_{t} u_{tx} 
              + 2 \eta_{uu}  u^2_{tx} + \eta_{ttu}   u_{xx} 
              + 2 \eta_{tu}  u_{txx} + \eta_{u}  u_{ttxx} + \eta_{uutt}   u^2_{x} 
              + 2 \eta_{xxtu}  u_{t} \\
              & + 2 \eta_{uu}  u_{ttx} u_{x} + 4 \eta_{uut}  u_{tx} u_{x} + 4 \eta_{utx}    u_{tx} + 2 \eta_{ux}  u_{ttx}  + \eta_{xxuu}  u^2_{t} + 2 \eta_{uuut}  u_{t} u^2_{x} + 4 \eta_{uuu} u_x u_{t} u_{tx} \\
              & + \eta_{uxx}  u_{tt}  + 2 \eta_{uttx}   u_{x} 
             + 2 \eta_{uux}  u_{tt} u_{x} + 2 \eta_{uuux}  u^2_{t} u_{x}                         
             -4 \xi^1_{tu} u_{ttx} u_x  - 6 \xi^1_{uu} u_{t} u^2_{tx} - 6 \xi^1_{u} u_{tx} u_{ttx} -  \xi^1_{uuu} u^3_{t} u_{xx} \\
             &- 2  \xi^1_{tuu} u^2_{t} u_{xx}
              - \xi^1_{u} u_{xx} u_{ttt} - 2 \xi^1_{tu} u_{tt} u_{xx} - \xi^1_{ttu} u_{t} u_{xx} - 3 \xi^1_{uu} u^2_{t} u_{txx} - 3 \xi^1_{u} u_{tt} u_{txx} 
              - 4 \xi^1_{tu} u_{t} u_{txx} \\
              &- 2 \xi^1_{uuux} u^3_{t} u_{x} - 2 \xi^1_{uttx} u_{t} u_{x} - 6 \xi^1_{uux} u^2_{t} u_{tx} - 6 \xi^1_{ux} u_{tt} u_{tx} 
              -8 \xi^1_{utx} u_{t} u_{tx} - 6 \xi^1_{ux} u_{t} u_{ttx} - 2 \xi^1_{ux} u_{x} u_{ttt} \\
              &- 4 \xi^1_{tu} u^2_{tx} - \xi^1_{tt} u_{txx} - 2 \xi^1_{x} u_{tttx}  
               - 2 \xi^1_{t} u_{ttxx}  - 2 \xi^1_{xxtu} u^2_{t} - \xi^1_{xxtt} u_{t} -3 \xi^1_{xxu} u_{t} u_{tt}  - \xi^1_{ttuu} u_{t} u^2_{x} - 2 \xi^1_{ttu} u_{x} u_{tx} \\ 
              & -2 \xi^1_{ttx} u_{tx} - 4 \xi^1_{tx}  u_{ttx} 
             -  6 \xi^1_{uux} u_{t} u_x u_{tt} - 3 \xi^1_{uuu} u_{t} u^2_x u_{tt} - 6 \xi^1_{uu} u_{t} u_x u_{ttx}  - 6 \xi^1_{uuu} u_x u^2_{t} u_{tx}  \\
              & -6 \xi^1_{uu} u_x u_{tt} u_{tx} - 8 \xi^1_{uut} u_x u_{t} u_{tx} - 3 \xi^1_{uu} u_t u_{tt} u_{xx} - 2 \xi^1_{uut} u^2_x u_{tt} 
              - 2 \xi^1_{uuut} u^2_x u^2_{t} - \xi^1_{uuuu} u^2_x u^3_{t} - \xi^1_{xx}  u_{ttt}  \\
              & - \xi^1_{xxuu}  u^3_{t} - 2 \xi^1_{u} u_x u_{tttx}  - 3  \xi^1_{u} u_t u_{ttxx} 
               -  \xi^1_{uu} u^2_x u_{ttt} - 4 \xi^1_{utx} u_x u_{tt} - 2 \xi^1_{xxt} u_{tt} - 4 \xi^1_{uutx} u^2_t u_{x} \\
             & - 2 \xi^2_{xxtu} u_{t} u_{x} - 3 \xi^2_{ttu} u_{xx} u_x - 6 \xi^2_{tu} u_{txx} u_x - 6 \xi^2_{u} u_{tx} u_{txx} - 6 \xi^2_{tu} u_{tx} u_{xx} - 3 \xi^2_{u} u_{xx} u_{ttx} \\
             & - \xi^2_{uu} u^2_{t} u_{xxx} - \xi^2_{u} u_{tt} u_{xxx}  - 2 \xi^2_{tu} u_{t} u_{xxx} - 2 \xi^2_{uux} u^2_{t} u_{xx}  - 2 \xi^2_{ux} u_{tt} u_{xx}  - 4 \xi^2_{utx} u_{t} u_{xx}  \\
             & - 4 \xi^2_{ux} u_{x} u_{ttx} - 4 \xi^2_{ux} u_{t} u_{txx} - \xi^2_{xxuu} u^2_{t} u_{x} - 2 \xi^2_{uuux} u^2_{t} u^2_{x} - 2 \xi^2_{uux} u_{tt} u^2_{x} - \xi^2_{uutt} u^3_{x}  \\
             & - \xi^2_{tt} u_{xxx} - 2 \xi^2_{x} u_{ttxx} - 2 \xi^2_{t} u_{txxx} - 2 \xi^2_{xxu} u_{t} u_{tx} - \xi^2_{xxu} u_{x} u_{tt} - \xi^2_{xx} u_{ttx} - 6 \xi^2_{uu} u_{x} u^2_{tx} \\ 
            & - 2 \xi^2_{xxt} u_{tx} - 4 \xi^2_{ux} u^2_{tx} - 2 \xi^2_{ttx} u_{xx} - 4 \xi^2_{tx} u_{txx} -8 \xi^2_{uux} u_{t} u_x u_{tx} - 6 \xi^2_{uuu} u_{t} u^2_x u_{tx}   \\
            &  - 3 \xi^2_{uuu} u^2_{t} u_x u_{xx}  - 3 \xi^2_{uu} u_{tt} u_x u_{xx} - 6 \xi^2_{uut} u_{t} u_x u_{xx}  - 6 \xi^2_{uu} u_{t} u_x u_{txx}  - 6 \xi^2_{uu} u_{t} u_{tx} u_{xx} \\
            & -2 \xi^2_{uuut} u_{t} u^3_{x}    - 6 \xi^2_{uut} u^2_{x} u_{tx}  - \xi^2_{uuuu} u^2_{t} u^3_{x} - \xi^2_{uuu} u^3_{x} u_{tt} - 2 \xi^2_{uttx} u^2_{x} - 3 \xi^2_{u} u_{x} u_{ttxx} \\
            &  - 2 \xi^2_{u} u_{t} u_{txxx} - 3 \xi^2_{uu} u^2_{x} u_{ttx} - 8 \xi^2_{utx} u_{x} u_{tx} - 4 \xi^2_{uutx} u_{t} u^2_{x} - \xi^2_{xxtt} u_{x} .
\end{eqnarray*}
The Lie infinitesimal criterion of PDE (\ref{bio3}) is given by 
\begin{equation*}
X^{[2]} [u_{tt} - \Phi^{'} u^2_x - \Phi u_{xx}] + X^{[4]}[h_1 u_{xxxx} + h_2 u_{xxtt}]  
{\LARGE|_{u_{tt} = ( \Phi(u)\cdot u_{x}  )_x + h_1 u_{xxxx} + h_2 u_{xxtt}}} = 0 
\end{equation*} 
Lie invariance criterion provide us the following expression
\begin{align*}
& \eta_{tt} + 2 \eta_{tu} u_t + \eta_{uu} u^2_t + \eta_u \cdot (\Phi^{'} u^2_x + \Phi u_{xx} + h_1 u_{xxxx} + h_2 u_{xxtt}) \\
& - 2(\Phi^{'} u^2_x + \Phi u_{xx} + h_1 u_{xxxx} + h_2 u_{xxtt} )(\xi^1_t + \xi^1_u \cdot  u_{t}) \nonumber \\
&  -2u_{tx}(\xi^2_t + \xi^2_u \cdot u_t)  - u_t(\xi^1_{tt} + 2 \xi^1_{tu} \cdot  u_t  + \xi^1_{uu} \cdot  u^2_t ) - u_t \xi^1_u ( \Phi^{'} u^2_x + \Phi u_{xx} + h_1 u_{xxxx} + h_2 u_{xxtt} )  \nonumber  \\
& -u_x (\xi^2_{tt} + 2 \xi^2_{tu}u_t + \xi^2_{uu} u^2_t )  -u_x \xi^2_u (\Phi^{'} u^2_x + \Phi u_{xx} + + h_1 u_{xxxx} + h_2 u_{xxtt}) - \eta \Phi^{''} u^2_x  \nonumber \\ 
& - 2 \Phi^{'} u_{x}[\eta_x + \eta_u u_x - u_t(\xi^1_x + \xi^1_u u_x) - u_x(\xi^2_x + \xi^2_u u_x) ] - \Phi^{'} u_{xx} \eta \nonumber \\
& -\Phi [ \eta_{xx} + 2 \eta_{xu} u_x + \eta_{uu} u^2_x + \eta_u u_{xx} - 2u_{tx}(\xi^1_x + \xi^1_u \cdot  u_{x}) \nonumber  \\
&  -2u_{xx}(\xi^2_x + \xi^2_u \cdot u_x)  - u_t(\xi^1_{xx} + 2 \xi^1_{xu} \cdot  u_x + \xi^1_{uu} \cdot  u^2_x + \xi^1_u \cdot u_{xx} )   \\
& -u_x (\xi^2_{xx} + 2 \xi^2_{xu}u_x + \xi^2_{uu} u^2_x + \xi^2_u u_{xx}) ] 
\end{align*}
\begin{align*}
& + h_1 [ \eta_{xxxx} + 4 \eta_{xxxu} u_x + 6 \eta_{xxuu} u^2_x +  6 \eta_{xxu} u_{xx} + 4\eta_{xuuu} u^3_x  + 12 \eta_{xuu} u_x u_{xx} \\
			 & +4 \eta_{xu} u_{xxx} + 6 \eta_{uuu} u^2_x u_{xx} + 3\eta_{uu} u^2_{xx} + 4\eta_{uu} u_x u_{xxx} + \eta_{uuuu} u^4_x  +  \eta_{u} u_{xxxx} \\
			   &  - \xi^1_{xxxx} u_{t}  -  4   \xi^1_{xxxu} u_{t} u_x - 6  \xi^1_{xxuu} u_{t} u^2_x -  6  \xi^1_{xxu} u_{t} u_{xx} - 4  \xi^1_{xuuu} u_{t} u^3_x  - 12  \xi^1_{xuu} u_{t} u_x u_{xx} \\
			 & - 4 \xi^1_{xu} u_{t}  u_{xxx} - 6 \xi^1_{uuu}  u_{t} u^2_x u_{xx} - 3  \xi^1_{uu} u_{t} u^2_{xx} - 4 \xi^1_{uu} u_{t}  u_x u_{xxx} -  \xi^1_{uuuu} u_{t} u^4_x  -  \xi^1_{u} u_{t}  u_{xxxx} \\
			   &  -  \xi^2_{xxxx} u_x - 4 \xi^2_{xxxu} u_x^2 - 6 \xi^2_{xxuu} u^3_x -  18   \xi^2_{xxu} u_x u_{xx} - 4  \xi^2_{xuuu} u^4_x  -24 \xi^2_{xuu} u^2_x u_{xx} \\
			 & -16 \xi^2_{xu} u_x u_{xxx} - 10 \xi^2_{uuu} u^3_x u_{xx} - 15 \xi^2_{uu} u_x u^2_{xx} - 10 \xi^2_{uu} u^2_x u_{xxx} - \xi^2_{uuuu} u^5_x  - 5  \xi^2_{u} u_x u_{xxxx} \\
			   & - 4 \xi^1_{xxx} u_{tx} -12 \xi^1_{xxu} u_{tx} u_x - 12 \xi^1_{xuu} u_{tx} u^2_x -12 \xi^1_{xu} u_{tx} u_{xx} - 12 \xi^1_{uu} u_{tx} u_x u_{xx} - 4 \xi^1_{uuu} u_{tx} u^3_x 
\end{align*}
\begin{align*}	 
			   &  - 4 \xi^1_{u} u_{tx} u_{xxx} - 4 \xi^2_{xxx} u_{xx} - 12  \xi^2_{xu} u^2_{xx} - 10 \xi^2_{u} u_{xx} u_{xxx}   - 6  \xi^1_{xx} u_{txx} - 12 \xi^1_{xu} u_{txx} u_x  \\ 
			   & - 6 \xi^1_{uu} u_{txx} u^2_x - 6 \xi^1_u u_{xx} u_{txx}  - 6  \xi^2_{xx} u_{xxx}  
			   - 4  \xi^1_x u_{xxxt} - 4 \xi^1_u u_{xxxt} u_x - 4  \xi^2_x u_{xxxx} ] \\
& + h_2 [ \eta_{xxtt} + \eta_{uuuu}  u^2_{t} u^2_{x} 
			 + \eta_{uuu}  (\Phi^{'} u^2_x + \Phi u_{xx} + h_1 u_{xxxx} + h_2 u_{xxtt}) u^2_x \\
             & + \eta_{uuu}  u^2_{t} u_{xx}  + \eta_{uu}  (\Phi^{'} u^2_x + \Phi u_{xx} + h_1 u_{xxxx} + h_2 u_{xxtt}) u_{xx}  + 2 \eta_{uut}  u_{t} u_{xx} + 2 \eta_{uu}  u_{t} u_{txx} \\
              &  + 4 \eta_{uutx}  u_{t} u_x + 4 \eta_{uux}  u_{t} u_{tx} 
              + 2 \eta_{uu}  u^2_{tx} + \eta_{ttu}   u_{xx} 
              + 2 \eta_{tu}  u_{txx} + \eta_{u}  u_{ttxx} + \eta_{uutt}   u^2_{x} 
              + 2 \eta_{xxtu}  u_{t} \\
              & + 2 \eta_{uu}  u_{ttx} u_{x} + 4 \eta_{uut}  u_{tx} u_{x} + 4 \eta_{utx}    u_{tx} + 2 \eta_{ux}  u_{ttx}  + \eta_{xxuu}  u^2_{t} + 2 \eta_{uuut}  u_{t} u^2_{x} + 4 \eta_{uuu} u_x u_{t} u_{tx} \\
              & + \eta_{uxx}  (\Phi^{'} u^2_x + \Phi u_{xx} + h_1 u_{xxxx} + h_2 u_{xxtt})  + 2 \eta_{uttx}   u_{x}  \\
             & + 2 \eta_{uux}  (\Phi^{'} u^2_x + \Phi u_{xx} + h_1 u_{xxxx} + h_2 u_{xxtt}) u_{x} + 2 \eta_{uuux}  u^2_{t} u_{x}       
\end{align*}
\begin{align*}                  
             & -4 \xi^1_{tu} u_{ttx} u_x  - 6 \xi^1_{uu} u_{t} u^2_{tx} - 6 \xi^1_{u} u_{tx} u_{ttx} -  \xi^1_{uuu} u^3_{t} u_{xx} \\
             &- 2  \xi^1_{tuu} u^2_{t} u_{xx}
              - \xi^1_{u} u_{xx} u_{ttt} - 2 \xi^1_{tu} (\Phi^{'} u^2_x + \Phi u_{xx} + h_1 u_{xxxx} + h_2 u_{xxtt}) u_{xx} \\
              &- \xi^1_{ttu} u_{t} u_{xx} - 3 \xi^1_{uu} u^2_{t} u_{txx} - 3 \xi^1_{u} (\Phi^{'} u^2_x + \Phi u_{xx} + h_1 u_{xxxx} + h_2 u_{xxtt}) u_{txx} 
              - 4 \xi^1_{tu} u_{t} u_{txx} \\
              &- 2 \xi^1_{uuux} u^3_{t} u_{x} - 2 \xi^1_{uttx} u_{t} u_{x} - 6 \xi^1_{uux} u^2_{t} u_{tx} \\
              & - 6 \xi^1_{ux} (\Phi^{'} u^2_x + \Phi u_{xx} + h_1 u_{xxxx} + h_2 u_{xxtt}) u_{tx} 
              -8 \xi^1_{utx} u_{t} u_{tx} - 6 \xi^1_{ux} u_{t} u_{ttx} - 2 \xi^1_{ux} u_{x} u_{ttt} 
\end{align*}
\begin{align*}
              &- 4 \xi^1_{tu} u^2_{tx} - \xi^1_{tt} u_{txx} - 2 \xi^1_{x} u_{tttx}  
               - 2 \xi^1_{t} u_{ttxx}  - 2 \xi^1_{xxtu} u^2_{t} - \xi^1_{xxtt} u_{t} \\
               & -3 \xi^1_{xxu} u_{t} (\Phi^{'} u^2_x + \Phi u_{xx} + h_1 u_{xxxx} + h_2 u_{xxtt})  - \xi^1_{ttuu} u_{t} u^2_{x} - 2 \xi^1_{ttu} u_{x} u_{tx} \\ 
              & -2 \xi^1_{ttx} u_{tx} - 4 \xi^1_{tx}  u_{ttx} 
             -  6 \xi^1_{uux} u_{t} u_x (\Phi^{'} u^2_x + \Phi u_{xx} + h_1 u_{xxxx} + h_2 u_{xxtt})\\
             & - 3 \xi^1_{uuu} u_{t} u^2_x (\Phi^{'} u^2_x + \Phi u_{xx} + h_1 u_{xxxx} + h_2 u_{xxtt})
             - 6 \xi^1_{uu} u_{t} u_x u_{ttx}  - 6 \xi^1_{uuu} u_x u^2_{t} u_{tx}  \\
              & -6 \xi^1_{uu} u_x (\Phi^{'} u^2_x + \Phi u_{xx} + h_1 u_{xxxx} + h_2 u_{xxtt}) u_{tx} - 8 \xi^1_{uut} u_x u_{t} u_{tx} \\
              & - 3 \xi^1_{uu} u_t (\Phi^{'} u^2_x + \Phi u_{xx} + h_1 u_{xxxx} + h_2 u_{xxtt}) u_{xx} 
\end{align*}
\begin{align*}
              & - 2 \xi^1_{uut} u^2_x (\Phi^{'} u^2_x + \Phi u_{xx} + h_1 u_{xxxx} + h_2 u_{xxtt})
              - 2 \xi^1_{uuut} u^2_x u^2_{t} - \xi^1_{uuuu} u^2_x u^3_{t} - \xi^1_{xx}  u_{ttt}  \\
              & - \xi^1_{xxuu}  u^3_{t} - 2 \xi^1_{u} u_x u_{tttx}  - 3  \xi^1_{u} u_t u_{ttxx} 
               -  \xi^1_{uu} u^2_x u_{ttt} - 4 \xi^1_{utx} u_x (\Phi^{'} u^2_x + \Phi u_{xx} + h_1 u_{xxxx} + h_2 u_{xxtt}) \\
               & - 2 \xi^1_{xxt} (\Phi^{'} u^2_x + \Phi u_{xx} + h_1 u_{xxxx} + h_2 u_{xxtt}) - 4 \xi^1_{uutx} u^2_t u_{x} \\
             & - 2 \xi^2_{xxtu} u_{t} u_{x} - 3 \xi^2_{ttu} u_{xx} u_x - 6 \xi^2_{tu} u_{txx} u_x - 6 \xi^2_{u} u_{tx} u_{txx} - 6 \xi^2_{tu} u_{tx} u_{xx} - 3 \xi^2_{u} u_{xx} u_{ttx} \\
             & - \xi^2_{uu} u^2_{t} u_{xxx} - \xi^2_{u} (\Phi^{'} u^2_x + \Phi u_{xx} + h_1 u_{xxxx} + h_2 u_{xxtt}) u_{xxx}             
\end{align*}             
\begin{align*}
& - 2 \xi^2_{tu} u_{t} u_{xxx} - 2 \xi^2_{uux} u^2_{t} u_{xx}  - 2 \xi^2_{ux} u_{tt} u_{xx}  - 4 \xi^2_{utx} u_{t} u_{xx}  \\
             & - 4 \xi^2_{ux} u_{x} u_{ttx} - 4 \xi^2_{ux} u_{t} u_{txx} - \xi^2_{xxuu} u^2_{t} u_{x} - 2 \xi^2_{uuux} u^2_{t} u^2_{x} \\
             & - 2 \xi^2_{uux} (\Phi^{'} u^2_x + \Phi u_{xx} + h_1 u_{xxxx} + h_2 u_{xxtt}) u^2_{x} - \xi^2_{uutt} u^3_{x}  \\
             & - \xi^2_{tt} u_{xxx} - 2 \xi^2_{x} u_{ttxx} - 2 \xi^2_{t} u_{txxx} - 2 \xi^2_{xxu} u_{t} u_{tx} - \xi^2_{xxu} u_{x} (\Phi^{'} u^2_x + \Phi u_{xx} + h_1 u_{xxxx} + h_2 u_{xxtt}) \\
             &  - \xi^2_{xx} u_{ttx} - 6 \xi^2_{uu} u_{x} u^2_{tx} 
             - 2 \xi^2_{xxt} u_{tx} - 4 \xi^2_{ux} u^2_{tx} - 2 \xi^2_{ttx} u_{xx} - 4 \xi^2_{tx} u_{txx} -8 \xi^2_{uux} u_{t} u_x u_{tx} - 6 \xi^2_{uuu} u_{t} u^2_x u_{tx}   \\
            &  - 3 \xi^2_{uuu} u^2_{t} u_x u_{xx}  - 3 \xi^2_{uu} (\Phi^{'} u^2_x + \Phi u_{xx} + h_1 u_{xxxx} + h_2 u_{xxtt}) u_x u_{xx} 
\end{align*}
\begin{align*}
            & - 6 \xi^2_{uut} u_{t} u_x u_{xx}  - 6 \xi^2_{uu} u_{t} u_x u_{txx}  - 6 \xi^2_{uu} u_{t} u_{tx} u_{xx} \\
            & -2 \xi^2_{uuut} u_{t} u^3_{x}    - 6 \xi^2_{uut} u^2_{x} u_{tx}  - \xi^2_{uuuu} u^2_{t} u^3_{x} - \xi^2_{uuu} u^3_{x} (\Phi^{'} u^2_x + \Phi u_{xx} + h_1 u_{xxxx} + h_2 u_{xxtt})\\
            & - 2 \xi^2_{uttx} u^2_{x} - 3 \xi^2_{u} u_{x} u_{ttxx} 
              - 2 \xi^2_{u} u_{t} u_{txxx} - 3 \xi^2_{uu} u^2_{x} u_{ttx} - 8 \xi^2_{utx} u_{x} u_{tx} - 4 \xi^2_{uutx} u_{t} u^2_{x} - \xi^2_{xxtt} u_{x} ]
 =0 . \nonumber 
\end{align*}
From the infinitesimal criteria and by comparing the monomials, we obtained the set of determining equations given by
\begin{eqnarray*}
u_{x x x x} &:&  h_{1} \eta_{u} -2\xi^1_{t} h_{1} - h_{1} \eta_{u} + 4 h_{1} \xi^2_{x} - h_{1} h_{2} \eta_{u x x} + 2 h_{1} h_{2} \xi^1_{x x t}=0, \\
u_{x x t t} &:& h_{2} \eta_{u} - 2\xi^1_{t} h_{2} - h_{2} \eta_{u} - h^2_{2} \eta_{u x x} + 2 h_{2} \xi^1_{t} + 2 h^2_{2} \xi^1_{x x t} + 2 h_{2} \xi^2_{x}=0, \\
u_{t} u_{x x x x} &:& -2 h_{1} \xi^1_{u} - h_{1} \xi^1_{u} + h_{1} \xi_{u}^{1} + 3 h_{1} \xi_{x x u}^{1} h_{2}=0, \\
u_{t} u_{x x t t} &:& -2 h_{2} \xi_{u}^{1} - h_{2} \xi_{u}^{1} + 3 h_{2}^{2} \xi_{x x u}^{1}+3 h_{2} \xi_{u}^{1}=0, \\
u_{x} u_{x x x x} &:& -h_{1} \xi_{u}^{2} - 5 h_{1} \xi_{u}^{2}-2 h_{1} h_{2}  \eta_{u u x} + 4 h_{1} h_{2} \xi_{u t x}^{1}+h_{1} h_{2} \xi_{x x u}^{2}=0, 
\end{eqnarray*}  
\begin{eqnarray*}
u_{x} u_{x x t t} &:& -h_{2} \xi_{u}^{2} - 2 h_{2}^{2} \eta_{u u x} + 4 h_{2}^{2} \xi_{u t x}^{1} + h_{2}^{2} \xi_{x x u}^{2} + 3 h_{2} \xi_{u}^{2}=0, \\
u_{x x x t} &:&  4 h_{1} \xi^1_{x}+ 2 h_{2} \xi_{t}^{2}=0, \\
u_{x} u_{x x x t} &:&  4 h_{1} \xi^1_{u}=0, \\
u_{x}^{2} u_{x x x x} &:& -h_{1} h_{2} \eta_{uuu }+ 2 \xi_{uut}^{1} h_{1} h_{2} + 2 h_{1} h_{2}  \xi_{u u x}^{2}=0, \\
u_{x}^{2} u_{x x t t} &:& -h_{2}^{2} \eta_{uuu }+ 2 \xi_{uut }^{1} h_{2}^{2} + 2 h_{2}^{2}  \xi_{uux }^{2}=0 ,
\end{eqnarray*}  
\begin{eqnarray*}
u_{x x} u_{x x x x} &:& -h_{1} h_{2} \eta_{uu } + 2 h_{1} h_{2} \xi_{t u}^{1}=0, \\
u_{x x} u_{x x t t} &:& -h_{2}^{2} \eta_{uu} + 2 h_{2}^{2} \xi_{t u}^{1}=0,  \\
u_{t x x} u_{x x x x } &:&  3 h_{1} h_{2} \xi_{u}^{1}=0, \\
u_{t x x} u_{x x t t} &:& 3 h_{2}^{2} \xi_{u}^{1}=0, \\
u_{t x} u_{x x x x} &:&  6 h_{1} h_{2} \xi_{u x}^{1}=0 ,
\end{eqnarray*}  
\begin{eqnarray*}
u_{t x} u_{x x t t} &:&  6 h_{2}^{2} \xi_{u x}^{1}=0, \\
u_{tttx}  &:&   2 h_{2} \xi_{x}^{1}=0, \\
u_{t} u_{x} u_{xxxx} &:&  6 h_{1} h_{2} \xi_{uu x}^{1}=0,  \  \quad \\
u_{t} u_{x} u_{x x t t} &:&  \  \quad \  \  6 h_{2}^{2} \xi_{uu x}^{1} =0 ,
\end{eqnarray*}  
\begin{eqnarray*}
u_{t} u^2_{x} u_{xxxx} &:&  3 h_{1} h_{2} \xi_{uuu}^{1}=0,  \  \quad 
u_{t} u^2_{x} u_{xxtt}  :  \  \quad \  \  3 h_{2}^{2} \xi^1_{uuu}=0, \\
u_x u_{tx}  u_{xxxx} &:&  6 \xi_{uu}^{1} h_{1} h_{2}=0,  \  \quad 
u_x u_{tx}  u_{xxtt} :  \  \quad \  \  6 h_{2}^{2} \xi_{ uu }^{1} =0, \\
u_t u_{xx}  u_{xxxx} &:&  3 h_1 h_{2}  \xi_{u u}^{1}=0,  \  \quad 
u_t u_{xx}  u_{xxtt} :  \  \quad \  \  3 h_{2}^{2}  \xi_{u u}^{1}=0 ,
\end{eqnarray*}  
\begin{eqnarray*}
u_x u_{tttx} &:& 2 \xi^1_u h_2 = 0, \quad \ \
u_{x x x} u_{xxxx} :  h_{1} h_{2} \xi_{u}^{2} =0,  \  \quad \\
u_{xxx} u_{xxtt} &:&  \  \quad \  \  h_{2}^{2} \xi_{u}^{2}=0, \quad  \ \
u_{x} u_{xx} u_{xxxx} :  3 \xi_{uu}^{2} h_{1} h_{2}=0,  \  \quad \\
u_{x} u_{x x} u_{x x t t} &:&   \  \quad \  \ 3 h_{2}^{2} \xi_{uu}^{2}=0, \quad \ \
u_{x}^{3} u_{xxxx} :  h_{1} h_{2} \xi_{uuu}^{2}=0,  \  \quad \\
u_{x}^{3} u_{xxtt}   &:&   \  \quad \  \ h^2_{2} \xi_{uuu}^{2}=0, \quad \ \ 
u_{t} u_{txxx} : 2 h_2 \xi^2_u = 0 .
\end{eqnarray*}  
 From the above set of determining equations, the following cases arise depending on the values of $h_1$ and $h_2$. Suppose
\begin{eqnarray}
4 h_{1} \xi^1_{x}+ 2 h_{2} \xi_{t}^{2}=0.
\end{eqnarray} 
\textbf{Case-1 : $\boldsymbol{h_1=0}$ and $\boldsymbol{h_2 \neq 0 }$} 
\begin{eqnarray}
u_{xxxt} &:& \xi_{t}^{2}=0  \  \quad \  \ \Rightarrow  \  \quad \  \ \xi^{2}=\xi^{2}(x,u) \\
u_{txx} u_{xxtt} &:& \xi_{u}^{1}=0  \  \quad \  \ \Rightarrow  \  \quad \  \ \xi^{1}=\xi^{1}(x,t) \\
u_{tttx} &:&  \xi_{x}^{1} = 0  \  \quad \  \ \Rightarrow  \  \quad \  \ \xi^{1}=\xi^{1}(t) \\
u_{x x x} u_{x x t t} &:& \xi_{u}^{2}=0  \  \quad \  \ \Rightarrow  \  \quad \  \ \xi^{2}=\xi^{2}(x) \\
u_{x x} u_{x x t t} &:& \eta_{uu}=0  \  \quad \  \ \Rightarrow  \  \quad \  \ \eta=\alpha(t, x) u + \beta(t, x)
\end{eqnarray}
\textbf{Case-2 : $\boldsymbol{h_2=0}$ and $\boldsymbol{h_1 \neq 0 }$} 
\begin{eqnarray}
u_{xxxt} &:&  \xi_{x}^{1}=0 \\
u_{x} u_{x x x t} &:& \xi_{u}^{1}=0 \  \quad \  \ \Rightarrow  \  \quad \  \  \xi^1 = \xi^1(t) \\
u_{x} u_{x x x x} &:& \xi_{u}^{2}=0 \  \quad \  \ \Rightarrow  \  \quad \  \ \xi^{2}=\xi^{2}(x,t) \\
u_{t}^{2}  &:&   \eta_{uu}=0 \  \quad \  \  \Rightarrow \  \quad \  \  \eta=\alpha(t, x) u + \beta(t, x)
\end{eqnarray}
\textbf{Case-3 : $\boldsymbol{h_1 h_2 \neq 0 }$} 
\begin{eqnarray}
u_{x} u_{xxtx} &:& \xi_{u}^{1}=0 \\
u_{x x} u_{xxxx} &:&  \eta_{uu} = 0  \  \quad \  \  \Rightarrow \  \quad \  \  \eta=\alpha(t, x) u+\beta(t, x)\\
u_{xttt} &:& \xi^1_x=0 \  \quad \  \  \Rightarrow \  \quad \  \ \xi^1 = \xi^1(t) \\
u_{x x x} u_{x x x x} &:& \xi_{u}^{2}=0 \\
u_{xxxt} &:&  \xi_{t}^{2}=0 \  \quad \  \  \Rightarrow \  \quad \  \  \xi^{2}=\xi^{2}(x)
\end{eqnarray}
By redefining the prolongation with respect to the infinitesimals, we obtain 
\begin{eqnarray}
 \zeta_{t}  = D_{t} \eta - u_{t} D_{t} \xi^1 - u_{x} D_{t} \xi^{2} = \alpha_{t} u + \alpha u_{t}+\beta_{t}-u_{t} \xi_{t}^{1},
\end{eqnarray}
\begin{eqnarray}
\begin{gathered}
\zeta_{tt} = D_{t} \zeta_{t} -  u_{tt} D_{t} \xi^{1} - u_{t x} D_{t} \xi^{2}, \\
         =  \alpha_{tt} u - 2 \alpha_{t} u_{t} + \alpha_{tt} + \beta_{tt} - u_{tt} \xi_{t}^{1} - u_{t} \xi^1_{t t} -u_{t t} \xi^1_{t}, \\ 
         = \alpha_{t t} u + 2 \alpha_{t} u_{t} + \alpha_{tt} + \beta_{tt} - 2 \xi_{t}^{1} u_{t t} - u_{t} \xi_{t t}^{1} .
\end{gathered}
\end{eqnarray}
Similarly, we can redefine $\zeta_x, \zeta_{xx}, \zeta_{xxx}, \zeta_{xxxx}, \zeta_{xxt}, \zeta_{xxtt} $ with respect to the infinitesimals given by  
\begin{eqnarray}
\zeta_x &=& \alpha_{x} u+\alpha u_{x}+\beta_{x}-u_{x} \xi_{x}^{2}, \\
\zeta_{xx} &=& \alpha_{xx} u + 2 \alpha_{x} u_{x} + \alpha u_{xx} + \beta_{xx} - 2 u_{xx} \xi_{x}^{2} -u_{x} \xi_{xx}^{2}, \\
\zeta_{xxx} &=& u \alpha_{xxx} + 3 \alpha_{xx} u_{x} + 3 \alpha_{x} u_{xx} + \alpha u_{xxx} + \beta_{xxx}\\
& &  -3 u_{xxx} \xi_{x}^{2} - 3 u_{xx} \xi_{xx}^{2} - u_{x} \xi_{xxx}^{2}, \nonumber \\
\zeta_{xxxx} &=& 4 u_{x} \alpha_{xxx} + u \alpha_{xxxx} + 6  u_{xx} \alpha_{xx} + 4  u_{xxx} \alpha_{x} + \alpha u_{xxxx} \\
& & + \beta_{xxxx} - 4 u_{xxxx} \xi^2_x - 6 u_{xxx} \xi^2_{xx} - 4 u_{xx} \xi^2_{xxx} - u_x \xi^2_{xxxx}, \nonumber 
\end{eqnarray}
\begin{eqnarray}
\zeta_{xxt} &=& \alpha_{t x x} u + \alpha_{xx} u_{t} + 2 \alpha_{t x} u_{x} + 2 \alpha_{x} u_{tx}+\alpha_{t} u_{xx} \\
& & +\alpha u_{t x x} +\beta_{t x x} - 2 u_{t x x} \xi_{x}^{2} -u_{t x} \xi_{x x}^{2}-u_{txx} \xi_{t}^{1}, \nonumber \\
\zeta_{xxtt} &=& \alpha_{t t x x} u + 2 u_{t} \alpha_{t xx} + \alpha_{xx} u_{t t} + 2 \alpha_{ttx} u_{x}
+4 \alpha_{tx} u_{t x} \\
& & + 2 \alpha_{x} u_{t t x} + \alpha_{t t} u_{xx} + 2 \alpha_{t} u_{t x x} + \alpha u_{t t x x} + \beta_{t t x x} \\
& & -2 u_{t t x x} \xi_{x}^{2} - u_{t t x} \xi_{x x}^{2} - 2 u_{ttx x} \xi_{t}^{1} - u_{t x x} \xi_{t t}^{1}.
\end{eqnarray}
From the invariance criteria (\ref{IC4th}) the set of determining equations reduces to  
\begin{eqnarray}
u_{xxtt} &:& \alpha h_{2} - 2 \xi^1_{t} h_{2} -h_{2}^{2} \alpha_{xx} - h_{2} \alpha + 2 h_{2} \xi_{x}^{2}+ 2 h_2 \xi^1_t =0 \\
u_{xxxx} &:& \alpha h_{1} - h_1 \alpha + 4 h_1 \xi_{x}^{2} - h_{2} h_1 \alpha_{xx} - 2 \xi_{t}^{1} h_{1}=0 \\
u_{t} &:&  2 \alpha_{t} - 2 h_{2} \alpha_{txx} - \xi_{tt}^{1}=0 \\
u^{2}_x &:& \alpha \Phi^{\prime} - 2 \xi^{1}_t \Phi^{\prime} - \eta \Phi^{\prime \prime} - 2 \alpha \Phi^{\prime} + 2 \Phi^{\prime} \xi_{x}^{2} - h_2 \alpha_{xx} \Phi^{\prime} =0 
\end{eqnarray}
\begin{eqnarray}
 u_{xx} &:& \alpha \Phi -2 \xi_{t}^{1} \Phi - \eta \Phi^{\prime} - \alpha \Phi  + 2 \xi_{x}^{2} \Phi \nonumber \\ & &  -6 h_{1} \alpha_{xx} + 4 h_{1} \xi_{xxx}^{2} - h_{2} \alpha_{xx} \Phi - h_{2} \alpha_{t t}=0  \\
 u_{xxx} &:&  -4 h_{1} \alpha_{x} + 6 h_{1} \xi_{xx}^{2}=0 \\
 u_{ttx} &:& -2 \alpha_{x} h_{2} + h_2 \xi_{xx}^{2}=0 \\
 u_{txx} &:& -2 h_2 \alpha_{t} + h_{2} \xi_{tt}^{1}=0 
\end{eqnarray}
\begin{eqnarray}
 u_{x} &:& -2 \Phi^{\prime} \alpha_{x} u - 2 \Phi^{\prime} \beta_{x} - 2 \Phi \alpha_{x} - 4 h_{1} \alpha_{xxx} \nonumber \\
 & & +  h_{1} \xi_{xxxx}^{2} - 2 h_{2} \alpha_{t t x}=0 \\
 u_{t x} &:& -4 h_{2} \alpha_{t x}=0 \\
 1 &:&  \alpha_{tt} u + \beta_{tt} - \Phi \alpha_{xx} u - \Phi \beta_{xx} - h_{1} u \alpha_{xxxx} \nonumber \\
  & & - h_1 \beta_{xxxx} - h_2 \alpha_{ttxx} u - h_2 \beta_{ttxx} = 0. 
\end{eqnarray}

\subsection{$\boldsymbol{h_1=0}$ and $\boldsymbol{h_2 \neq 0} $}
The set of determining equations reduces to 
\begin{eqnarray}
u_{ttx} &:&  \xi_{x x}^{2} = 2 \alpha_{x} \ \ \Rightarrow \ \ (\xi_{t}^{2}=0) \\
u_{txx} &:&  \xi_{tt}^{1} = 2 \alpha_{t} \ \ \Rightarrow \ \  (\xi_{x}^{1}=0 ) \\
u_{t x} &:&  \alpha_{t x}=0 \\
u_{x x t t} &:&  2 \xi_{x}^{2} = h_{2} \alpha_{x x} \\
u_{xx} &:& -2 \xi_{t}^{1} \Phi - \Phi^{\prime} (\alpha u + \beta ) - h_{2} \alpha_{tt}=0 \label{Equxx} \\
u_x   &:&  - \Phi^{\prime} (\alpha_{x} u+\beta_{x})=0 \ \ \Rightarrow \ \ \alpha_{x} =0= \beta_x \ \ \mbox{As,} \ \ \Phi^{\prime} \neq 0  \\
1 &:& \alpha_{tt} u + \beta_{tt} = 0  \ \ \Rightarrow \ \   \alpha_{tt} =0= \beta_{tt} .
\end{eqnarray}
As ${\xi}_{x}^{2}=0$ and $\eta = \alpha u + \beta$, the equation (\ref{Equxx}) implies that 
\begin{eqnarray} \label{Equxx2}
u_{xx} &:& +2 \xi_{t}^{1} \Phi - \Phi^{\prime} (\alpha u + \beta ) =0.
\end{eqnarray}
Differentiating with respect to $t$, we get $u_tt=0$ so 
\begin{eqnarray}
\dot{\alpha}=0=\beta^{\prime}.
\end{eqnarray}
This further implies that  
\begin{eqnarray}
\xi^{1} &=& c_{1}+c_{2} t \ \ \Rightarrow \ \ \xi_{t}^{\prime}=c_{2} \\
\alpha &=& c_3  \\
\beta &=& c_4  \\
\xi^2 &=& c_5 \\
\eta &=& c_{3} u+c_{4}.
\end{eqnarray}
Now, equation (\ref{Equxx2}) implies that 
\begin{eqnarray}
2 c_{2} \Phi + \Phi^{\prime} (c_{3} u + c_{4} ) =0 .
\end{eqnarray}
As $\Phi \neq 0 $, so we get 
\begin{eqnarray} \label{6Eq1}
\frac{\Phi^{\prime}}{\Phi} + \frac{2 c_{2}}{c_{3} u + c_{4}} = 0 .
\end{eqnarray}
\subsection{ $\boldsymbol{\Phi = \epsilon (u+k_1)^{k_2} }$, where $\boldsymbol{\epsilon = \pm 1}$}
From equation (\ref{6Eq1}), we obtain 
\begin{eqnarray}
\ln \Phi + \frac{2c_2}{c_3} \ln (c_3 u + c_4) = \ln A , \\
\Phi (c_{3} u+c_{4})^{\frac{2 c_{2}}{c_{3}}}=A, \\
\Phi(u+\frac{c_{4}}{c_{3}})^{\frac{2 c_{2}}{c_{3}}}=\bar{A}.
\end{eqnarray} 
where $\frac{c_{4}}{c_{3}}=k_{1} \Rightarrow c_{4}=k_{1} c_{3}$ and -$\frac{2c_{2}}{c_{3}}=k_{2}$. So the infinitesimals are
\begin{eqnarray}
\xi^1 = c_1 + \frac{k_2}{2} c_3 t , \quad
\xi^2 = c_5, \quad
\eta = c_3 u + k_1 c_3.
\end{eqnarray}
From the infinitesimals, we obtain three symmetries which form a three dimensional algebra spanned by 
\begin{eqnarray*}
X_1 = \frac{\partial}{\partial t}, \quad
X_2 = \frac{\partial}{\partial x} , \quad
X_3 = \frac{k_2}{2} t \frac{\partial}{\partial t} + (k_1 + u )  \frac{\partial}{\partial u}.
\end{eqnarray*}
\subsubsection{ $\boldsymbol{\Phi = \epsilon u^{k_2} }$, where $\boldsymbol{\epsilon = \pm 1}$}
Equivalence transformations are given by 
\begin{eqnarray}
\bar{t}=a_{1} t+a_{2}, \ \  \ \bar{x}=b_{1} x+b_{2}, \ \  \ \bar{u}=c_{1} u +c_{2}, \\
\bar{h}_{1}=\frac{b_{1}^{2}}{a_{1}^{2}} h_{1}, \ \  \  \bar{h}_{2} =b_{1}^{2} h_{2}, \ \  \ \bar{\Phi}=\frac{b_{1}^{2}}{a^{2}} \Phi.
\end{eqnarray}
Using these transformations, we obtain 
\begin{eqnarray*}
X_1 = \frac{\partial}{\partial t},  \quad
X_2 = \frac{\partial}{\partial x} ,  \quad
X_3 = \frac{k_2}{2} t \frac{\partial}{\partial t} + u \frac{\partial}{\partial u}.
\end{eqnarray*}

\subsection{ $\boldsymbol{\Phi = \epsilon \exp (ku) }$, where $\boldsymbol{\epsilon = \pm 1}$}
From equation (\ref{6Eq1}), If 	$c_3=0$, we deduce
\begin{eqnarray}
\frac{\Phi^{\prime}}{\Phi} = -2 \frac{c_2}{c_4}, \\
\Phi = A \exp ( \frac{-2c_2}{c_4}  u). 
\end{eqnarray}
Let $\frac{-2c_2}{c_4} = k $ and this implies to $c_2 = \frac{-k}{2} c_4$. So we obtain infinitesimals in this case as 
\begin{eqnarray}
\xi^{1} = c_{1}+c_{2} t,  \quad
\xi^{2} = c_{5} ,  \quad
\eta = c_4. 
\end{eqnarray}
So the the symmetries we obtain in this case are 
\begin{eqnarray*}
X_1 = \frac{\partial}{\partial t},  \quad
X_2 = \frac{\partial}{\partial x},  \quad
X_3 = \frac{k}{2} t \frac{\partial}{\partial t} +  \frac{\partial}{\partial u}.
\end{eqnarray*}

\subsection{ $\boldsymbol{h_2=0}$ and $\boldsymbol{h_1 \neq 0} $}
The monomials in this case are 
\begin{eqnarray}
u_{xxxx} &:&   2 \xi^2_x = \xi^1_t \\
u_t      &:&   2 \alpha_t = \xi^1_{tt} \\
u_{x}^{2} &:& -2 \xi^1_{t} \Phi^{\prime} - \Phi^{\prime \prime} (\alpha u+\beta)- \alpha \Phi^{\prime}+2 \Phi^{\prime} \xi_{x}^{2}=0 \label{6Equ2} \\
 u_{x x}  &:&  -2 \xi^1_{t} \Phi -\Phi^{\prime}(\alpha u+\beta)+2 \xi_{x}^{2} \Phi - 6 h_{1} \alpha_{xx} + 4 h_{1} \xi_{xxx}^{2}=0 \\
 u_{xxx} &:& 4 \alpha_x = 6 \xi^2_{xx} \\
 u_{x} &:& -2 \Phi^{\prime} \alpha_x u - 2 \Phi^{\prime} \beta_{x} - 2 \Phi \alpha_{x} - 4 h_{1} \alpha_{xxx} + h_{1} \xi_{xxxx}^{2}=0.
\end{eqnarray}
Now, let us simplify the set of determining equations to obtain infinitesimals
\begin{eqnarray}
\xi^1_{t} &=& c_{1} \ \ \   \ \Rightarrow \ \ \   \ \xi^1=c_{1} t +c_{2} \\
2 \xi_{x}^{2} &=& c_{1} \  \ \Rightarrow \  \ 2 \xi^{2}= c_{1} x+ 2 c_{3}  \  \ \Rightarrow \  \ \xi^{2}=\frac{1}{2} c_{1} x + c_{3} \\
\xi_{t t}^{1} &=& 0 \ \ \Rightarrow \ \ \alpha_{t}=0 \ \ \Rightarrow \ \ \alpha=\alpha(x) \\
\alpha_{x} &=& 0 , \ since \ \ \xi^2_{xx}=0 \ \ \Rightarrow \ \ \alpha = c_4.
\end{eqnarray}
Now, from equation (\ref{6Equ2}), we get 
\begin{eqnarray}
-2 c_{1} \Phi^{\prime} -\Phi^{\prime \prime} (c_{4} u+ \beta ) - c_{4} \Phi^{\prime} + 2 \Phi^{\prime} \frac{1}{2} c_{1}=0 ,\\
 c_{1} \Phi^{\prime} +  \phi^{\prime \prime} (c_{4} u+\beta ) +c_{4}  \Phi^{\prime}=0, \\
 -2 c_{1} \Phi - \Phi^{\prime} (c_{4} u +  \beta ) + c_{1} \Phi=0, \\
 \Rightarrow c_{1} \Phi + \Phi^{\prime} (c_{4} u + \beta )=0, \label{6Equ3} \\
 -2 \Phi^{\prime} \beta_{x}=0 \ \  \Rightarrow \ \  \beta_{x}=0 \ \  \Rightarrow \ \  \beta=\beta(t).
\end{eqnarray}
Further, equation (\ref{6Equ3}) can be written as 
\begin{eqnarray} \label{6Equ4}
 \frac{\Phi^{\prime}}{\Phi} +  \frac{c_1}{(c_{4} u + c_5 )}=0 .
\end{eqnarray}
\begin{eqnarray}
\Rightarrow \ln \Phi + \frac{c_{1}}{c_{4}} \ln (c_{4} u+ c_{5} )= \ln A, 
\end{eqnarray}
\subsection{ $\boldsymbol{\Phi = \epsilon (u)^{k_1} }$, where $\boldsymbol{\epsilon = \pm 1}$}
When $c_4 \neq 0$, then equation (\ref{6Equ4}) implies that 
\begin{eqnarray}
\ln \Phi + \frac{c_{1}}{c_{4}} \ln (c_{4} u+ c_{5} )= \ln A, \\
\Phi (c_{4} u+ c_{5} )^{\frac{c_1}{c_4}}=A, \\
\Phi ( u+ \frac{c_{5}}{c_{4}}  )^{\frac{c_1}{c_4}}= \bar{A} .
\end{eqnarray}
Let $ \frac{-c_{1}}{c_{4}}=k_{1} $ and $ \frac{c_{5}}{c_{4}}=k_{2}$. Therefore $\Phi=\bar{A} (u+k_{2})^{k_{1}}$. This provides us the set of infinitesimals given by 
\begin{eqnarray}
\xi^1 = -k_{1} c_{4} t+c_{2} ,  \quad
\xi^{2} = -\frac{1}{2} k_{1} c_{4} x + c_{3},  \quad
\eta = c_{4} u + k_{2}.
\end{eqnarray}
Further, we obtain the set of symmetries given by 
\begin{eqnarray*}
X_1 = \frac{\partial}{\partial t},  \quad
X_2 = \frac{\partial}{\partial x},  \quad
X_3 = -k_1 t \frac{\partial}{\partial t}  - \frac{1}{2} k_1 x \frac{\partial}{\partial x} + (u+k_2) \frac{\partial}{\partial u}.
\end{eqnarray*}
By equivalence transformation, symmetries simplify to  
\begin{eqnarray*}
X_1 =\frac{\partial}{\partial t},  \quad
X_2 = \frac{\partial}{\partial x} ,  \quad
X_3 = -k_1 t \frac{\partial}{\partial t}  - \frac{1}{2} k_1 x \frac{\partial}{\partial x} + u \frac{\partial}{\partial u}.
\end{eqnarray*}
\subsection{ $\boldsymbol{\Phi = \epsilon \exp (u) }$, where $\boldsymbol{\epsilon = \pm 1}$}
When $c_4 = 0$, then equation (\ref{6Equ4}) leads us to the following result 
\begin{eqnarray} \label{6Equ4}
 \frac{\Phi^{\prime}}{\Phi} +  \frac{c_1}{c_5 }=0, \\
 \Rightarrow \ \ \Phi = A \exp (- \frac{c_1}{c_5 } u).
\end{eqnarray}
Let $\frac{-c_{1}}{c_{5}}=k$ and $c_{1}=-k c_{5}$ and we get the set of infinitesimals given by 
\begin{eqnarray}
\xi^1 = -k c_{5} t + c_{2},  \quad
\xi^{2} = -\frac{1}{2} k c_5 x + c_{3},  \quad
\eta = c_5 .
\end{eqnarray}
This provides us the set of symmetries 
\begin{eqnarray*}
X_1 = \frac{\partial}{\partial t},  \quad
X_2 = \frac{\partial}{\partial x},  \quad
X_3 =-k t \frac{\partial}{\partial t}  - \frac{1}{2}  x \frac{\partial}{\partial x} +  \frac{\partial}{\partial u}.
\end{eqnarray*}

\subsection{ $\boldsymbol{h_1 h_2 \neq 0} $}
For this case, we will obtain the monomials given by  
\begin{eqnarray}
u_{ttx} &:& \xi^2_{xx} = 2 \alpha_x \\
u_{xxx} &:& - 4 \alpha_{x} + 6 \xi_{x x}^{2}=0 \\
        & &  \Rightarrow 8 \alpha_{x}=0 \ \ \Rightarrow \ \ \alpha_{x}=0 \ \ \Rightarrow \ \ \alpha=\alpha(t) As \xi_{x x}^{2}=0 \\
u_{txx} &:& \xi_{t t}^{1} = 2 \dot{\alpha} \\
u_x   &:& -2 \Phi^{\prime} \beta_{x}=0 \\
1     &:&   \alpha_{t t} u + \beta_{t t} - \Phi \beta_{x x} - h_{1} \beta_{xxxx}-h_{2} \beta_{t t x x}=0 \\
u_{xx}	 &:&   -2 \Phi \xi^1_{t} -\Phi^{\prime}(\alpha u+\beta)+2 \Phi \xi_{x}^{2} - h_2 \ddot{\alpha}=0\\
u^2_x   &:&   \alpha \Phi^{\prime} - 2 \Phi^{\prime} \xi_{t}^{1} - \Phi^{\prime \prime}(\alpha u+\beta)-2 \Phi^{\prime} \alpha + 2 \Phi^{\prime} \xi_{x}^{2}=0 \\
u_{xxxx}    &:& 4 \xi_{x}^{2} - 2 \xi_{t}^{1}=0 \ \  \Rightarrow \ \ 2 \xi_{x}^{2}=\xi_{t}^{1} \\
  u_{xxtt}    &:&  \xi_{x}^{2}=0 \ \ \quad \ \ \xi_{t}^{1}=0 .
\end{eqnarray}
Also, we get 
\begin{eqnarray}
\xi_{x}^{1}=0=\xi_{u}^1, \\
\xi_{t}^{2}=0=\xi_{u}^{2},  \\
 \eta_{uu}=0.
\end{eqnarray} 
This provides us $\alpha=c_1$, $\xi^1 = c_2$ and $\xi^2 = c_3$. The set of determining equations becomes
\begin{eqnarray}
\beta_{tt} - \Phi \beta_{xx} - h_{1} \beta_{xxxx} - h_{2} \beta_{ttxx} = 0 , \\
- \Phi^{ \prime }( \alpha u + \beta) = 0, \\
c_{1} \Phi^{\prime} - \Phi^{\prime \prime}(\alpha u+\beta)-2 \Phi^{\prime} c_{1}=0 ,\\
\Rightarrow  \ \ \Phi^{\prime} c_{1} + \Phi^{\prime \prime} (\alpha u+\beta)=0. \label{6Equ5}
\end{eqnarray}
Equation (\ref{6Equ5}) can be written as \\
\begin{eqnarray}
\frac{\Phi^{\prime \prime}}{\Phi^{\prime }} + \frac{c_1}{\alpha u+\beta} = 0 .
\end{eqnarray}

\begin{align}
\int \frac{\Phi^{\prime \prime}}{\Phi^{\prime}} du + \int \frac{c_1}{\alpha u+\beta}  du &= 0 \\
\ln |\Phi^{\prime}| + c_1 + \int \frac{c_1}{\alpha u+\beta} du  &= 0 
\end{align}

\begin{align}
\ln |\Phi^{\prime}| &= -\frac{c_1}{\alpha} \ln (\alpha x+\beta) + {c_2}^{\prime} \\
|\Phi^{\prime}| &= c_2 e^{-\frac{c_1}{\alpha} \ln (\alpha x+\beta) } \\
\end{align}

\subsection{ $\boldsymbol{\Phi^{\prime} = 0}$}
If $\Phi^{\prime} = 0$, then we have 
\begin{eqnarray}
2(\beta_{tt} - A \beta_{xx} - h_1 \beta_{xxxx} - h) \beta_{ttxx} = 0 .
\end{eqnarray}
The infinitesimals in this case are 
\begin{eqnarray}
\xi^1 = c_2,  \quad
\xi^2 = c_3,  \quad
\eta = c_1 u + \beta(x,t).
\end{eqnarray}
Thus, the symmetries we obtain are
\begin{eqnarray*}
X_1 = \frac{\partial}{\partial t},  \quad
X_2 = \frac{\partial}{\partial x},  \quad
X_3 =  u \frac{\partial}{\partial u},  \quad
X_{\beta} = \beta  \frac{\partial}{\partial u}.
\end{eqnarray*}
For this case, we have infinite symmetries, where $\beta(x,t)$ satisfies the wave equation.

\subsection{Principal Case}
For arbitrary $\Phi$, we obtain $\alpha = \beta = 0 $ and the infinitesimals are given by 
\begin{eqnarray}
\xi^1 = c_2,  \quad
\xi^2 = c_3,  \quad
\eta = 0.
\end{eqnarray}
and the corresponding symmetries are 
\begin{eqnarray*}
X_1 = \frac{\partial}{\partial t},  \quad
X_2 = \frac{\partial}{\partial x} .
\end{eqnarray*}
This is the case when $h_1 h_2 \neq 0 $.

\subsection{Detailed explanation for computing multiplier} 
Consider the multiplier $\mathcal{M}(x,t,u)$ and substitute it in expression (\ref{E13aa}). Its simplification yields the following system of determining equations given by 
\begin{eqnarray}
\mathcal{E} [ \mathcal{M}  (u_{t t}-\phi_u u_x^2-\phi u_{x x} - h_1 u_{x x x x} - h_2 u_{x x t t} ) ]=0
\end{eqnarray}
This provides us with the following expression given by 
\begin{align*}
= & \mathcal{M}_u (u_{t t}-\phi_u u_x^2-\phi u_{x x}-h_1 u_{x x x x} - h_2 u_{x x t t} ) -D_x (-2 \mathcal{M} \phi_u u_x ) \\
& + D_t^2(\mathcal{M})+ D_x^2(-\phi \mathcal{M}) -h_1 D_x^4( \mathcal{M} )+ h_2 D_x^2 D_t^2( \mathcal{M} ).
\end{align*}
Applying the total derivative operator $D_x$ and $D_t$ we get 
\begin{align*}
 = & \mathcal{M}_u (u_{t t}-\phi_u u_x^2-\phi u_{x x}- h_2 u_{x x x x} -h_2 u_{x x t t}) \\ & +(2 \mathcal{M}_x \phi_u u_x + 2 \mathcal{M}_u \phi_u u_x^2 
  + 2 \mathcal{M} \phi_{u u} u_x^2 + 2 M \phi_u u_{x x} ) \\
  & + (\mathcal{M}_{t t}+2 \mathcal{M}_{t u} u_t + M_{u u} u_t^2 +  \mathcal{M}_u u_{t t} )\\
    & - (\phi \mathcal{M}_{x x}+\phi  \mathcal{M}_{x u} u_x +  \mathcal{M}_x \phi_u u_x + \phi \mathcal{M}_{x u} u_x )  \\
   & + \mathcal{M}_x \phi_u u_x+\phi \mathcal{M}_{u u} u_x^2 + 2 \phi_u M_u u_x^2 + \mathcal{M} \phi_{u u} u_x^2
\end{align*}

\noindent We can construct the set of determining equations by comparing the coefficients of the derivatives of $u$ and their products given by 
\begin{align*}
u_{x x x x} &: -h_1 \mathcal{M}_u - h_1 \mathcal{M}_u, \quad
u_{x x t t} : -h_2 \mathcal{M}_u - h_2 \mathcal{M}_u\\
u_{x x x} &: -h_1 \mathcal{M}_{x u}-3 h_1 \mathcal{M}_{x u}, \quad
u_x u_{x x x} : -4 h_1 \mathcal{M}_{u u} \\
u_{x x t} &:  -h_2 \mathcal{M}_{t u} - h_2 \mathcal{M}_{t u}, \quad
u_{x tt} : -2 h_2 \mathcal{M}_{x u},
\end{align*}
\begin{align*}
u_t u_{x x t} &: -h_2 \mathcal{M}_{x u}, \quad 
u_x u_{x t t} : -2 h_2 \mathcal{M}_{u u} \\
u_t^2 u_{x x t} &: -h_2 \mathcal{M}_{u u}, \quad
u_x u_t u_{x t} : -4 h_2 \mathcal{M}_{u u u}  \\
u_{x x} u_{t t} &: -h_2 \mathcal{M}_{u u}, \quad
u_x^2 u_{t t} : -h_2 \mathcal{M}_{u u u}  \\
\end{align*}
\begin{align*}
u_t^2 u_{x x} &: -h_2 \mathcal{M}_{u u u}, \quad
u_x^2 u_{x x} : -2 h_1 \mathcal{M}_{u uu }-h_1 \mathcal{M}_{uuu}-h_1 \mathcal{M}_{uuu}-2 h_1 \mathcal{M}_{uuu} \\
u_{x x}^2  &: -3 h_1 \mathcal{M}_{u u}, \quad
u_x u_{x x} : -12 h_1 \mathcal{M}_{x u u } \\
u_t u_{x x} &: -2 h_2 M_{t u u}, \quad
u_x u_{x t} : -2 h_2 \mathcal{M}_{tuu}-2 h_2 \mathcal{M}_{tuu} \\
u_t u_{x t} &: -2 h_{2} \mathcal{M}_{x u u}-2 h_2 \mathcal{M}_{xuu}, \quad
u_x u_{t t} :  -2 h_2 \mathcal{M}_{x u  u}, 
\end{align*}
\begin{align*}
u_{x t}^2 &:  -2 h_2 \mathcal{M}_{u u}, \quad
u_x u_t  :  -3 h_2 \mathcal{M}_{x t uu} -h_2 \mathcal{M}_{x t uu}  \\
 u_x u_t^2  &:  -2 h_2 \mathcal{M}_{x u u u}, \quad 
 u_x^2 u_t^2  :  -2 h_2 \mathcal{M}_{{uuuu }} \\ 
 u_{x t} &:  -2 h_2 \mathcal{M}_{x t u}-2 h_2 \mathcal{M}_{xtu}, \quad
  u_{t t} :  -2 h_2 \mathcal{M}_{x x u}+2 \mathcal{M}_u \\
  u_x^2 u_t &:  -2 h_2 \mathcal{M}_{t u u u}, \quad
 u_{x x} : -h_2 \mathcal{M}_{t t u}-6 h_1 \mathcal{M}_{x x u} + 2 \mathcal{M} \phi_u-\phi \mathcal{M}_u \\ 
 u_t^2 &: -h_2 \mathcal{M}_{x x u u}+ \mathcal{M}_{u u}, \quad
 u_t :  -2 h_2 \mathcal{M}_{x x t u}+2 \mathcal{M}_{t u},
\end{align*}
\begin{align*}
u_x^2 &:  -M_u \phi_u+ M \phi_{u u}-\phi M_{u u}- 6h_1 M_{x x u u} -h_2 \mathcal{M}_{t t u u} \\
u_x &: -4 h_1 \mathcal{M}_{x x x u}-2 h_2 M_{xtt u}-2 \phi M_{x u} \\
u_x^3 &: -4 h_1 \mathcal{M}_{x u u u}, \quad
u_x^4 : -h_1 \mathcal{M}_{{uuuu }} \\
\text{Const.} &: \mathcal{M}_{t t} - \phi \mathcal{M}_{x x}- h_1 \mathcal{M}_{x x x x} - h_2 \mathcal{M}_{x x t t}
\end{align*}
After simplifying the determining equations, we get 
\begin{align}
\mathcal{M}_u &=0 \\
2 \mathcal{M} \phi_u &=0 \\
\mathcal{M} \phi_{u u} &=0 \\
\mathcal{M}_{t t}-\phi \mathcal{M}_{x x} - h_1 \mathcal{M}_{x x x x} - h_2 M_{x x t t} &= 0
\end{align}
We obtained the following multipliers $\{ \mathcal{M}_1, \mathcal{M}_2, \mathcal{M}_3, \mathcal{M}_4 \} = \{ xt, x, t, 1 \}$ by solving above system. The multipliers of wave equation (\ref{Eq_3}) for an arbitrary function $\Phi(u)$ indicating that the wave equation (\ref{Eq_3}) possesses four conservation laws.

\pagebreak

\subsection{Soliton Solutions of Second Order Nerve Equation}

The 3D illustration of soliton solutions $u_1(x, t)$ and $u_2(x, t)$ are presented in Figures (\ref{fig2za}) and (\ref{fig2zb}). 3D Plots for soliton solutions $u_1(x, t)$ and $u_2(x, t)$ of second order nerve equation are visualised for particular values of parameters:

 \begin{figure}[h]
    \centering
    \begin{minipage}{0.4\textwidth}
        \includegraphics[width=\linewidth]{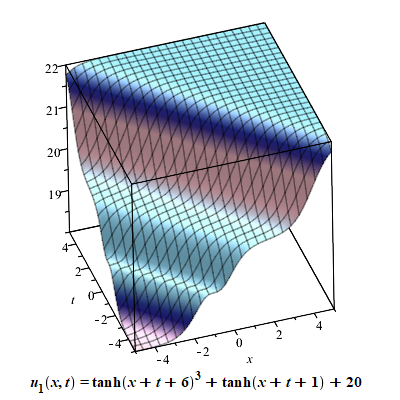}
        \includegraphics[width=\linewidth]{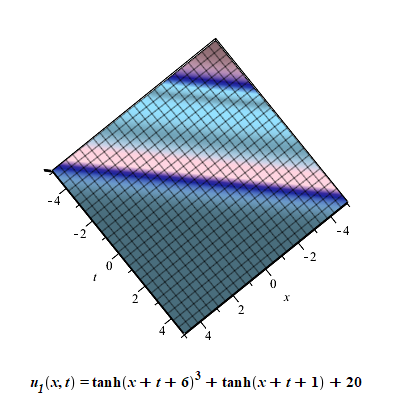}
        \caption{ Soliton solution of second order nerve equation $u_1(x, t)$}
         \label{fig2za}
    \end{minipage}
    \hfill
     \begin{minipage}{0.4\textwidth}
        \includegraphics[width=\linewidth]{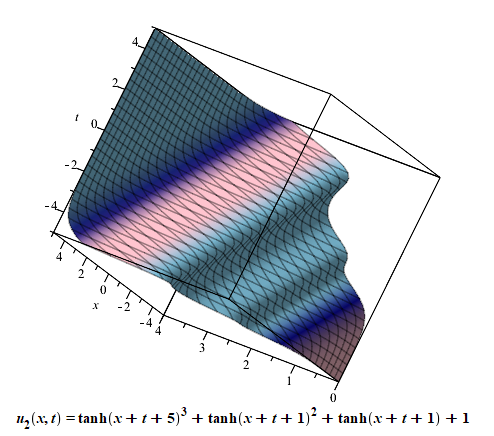}
        \includegraphics[width=\linewidth]{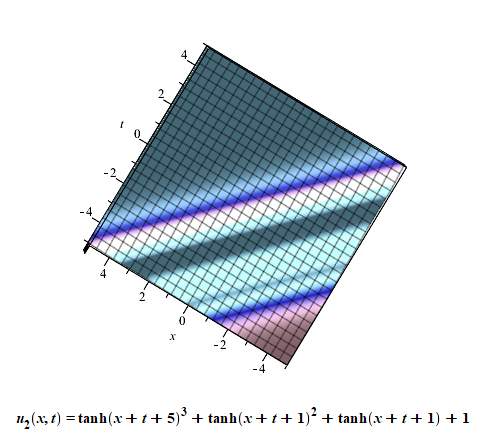}
        \caption{ Soliton solution of second order nerve equation $u_2(x, t)$}
         \label{fig2zb}
    \end{minipage}
\end{figure}  

\pagebreak

\subsubsection{Kink-Type Solutions of Fourth Order Nerve Equation}

\noindent \textbf{Principal Case: For arbitrary values of parameters ${h, k, p, q}$ and $r$}\\

\begin{figure}[h]
    \centering
    \begin{minipage}{0.45\textwidth}
        \includegraphics[width=\linewidth]{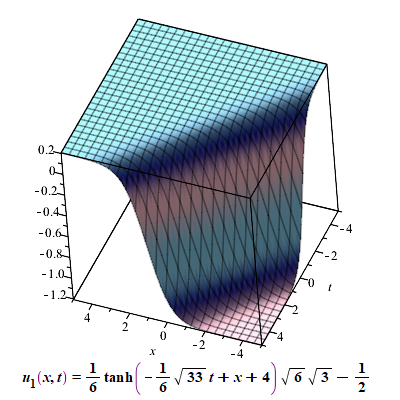}
    \end{minipage}
    \hfill
     \begin{minipage}{0.45\textwidth}
        \includegraphics[width=\linewidth]{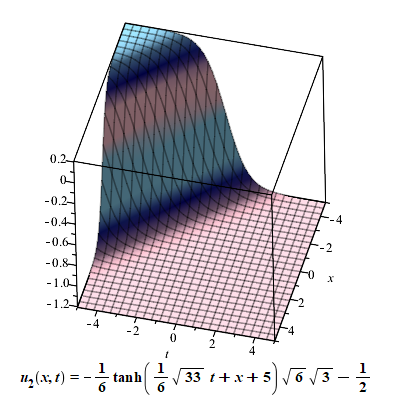}
    \end{minipage}
    \hfill
     \begin{minipage}{0.45\textwidth}
        \includegraphics[width=\linewidth]{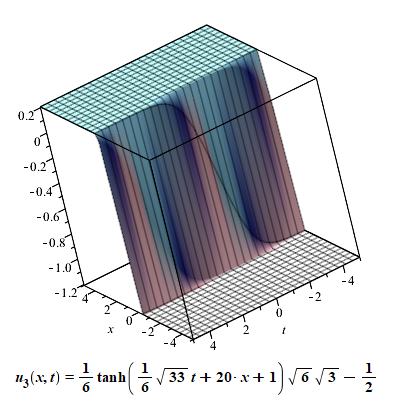}
    \end{minipage}
    \hfill
     \begin{minipage}{0.45\textwidth}
        \includegraphics[width=\linewidth]{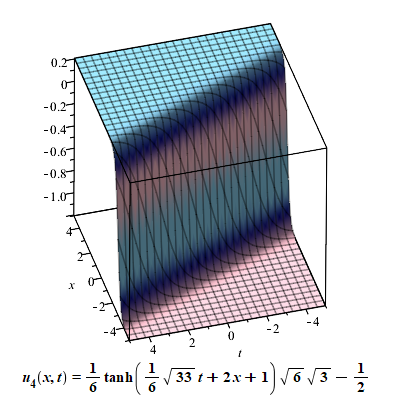}
    \end{minipage}
     \caption{3D Plots for Kink-Type solutions $u_1(x, t)$, $u_2(x, t)$, $u_3(x, t)$ and $u_4(x, t)$ of the fourth order nerve equation, Kink-Type solutions: Visualised for particular values of parameters}
    \label{fig2b}
\end{figure}

Four Kink-Type solutions $u_1(x, t), u_2(x, t), u_3(x, t)$ and $u_4(x, t)$ of equation (\ref{bio3}) is presented. For the better physical illustration and also for understanding the phenomenon of dynamics of the materials through the biomembranes, we discuss the geometrical explanation of the Kink-Type solution of the nerve equation (\ref{bio3}). The 3D illustration  of Kink-Type solutions $u_1(x, t), u_2(x, t), u_3(x, t)$ and $u_4(x, t)$ are given in Figure-(\ref{fig2b}).\\

\pagebreak

\begin{figure}[h]
    \centering
    \begin{minipage}{0.45\textwidth}
        \includegraphics[width=\linewidth]{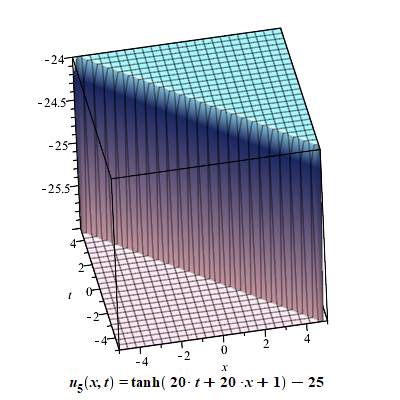}
    \end{minipage}
    \hfill
     \begin{minipage}{0.45\textwidth}
        \includegraphics[width=\linewidth]{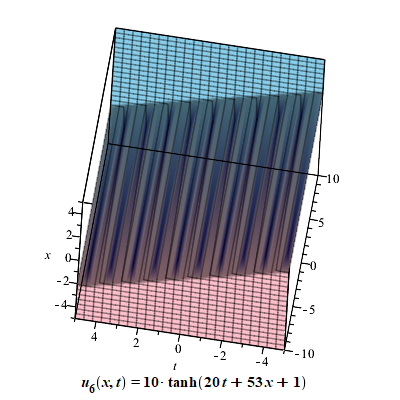}
    \end{minipage}
    \hfill
     \begin{minipage}{0.45\textwidth}
        \includegraphics[width=\linewidth]{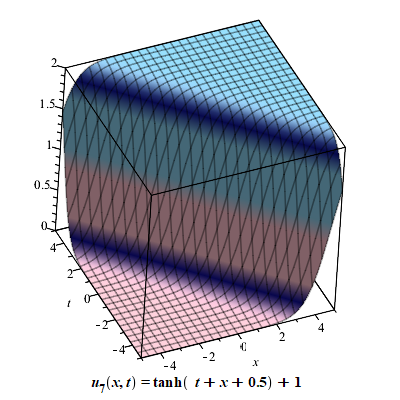}
    \end{minipage}
     \caption{3D Plots for Kink-Type solutions $u_5(x, t)$, $u_6(x, t)$ and $u_7(x, t)$ of the fourth order nerve equation, Kink-Type solutions: Visualised for particular values of parameters}
    \label{fig2c}
\end{figure}

\noindent The 3D graphs of Kink-Type solutions $u_5(x, t), u_6(x, t)$ and $u_7(x, t)$ are illustrated in Figure (\ref{fig2c}). The type of sech-Kink-Type solutions of the form (\ref{sol1s}) based on the values of parameters arises in this analysis.\\

\pagebreak

\begin{figure}[h]
    \centering
    \begin{minipage}{0.45\textwidth}
        \includegraphics[width=\linewidth]{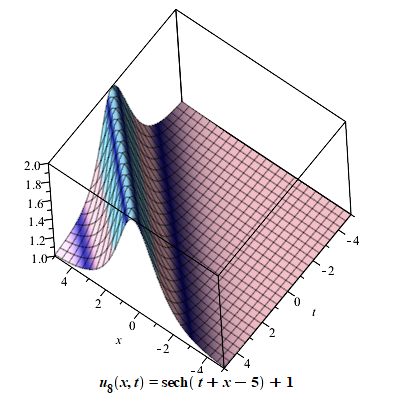}
        \includegraphics[width=\linewidth]{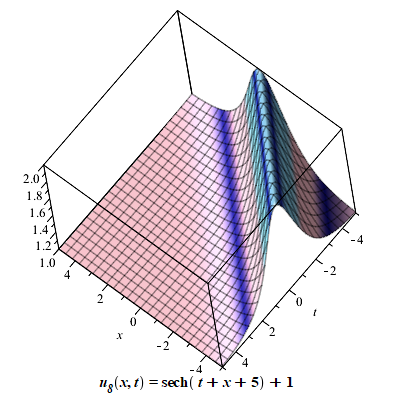}
        \caption{ Plot of $u_8(x, t)$}
            \label{fig2d1}
    \end{minipage}
    \hfill
     \begin{minipage}{0.45\textwidth}
        \includegraphics[width=\linewidth]{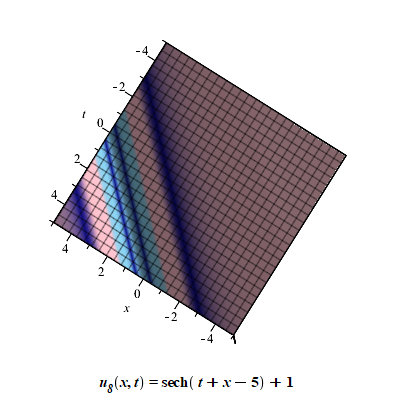}
         \includegraphics[width=\linewidth]{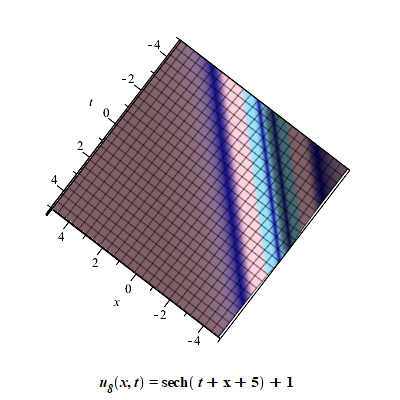}
        \caption{ Plot of $u_8(x, t)$}
            \label{fig2d2}
    \end{minipage}
\end{figure}

\noindent The 3D graphs of soliton solution $u_8(x, t)$ is illustrated in Figure-(\ref{fig2d1}),(\ref{fig2d2}).\\

 \end{document}